\documentclass[graybox]{svmult}

\usepackage{hyperref}
\usepackage{type1cm}
\usepackage{makeidx}
\usepackage{graphicx}
\usepackage{multicol}
\usepackage[bottom]{footmisc}
\usepackage{newtxtext}
\usepackage[varvw]{newtxmath}
\usepackage{hyperref}
\usepackage{listings}
\usepackage{tcolorbox}
\usepackage{xcolor}
\makeindex

\newcommand{\OmegaGas}{\ensuremath{\Omega_{\rm neutral \ gas}}}

\newcommand{\OmegaDust}{\ensuremath{\Omega_{\rm dust}}}

\newcommand{\rhocrit}{\ensuremath{\rho_{\rm crit, 0}}}

\newcommand{\rhogas}{\ensuremath{\rho_{\rm neutral \ gas}}}

\newcommand{\dtg}{\ensuremath{{\rm DTG}}}

\newcommand{\la}{\lesssim}
\newcommand{\ga}{\gtrsim}

\def\lya{Ly-$\alpha$}

\def\mgii{Mg~{\sc ii}}

\begin{document}

\title*{The Multi-Scale Multi-Phase Circumgalactic Medium: Observed and Simulated \\
\small{Lecture notes for the 52$^{\rm nd}$ (March 2023) Saas-Fee Advanced School, Switzerland. 
%See https://www.astro.unige.ch/saasfee2023/
}}
\titlerunning{The Multi-Scale Multi-Phase Circumgalactic Medium: Observed and Simulated}

\author{C\'eline P\'eroux \& Dylan Nelson}
\institute{C\'eline P\'eroux \at European Southern Observatory, Karl-Schwarzschild-Str. 2, 85748 Garching-bei-M\"unchen, Germany \& \at Aix Marseille Universit\'e, CNRS, LAM (Laboratoire d'Astrophysique de Marseille) UMR 7326, 13388, Marseille, France \email{celine.peroux@gmail.com} 
\and Dylan Nelson \at Universit\"{a}t Heidelberg, Zentrum f\"{u}r Astronomie, ITA, Albert-Ueberle-Str. 2, 69120 Heidelberg, Germany \email{dnelson@uni-heidelberg.de}}

\maketitle

\abstract{These are exciting times for studies of galaxy formation and the growth of structures. New observatories and advanced simulations are revolutionising our understanding of the cycling of matter into, through, and out of galaxies. This chapter first describes why baryons are essential for galaxy evolution, providing a key test of $\Lambda$-Cold Dark Matter cosmological model. In particular, we describe a basic framework to convert measurements of the gas properties observed in absorption spectra into global estimates of the condensed (stars and cold gas) matter mass densities. We then review our current understanding of the cycling of baryons from global to galactic scales, in the so-called circumgalactic medium. The final sections are dedicated to future prospects, identifying new techniques and up-coming facilities as well as key open questions. This chapter is complemented with a series of hands-on exercises which provide a practical guide to using publicly available hydrodynamical cosmological simulations. Beyond providing a direct connection between new observations and advanced simulations, these exercises give the reader the necessary tools to make use of these theoretical models to address their own science questions. Ultimately, our increasingly accurate description of the circumgalactic medium reveals its crucial role in transforming the pristine early Universe into the rich and diverse Universe of the present day.
%Each chapter should be preceded by an abstract (no more than 200 words) that summarizes the content.
}

\vspace{2cm}
\newpage

\begin{overview}{Overview}
\begin{enumerate}
\item{The Baryon Census }
    \begin{enumerate}
    \item{Why is Baryon Physics Key? - {\it Hands-on \#1}}
    \item{The Baryon Cycle - {\it Hands-on \#2}}
    \item{The Circumgalactic Medium}
    \end{enumerate}
\item{How to Observationally Probe the Baryons?}
    \begin{enumerate}
    \item{Physics of Quasar Absorbers - {\it Hands-on \#3}}
    %col dens, EW, Voigt Profile, CoG
    \item{Quasar Absorbers in a Cosmological Context}
    %n(z), dz, g(z), f(N)
    \item{Mass Density}
    % Omega, intro sub-DLAs, 
    \end{enumerate}
\item{Cosmic Evolution}
    \begin{enumerate}
    \item{Evolution of Cold Gas}
    \item{Evolution of Metals}
    \item{Evolution of Dust}
    \end{enumerate}
\item{The Galactic Baryon Cycle}
    \begin{enumerate}
    \item{Observational Techniques}
    \item{Metal Abundance Determination}
    \item{Geometrical Argument}
    \item{Mass Loading Factor - {\it Hands-on \#4}}
    \item{Elusive Accretion}
    \end{enumerate}
\item{Future Avenues}
    \begin{enumerate}
    \item{The Hot CGM Gas }
    \item{The Power of Statistics}
    \item{IGM Tomography}
    \item{Intensity Mapping}
    \end{enumerate}
\item{Open Questions \& Conclusions}
    \begin{enumerate}
    \item{The Global Baryon Cycle}
    \item{The Galactic Baryon Cycle}
    \item{Conclusions}
    \end{enumerate}
\end{enumerate}
\end{overview}

\clearpage
%%%%%%%%%%%%%%%%%
%%%%%%%%%%%%%%%%%
%SECTION 1
%%%%%%%%%%%%%%%%%
%%%%%%%%%%%%%%%%%

\section{The Baryon Census}
\label{sec:census}

\begin{figure}[t]
\centering
\includegraphics[width=0.9\linewidth]{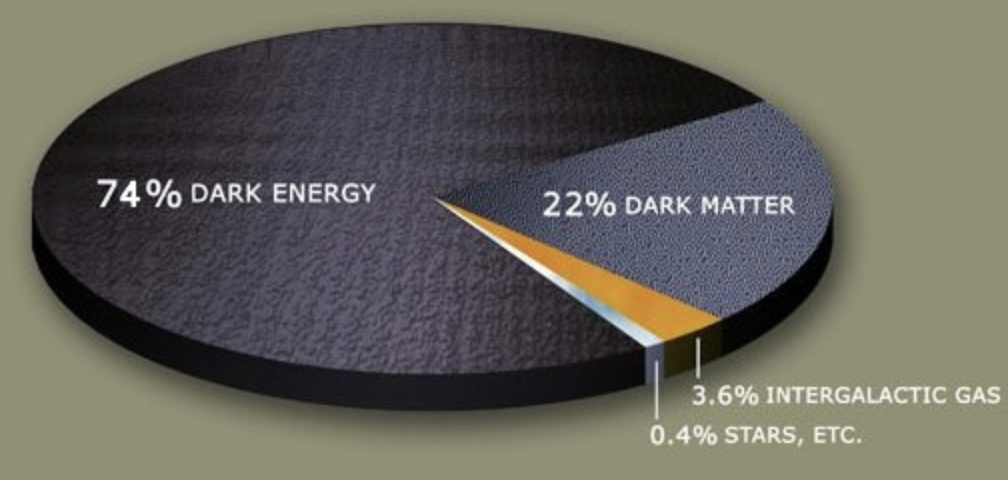}
\caption{A visual representation of the Universe's matter-energy budget. Dark energy is an unknown form of energy which affects the largest scales of the Universe. The $\Lambda$-CDM model postulates the existence of an elusive matter component, dubbed dark matter, which together with the dark energy dominate the Universe's component budget. Baryonic matter refers to the ordinary, non-dark matter, material. Importantly, only a minority of the baryonic matter can be probed by the observations of the light from stars within galaxies. A central tenet is that the vast majority (90\%) of the normal matter is in the form of intergalactic gas. }
\label{f:baryon_census}
\end{figure}

While tremendously successful, the $\Lambda$-Cold Dark Matter ($\Lambda$-CDM) model of cosmology remains incomplete. Indeed, 95\% of the expected matter-energy content is in a form which is currently unknown. The $\Lambda$-CDM model posits the need of an elusive matter component, coined dark matter, which together with the dark energy dominate the Universe's matter-energy budget. Figure~\ref{f:baryon_census} provides a visual representation of this. 

Another challenge comes from the formation of structure in the Universe which is based on a well accepted cosmological framework. In a nutshell, matter and radiation initially coupled in the hot, post big-bang plasma where matter was distributed evenly throughout the cosmos. Small primordial density fluctuations at the epoch of the Cosmic Microwave Background (CMB) are amplified with cosmic time through gravitational instability. We currently have limited empirical information about this epoch dubbed ‘dark ages’. This is the period, however, at which the framework for the large-scale structures of the Universe begins to assemble. Dark matter produced the first cosmic web structures through collapse, while ordinary baryonic matter is able to gravitationally follow this structure. As fluctuations grow, they create a network of filaments of dark matter and an overdense gaseous medium. With only hydrogen and helium in atomic form and no metals yet, this epoch of structure formation gives rise to the first (Population III) stars which, to this date, remain challenging to observe. These stars are however important both because they likely produce the first ionising photons that would then reionise the Universe, but also because they pollute the pristine interstellar media with metals \cite{Welsh23}. 

\begin{figure}[h]
\centering
\includegraphics[width=0.75\linewidth]{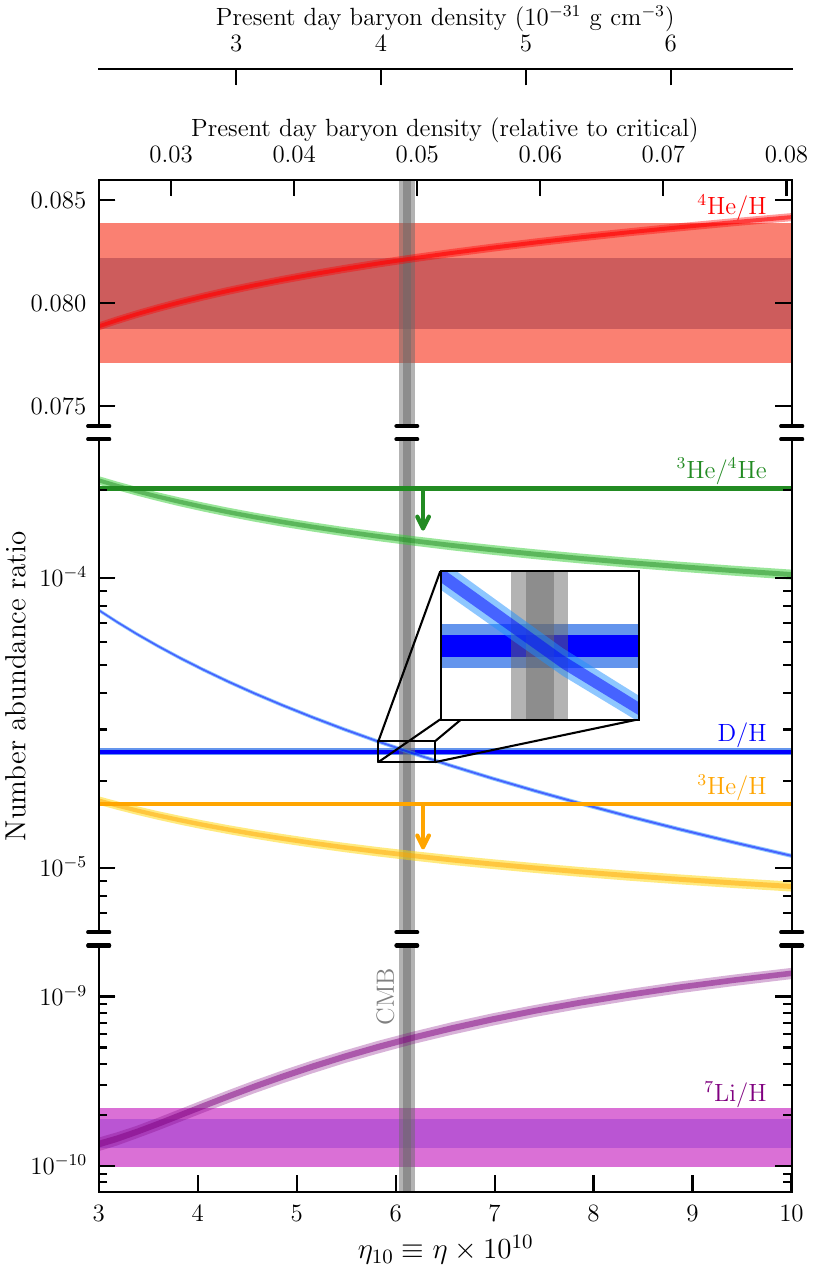}
\caption{Observations of light element abundances. These measurements, together with Big Bang nucleosynthesis, provide an estimate of baryonic matter density of $\Omega_{\rm baryons} \approx 0.0490$, in agreement with both Cosmic Microwave Background (CMB) anisotropies \cite{planck2016} and dispersion measures in Fast Radio Bursts (FRBs) estimates \cite{Macquart20}. The total amount of baryons in the Universe is thus well-constrained \cite{Cooke24}.}
\label{f:nucleosynthesis}
\end{figure}

Galaxy formation continues through Dark Matter halos growth and merging; baryonic matter accretes through cosmic web filaments, and fuels star formation and galaxy formation to make the first objects already observable at z$\sim$10 \cite{Carniani24}. The gas is of prime interest since it will flow along the filaments and feed the formation of galaxies, groups and clusters. However, several physical processes affect the dynamics, thermodynamics and composition of the gas. First of all, filaments are heated by gravitational contraction and shocks \cite{Kang05}. Second, the ultraviolet background from young stars and quasars photoionises this gas and heats it \cite{Gnedin10, Hoeft06}. At the same time, cooling effects in the densest regions, radiate the thermal energy of the gas. Finally, supernovae-driven winds enrich and heat this gas in the neighbourhood of star-forming galaxies \cite{Cen06}. Both hydrodynamical cosmological simulations and observations indicate that the rate of star formation is to first order predicted by the amount of cold, dense gas available at any cosmic epoch (see Section \ref{subsec:baryons}). These results help us understand how efficiently (in terms of conversion of gas into stars) and where exactly stars form.

\subsection{Why is Baryon Physics key?}
\label{subsec:why}

\begin{figure}[t]
\includegraphics[width=0.55\linewidth]{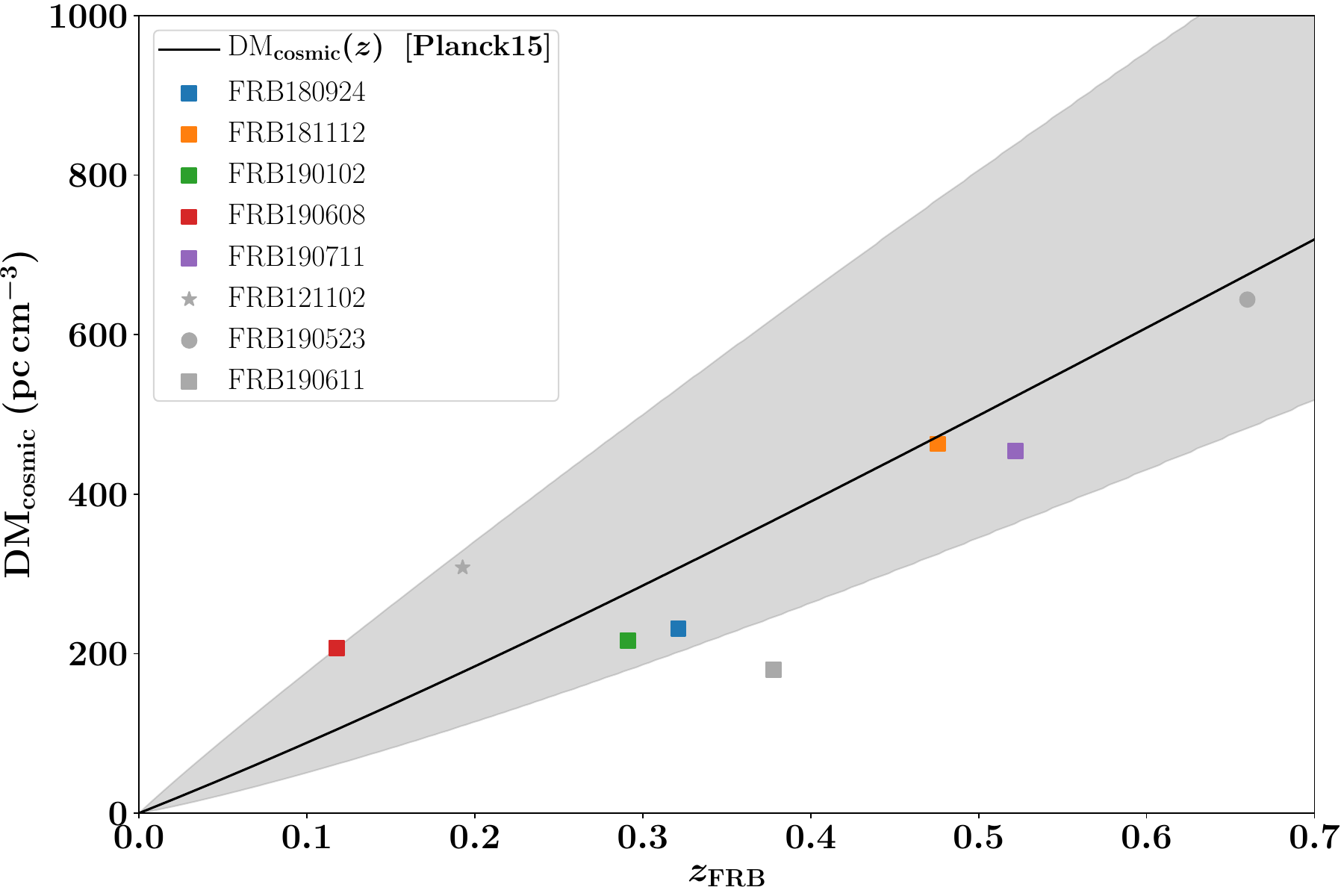}
\includegraphics[width=0.45\linewidth]{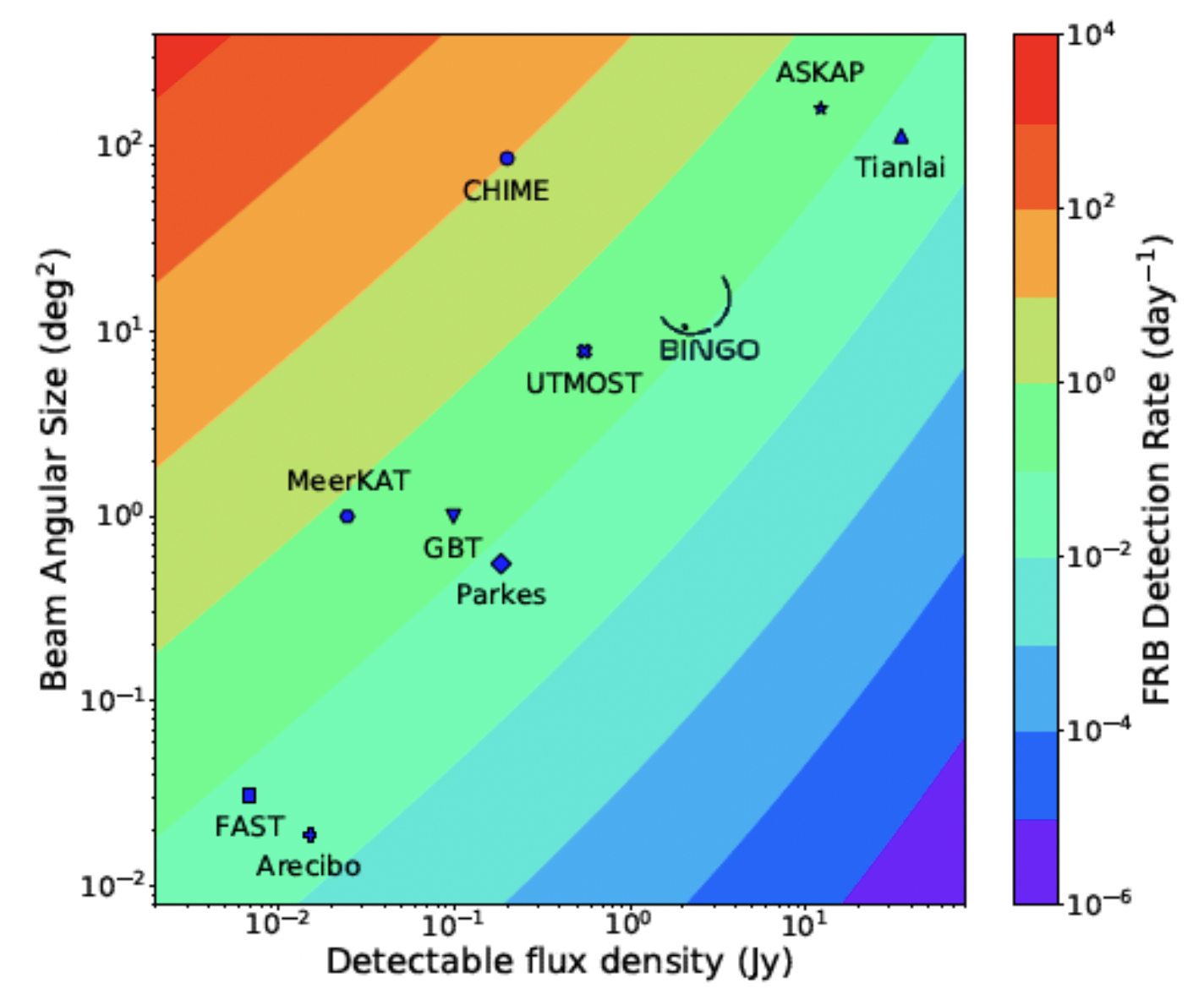}
\caption{{\bf Left:} Recently, estimates of Dispersion Measures (DM) of Fast Radio Bursts (FRBs) with known redshifts has resulted in a third independent estimate of $\Omega_{\rm baryons}$ \cite{Macquart20}. {\bf Right:} Such studies will flourish as more observations are accumulating fast with upcoming facilities including CHIME which leads to tens of FRBs detections a day \cite{Abdalla22}.}
\label{f:FRB}
\end{figure}

Baryonic matter refers to the ordinary, non-dark matter, material. The total amount of baryons in the Universe has been quantified by three independent sets of observations, all of which converge to similar value of $\Omega_{\rm baryons} \approx 0.0490$. On one hand, Cosmic Microwave Background (CMB) anisotropies \cite{planck2016} provides a measure of $\Omega_{\rm baryons}$. Additionally, observations light element abundances together with Big Bang nucleosynthesis \cite{cooke2018,Mossa20, Cooke24} lead to a measure of the baryon density (Figure~\ref{f:nucleosynthesis}). Indeed, the first
synthesis of light elements (such as deuterium, helium and lithium) happened in the early Universe while heavier elements have been produced through stellar
nucleosynthesis. Observations can
be used to determine the primordial abundances of elements formed in
the Big Bang, which provides a unique measure of the baryonic density of the
Universe, $\Omega_{\rm baryons}$. Recently, \cite{Macquart20} estimated Dispersion Measures (DM) of Fast Radio Bursts (FRBs) with known redshifts resulting in a third independent estimate of $\Omega_{\rm baryons}$ (Figure~\ref{f:FRB} left panel). Such studies will flourish as more observations are accumulating fast with upcoming facilities including CHIME which will lead to tens of FRB detections a day (Figure~\ref{f:FRB} right panel, \cite{Abdalla22}).

\begin{figure}[t]
\centering
\includegraphics[width=0.7\linewidth]{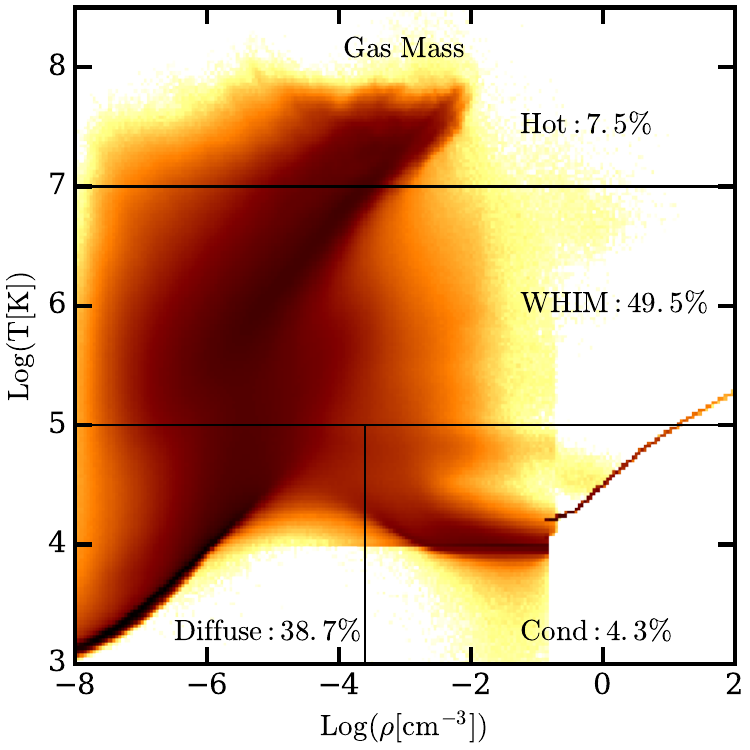}
\caption{Simulated phase-diagram at redshift z=0. The figure displays the density distribution of simulated particles in a temperature-density parameter space. Star formation is labeled "cond" for condensed material, the cold diffuse gas as "diffuse", the Warm-Hot Intergalactic Medium is refered to as WHIM, while hot gas in groups of galaxies and galaxy clusters are labeled "hot". The fractions of gas-mass within each region, whose boundaries are indicated with black lines, is provided on the plot. This figure evidences the large contribution of the so-called WHIM to the matter budget \cite{torrey19}.}
\label{f:phase_diag}
\end{figure}

\begin{figure}[h]
\centering
\includegraphics[width=0.7\linewidth]{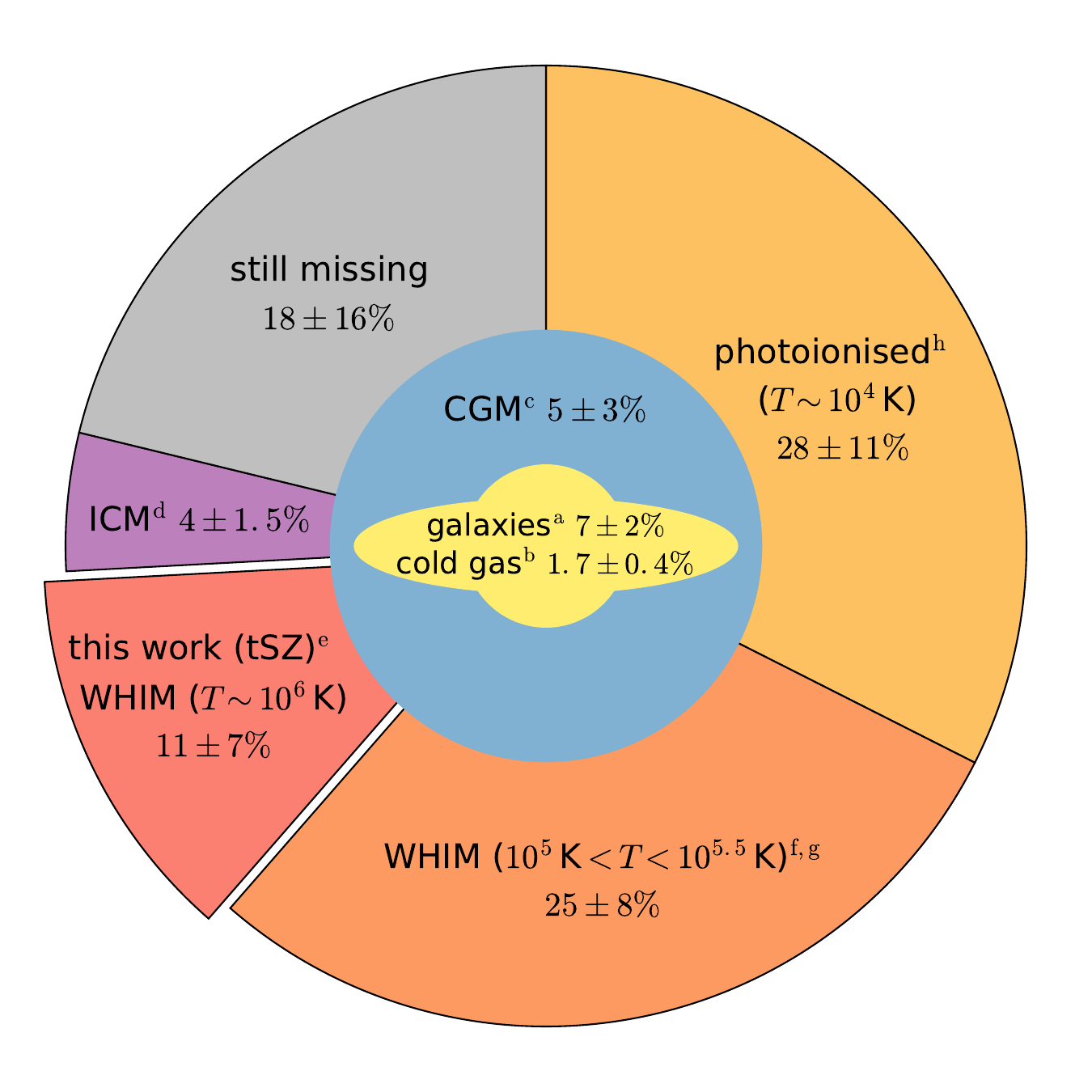}
\caption{Current census of baryons. Low-redshift budget of the baryonic matter probed by observations. About 20\% of baryons have not yet been located or most importantly, their physical properties have not been characterised. These are referred to as the “missing baryons” \cite{DeGraaff19}.}
\label{f:missing_baryons}
\end{figure}

Importantly, only a minority of the baryonic matter can be probed by the observations of light in galaxies. The vast majority (90\%) of the normal matter is in the form of gas as predicted by simulations in a typical phase diagram (Figure~\ref{f:phase_diag}). This gas, notably the cold gas, provides the reservoir of fuel for forming stars and ultimately planets. A large reservoir of this gas still remains elusive and up to 20-50\% of the baryons are thought to be hidden into the so-called Warm-Hot Intergalactic Medium (WHIM) often referred to as the “missing baryons” \cite{Cen99, Dave01, Driver21} as illustrated in \ref{f:missing_baryons}. Accompanying {\it hands-on \#1} provides a starter guide to access similar particle properties in the TNG simulations \cite{Nelson13}. It also proposes a number of exercises to naviguate the interface to the simulations for the reader to get familiar with the working framework.

The current $\Lambda$-CDM paradigm successfully describes cosmological evolution of the large-scale structures of the Universe, organised in galaxies, clusters, sheets, and filaments separated by voids. However, $\Lambda$-CDM over-predicts the amount of small-scale structures, including the number of dwarf galaxies, a problem referred to as the “missing satellites problem” \cite{Bullock10, Perivolaropoulos22}. The lack of observed small-scale structures may imply the existence of physical mechanisms which would suppress small-scale fluctuations in particular. The missing dwarf galaxies in cosmological simulations \cite{Bullock10} relates to the poorly resolved baryonic processes on the small-scales, including gas inflows and outflows. 

\begin{figure}[h]
\centering
\includegraphics[width=0.75\linewidth]{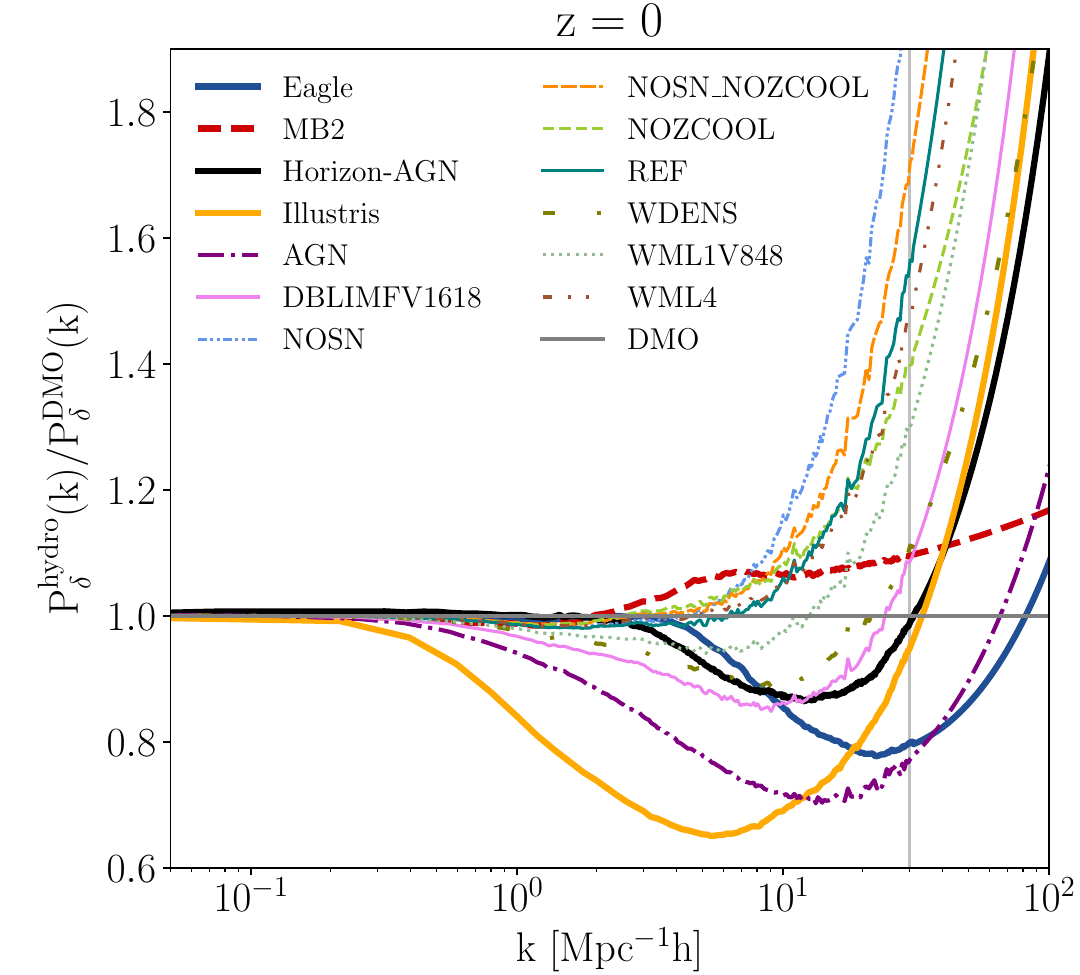}
\caption{Dark Matter power spectrum. The figure presents the ratio of the predictions from models including baryonic physics (labelled "hydro") to Dark Matter Only (labelled "DMO"). The lines display various state-of-the-art cosmological hydrodynamical simulations. Both the sign and amplitude of the difference between these predictions is important, especially at large k-values, corresponding to small physical scales where baryonic physics dominate \cite{Huang19}.}
\label{f:power_spectrum}
\end{figure}

Furthermore, constraining the matter distribution in the Universe and its evolution with cosmic time from the next generation of galaxy surveys will rely on various type of measurements. Weak gravitational lensing of galaxies in particular offers a promising avenue by measuring per-cent-level distortions of galaxies' ellipticities which are caused by bending of the path of photons due to the effects of gravity \cite{Mandelbaum18, Mellier24}. These distortions map the distribution of matter in the Universe at various epochs and thus set new constraints on the physical properties of dark energy, theories of gravity, as well as the nature of dark matter. \cite{VanDaalen11} showed that the distribution of matter, which an inherent input to weak lensing analyses, is significantly affected by baryonic effects. The major impact takes place at Mpc-scales, where the power is suppressed due to active galactic nuclei-driven gas ejection. Cosmological hydrodynamical simulations make predictions of these physical processes at the relevant scales and their impact on the total matter power spectrum. Predictions vary vastly from one model to another, in part because the numerical methods differ but also because of the various implementation of baryonic (‘sub-grid’) processes (Figure~\ref{f:power_spectrum}). These effects have been major topic of research in recent years \cite{Semboloni11, Chisari18, Foreman20, Chisari18, Schneider15, Schneider19, Huang19, Debackere20, vanDaalen20,  Amon22, Salcido23, Arico23}. Indeed, baryons are known to have effect on haloes: they compress the mass near the center, but suppress the density beyond the core which results in a reduction of the halo mass (and hence mass function) that becomes weaker towards higher masses \cite{Velliscig14, Cui14, Cusworth14, Bocquet16, HernandezAguayo23}. Currently, there is relatively little quantitative agreement among simulations regarding a robust description of the properties of gas flows, and specifically the implementation of feedback processes. Only detailed observations of star-forming and active galactic nuclei-driven winds (leading to measurements of velocity, gas phase, mass loading factor) will better constrain these complex physical processes. Observations of the baryonic matter are thus a unique means of tackling this degeneracy and providing new constraints on feedback models. Therefore, the modelled baryonic physics impact the properties of warm dark matter and thus the inference of cosmological parameters from weak lensing measurements from KIDS \cite{Janis19}, DES \cite{Abbott22} and the next generation of surveys such as LSST \cite{Ivezic19}, Euclid \cite{Amendola18} and Roman/WFIRST \cite{Spergel15}. 

%%%%%%%%%%%%%%%%%

\clearpage
\begin{trailer}{Hands-on to analyzing cosmological galaxy formation simulations}

These hands-on sections walk through a ``getting started'' guide for analyzing cosmological hydrodynamical simulations of galaxy formation -- like IllustrisTNG. Additional documentation and further information is available at \url{www.tng-project.org/data}.

The content is split into four sections, with a brief setup first.

\begin{enumerate}
\item A first plot from the Group Catalog
\item Galaxy population relations and integral properties
\item Observables: predicting gas absorption/emission
\item The baryon cycle: measuring mass flow rates
\end{enumerate}

\subsection*{[0] Setup}

First, we import the helper scripts for loading the simulation data:

\begin{tcolorbox}
\begin{verbatim}
import illustris_python as il
import matplotlib.pyplot as plt
import numpy as np
\end{verbatim}
\end{tcolorbox}

\subsection*{[Hands-on \#1] A first plot from the Group Catalogs}

Define the path for the data, i.e. choose the simulation we want to work with. Always start with a low resolution simulation for testing, as the size of data (i.e. time needed to load and plot) is smaller. Once you have a piece of code working, you can always change the path to a higher resolution simulation.

\begin{tcolorbox}
\begin{verbatim}
basePath = 'sims.TNG/TNG100-3/output/'
\end{verbatim}
\end{tcolorbox}

We can then load fields from the $z=0$ group catalog. We specify the redshift by the snapshot number:

\begin{tcolorbox}
\begin{verbatim}
snap_number = 99
\end{verbatim}
\end{tcolorbox}

We can load any field which is available (i.e. pre-computed) in the group catalogs. In particular, plot the relationship between star formation rate and stellar mass for galaxies. To do so, we will load both quantities, for all subhalos:

\begin{itemize}
  \item  masses "by type" (SubhaloMassInRadType),
  \item  star formation rates (SubhaloSFRinRad).
\end{itemize}

Note: the "InRad" suffix indicates that both measurements are considering only particles/cells within twice the stellar half mass radius, which is an OK first definition for the size of a galaxy. Other measurements of these same quantities, following other definitions, are also available.

\begin{tcolorbox}
\begin{verbatim}
fields = ['SubhaloMassInRadType','SubhaloSFRinRad']

subhalos = il.groupcat.loadSubhalos(basePath, snap_number, 
                                    fields=fields)
\end{verbatim}
\end{tcolorbox}

Then we can inspect the result:

\begin{tcolorbox}
\begin{verbatim}
subhalos.keys()
dict_keys(['count', 'SubhaloMassInRadType', 
           'SubhaloSFRinRad'])

subhalos['count']
118820

subhalos['SubhaloMassInRadType'].shape
(118820, 6)
\end{verbatim}
\end{tcolorbox}

Inspecting the return, we see it is a normal dictionary. The "count" indicates that there are about 100k total subhalos (in TNG100-3, versus 4.4 million in TNG100-1). Each requested field is returned as a numpy array.

We can now make a plot of star formation rate versus stellar mass. To do so, we need to select the values of interest from the catalog arrays, and be careful of physical unit conventions. (Always check the units and definition of fields before use).

\begin{tcolorbox}
\begin{verbatim}
# select which particle type we are interested in
mass_stars_code = subhalos['SubhaloMassInRadType'][:,4]

# careful of units!
mass_stars_msun = mass_stars_code * 1e10 / 0.6774

sfr_msun_per_yr = subhalos['SubhaloSFRinRad']

# plot
fig, ax = plt.subplots()
ax.plot(mass_stars_msun, sfr_msun_per_yr, '.', ms=1.5)
ax.set_xscale('log')
ax.set_yscale('log')
ax.set_xlabel('Galaxy Stellar Mass [M$_\odot$]')
ax.set_ylabel('Star Formation Rate [M$_\odot / yr$]');
\end{verbatim}
\end{tcolorbox}

\begin{figure}
\centering
\includegraphics[width=0.9\textwidth]{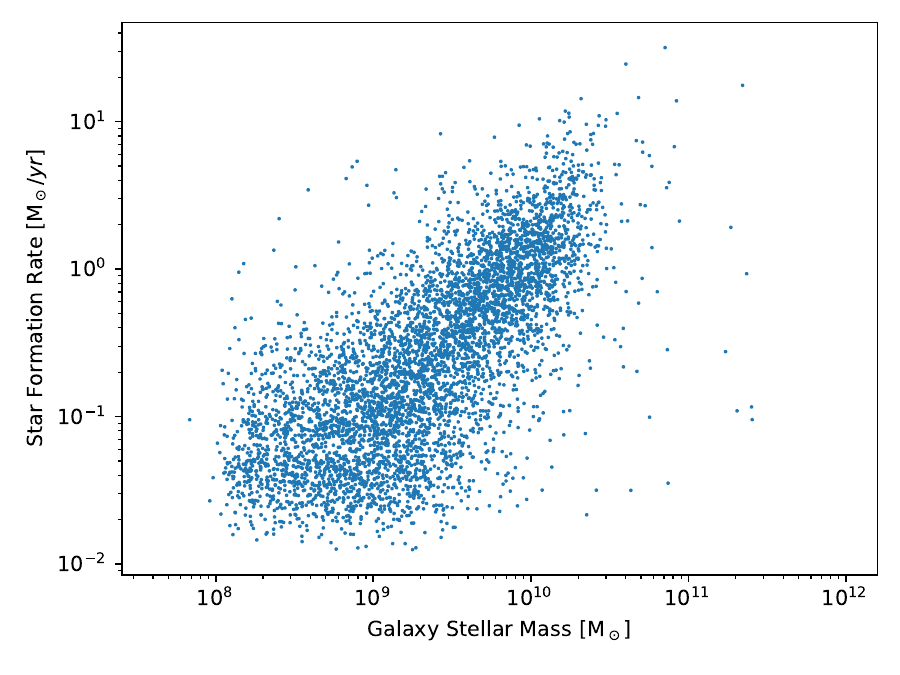}
\caption{\textbf{Hands-on plot:} the relationship between galaxy star formation rate and galaxy stellar mass, for the TNG100-3 simulation at $z=0$.}
\label{fig_handson_1}
\end{figure}

\subsection*{Exercise}

* 1. Change the simulation above (e.g. to a higher resolution, or to a different TNG box), and remake the plot. Do you see any different features? What is your interpretation?

Caution: are there any subhalos with `SFR == 0`? They would have gotten lost due to the y-log axis. Find them, and include them on the plot. What is your interpretation of these subhalos?

* 2. Quenched galaxies have largely stopped forming new stars. There are many definitions of quenched: one is that the specific star formation rate $\rm{sSFR} = \rm{SFR} / M_\star$ is below a constant threshold value of $10^{-11} \rm{yr}^{-1}$ (at $z=0$). Compute and plot sSFR versus stellar mass for all galaxies. Add a horizontal line for this threshold. Then, measure the fraction of quenched galaxies as a function of mass (i.e. in $N$ bins of stellar mass) and plot quenched fraction versus stellar mass. Does it make sense?

\end{trailer}

\clearpage

%%%%%%%%%%%%%%%%%

\subsection{The Baryon Cycle}
\label{sec:cycle}

Galaxies are not isolated islands. Their interactions with their environment profoundly influence their evolution. They form as gas cools and condenses at the centres of a population of massive halos growing by gravitational amplification of fluctuations in an initially near-uniform distribution of pre-existing dark matter. The canonical picture has galaxy growth being fed by inflows of gas from the intergalactic medium, IGM \cite{Dekel09, vandeVoort11}. Gas acquired through mergers or through accretion of IGM replenishes the fuel needed for star formation. These baryons from the cosmic web cool into a dense atomic then a molecular phase, which fuels star formation. These processes are collectively described as a self-regulation, where part of the gas is consumed in star formation while outflows eject some other quantity of gas from the system. Molecular gas is formed by complex small-scale physics including cloud collisions, dynamical/orbital features, or turbulence making the clouds become self-gravitating, and initiating star formation \cite{Gronke2017, gronke18, Gronke2022}. Another important component of these processes is the metal enrichment of the interstellar medium which in turns will affects
the efficiency of star formation since not only do metals
act as coolants but they also are fundamental to dust production by shielding molecules from dissociating radiation. 

\begin{figure}[t]
\centering
\includegraphics[width=0.60\linewidth]{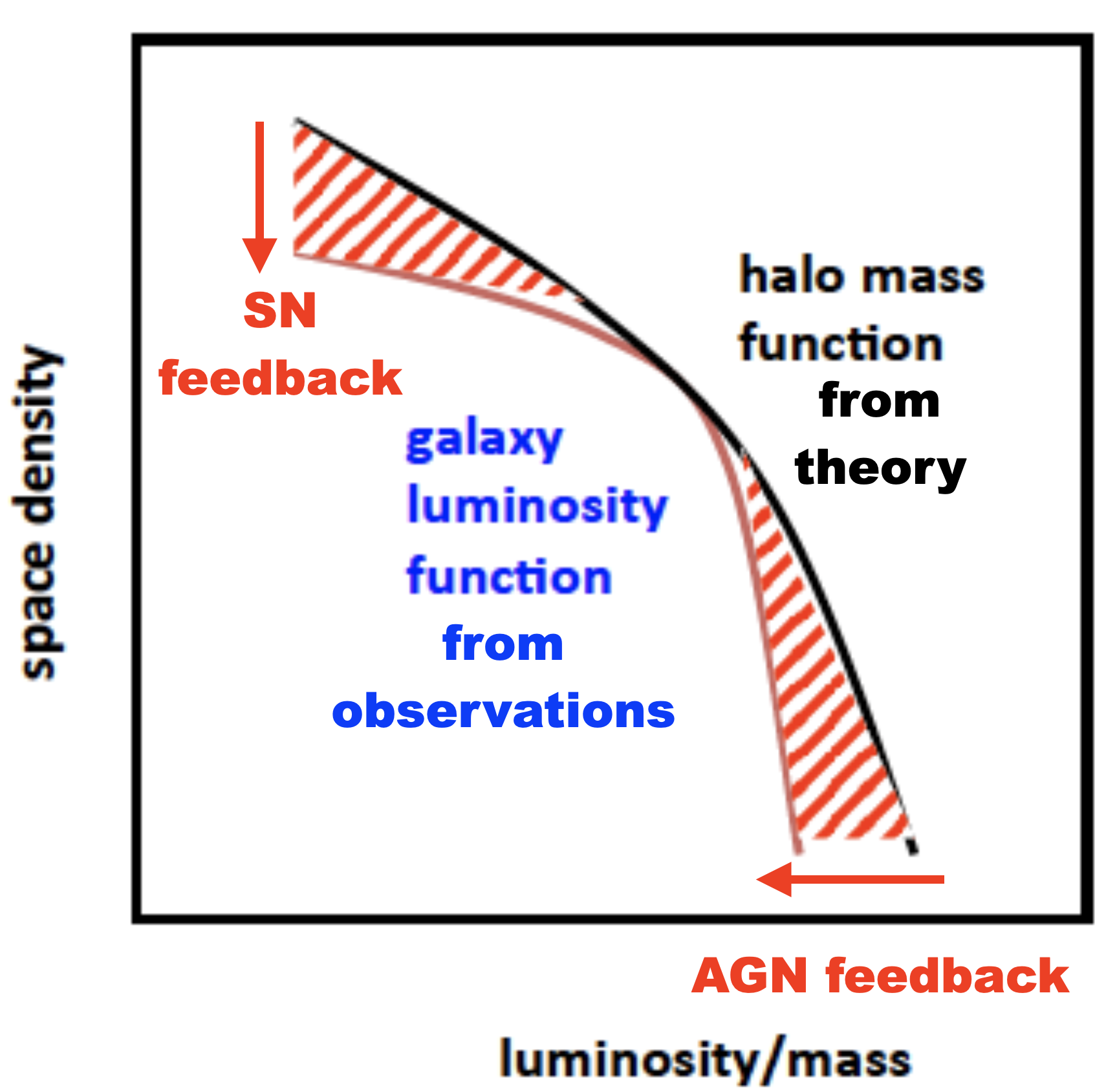}
\caption{The expected (black line) and observed (red line) galaxy luminosity function. The discrepancies in the low- and high-mass ends is related to Supernovae and Active Galactic Nuclei feedback, respectively \cite{Silk12}.}
\label{f:LumFctFdbck}
\end{figure}

The $\Lambda$-CDM cosmology also makes strong predictions on the halo mass function. Departures from these expectations with respect to the observed stellar mass function of galaxies are apparent at both the low and the high mass ends \cite{Silk12}. These deviations indicate a lower star formation efficiency in these two different mass regimes (Figure~\ref{f:LumFctFdbck}). To reconcile predictions from theory with observations, current simulations invoke feedback from various astrophysical processes at both mass scales. In low-mass galaxies, models include supernovae feedback to reproduce the observations while at high masses, feedback from an accreting super-massive black holes is required. The canonical picture is thus that once stars are formed, galaxies enrich the intergalactic medium with ionising photons and heavy elements formed in stars and supernovae, by driving galactic and active galactic nuclei-driven winds into the surrounding \cite{Pettini03}, some of which will fall back onto the galaxies in so-called galactic fountains \cite{Fraternali17, Bish19}. Gas flowing out of the galaxy (winds driven by supernovae or a super-massive black hole in the galaxy center) quenches star formation. Yet another important open question for galaxy formation models is to reproduce the low star formation efficiency observed in galaxies and dark matter halos. 

\begin{figure}[t]
\includegraphics[width=0.33\linewidth]{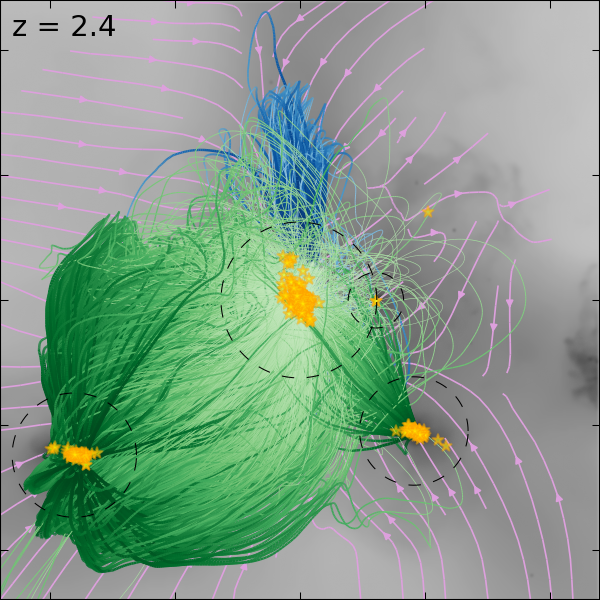}
\includegraphics[width=0.33\linewidth]{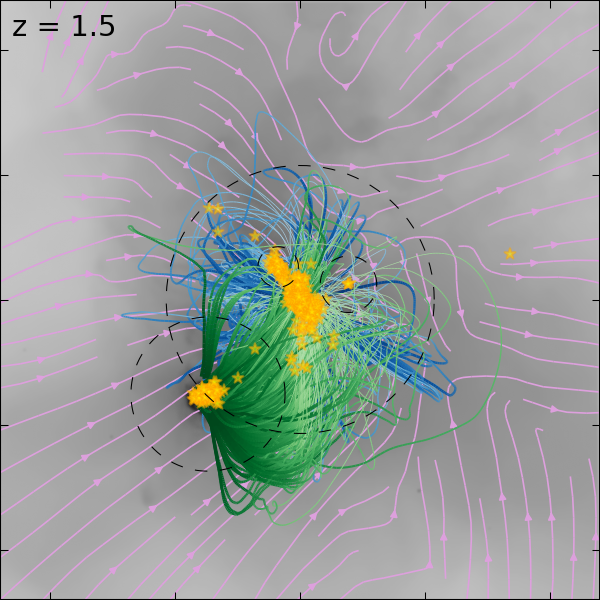}
\includegraphics[width=0.33\linewidth]{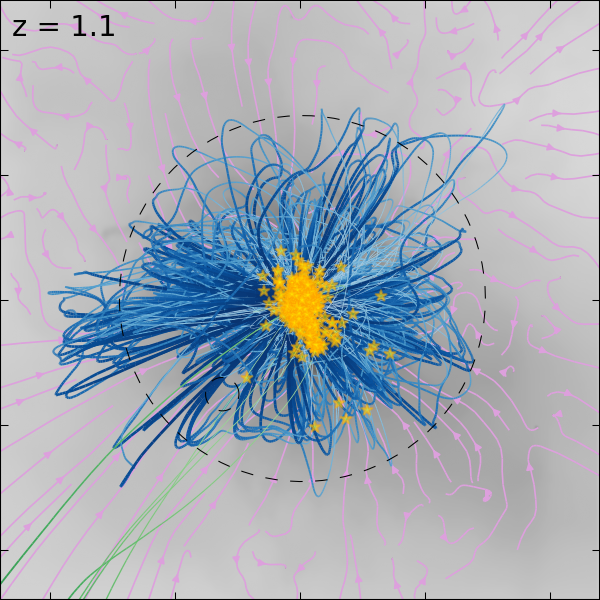}
\caption{Illustration of intergalactic transfer. Hydrodynamical FIRE simulations predict the amount of baryonic material acquired through fresh accretion (purple), wind recycling (blue) or transfer from another haloe (green) as a function of redshift \cite{anglesalcazar17}.}
\label{f:intergalactic_transfer}
\end{figure}

From a theoretical point of view, the difficulty is related to the various scales involved: simulating the large cosmological scales together with the pc-scale typical of interstellar physics poses an major dynamical scale computing challenge \cite{Crain23}. Recently, advances in numerical methods and computing capabilities have enabled extraordinary progress in the simulation of structure formation. Accompanying chapter by Jeremy Blaizot covers most of the fundamental processes implemented in current state-of-the-art simulations.
%{\blue Jeremy's chapter covers most of the fundamental processes implemented in current state-of-the-art simulations.} 
It is a remarkable achievement that the overall physical properties (density and temperature) of the gas (specially HI) in these models reproduce observations of the absorbers column densities (i.e. the number of atoms along the line-of-sight) and line widths \cite{Fumagalli2011, Rahmati2014, gaikwad2017}. However, simulating the multiphase interstellar medium still remain challenging even in “zoom-in” simulations even with modern powerful computers. To overcome this challenge, it is necessary to make use of sub-grid modules to model unresolved physical processes, such as winds from dying stars, and supernovae \cite{teyssier2019, Maio22, Butsky24}. Only by implementing the most realistic physics will simulations be able to interpret contemporary observations. Important questions still to answer include which objects and media contribute to the global quantities. The question remains of the relation between the intergalactic gas relation and the interstellar medium of galaxies, including small-scale cold clumps and eventually molecular gas in the ISM of galaxies. Reaching a full understanding of the cycling of baryons between HI, H$_{\rm 2}$ and the ionised phase of the gas requires even more advanced simulations. {\it Hands-on \#2} provides an opportunity to look at the gas properties in TNG simulations. 

A description of galaxy evolution thus requires to depict fully the ‘baryon cycle’, i.e., how material cycles through different phases (hot, warm and cold) and locations from outside galaxies into the interstellar medium and back \cite{Tumlinson17, FaucherGiguere23}. This baryon cycle is instrumental to the fueling and the regulation of star formation through and accretion and outflowing processes. The availability of the gas reservoir drives the instantaneous star formation rate and therefore directly governs the growth of galaxies and super massive black holes. More globally, the temporal and spatial evolution of material describes these processes of motion and transformation of the baryons \cite{Walter2020, Tacconi2020, PerouxHowk20}. The inflows and outflows exchange gas, metals, and angular momentum between the galaxy and its atmosphere. A detailed probe of baryons exchanges is of paramount importance for understanding these processes (Figures~\ref{f:intergalactic_transfer}). Since gas, stars, and metals are intimately connected, gas flows affect the history of star formation and chemical enrichment in galaxies. Understanding this cosmic {\it baryon cycle} is crucial for understanding galaxy evolution. A comprehensive model of structure formation must thus consider our Universe as one large, complex ecosystem in which galaxies, stars and ultimately planets can form on different scales.

%%%%%%%%%%%%%%%%%

\subsection{The Circumgalactic Medium}
\label{sec:CGM}

In this context, the enriched circumgalactic gas, or CGM, surrounding galaxies provides the most direct
probe of inflows and winds, which are driven
by either stars, supernovae, active galactic nuclei and/or cosmic rays. Ultimately, we would like to measure the mass moves in these processes and how this evolves with time as well as obtain a description on how material enriched in metals, dust and energy eventually cycles back onto the galaxy. New observations and modern simulations reveal that the large, diffuse gas reservoir of the CGM is both multi-scale and multi-phase \cite{FaucherGiguere23}. Recently, it has been suggested the cloudlets of cold gas could be entrained in hotter medium those providing a theoretical framework for the recent observational results \cite{Gronke2022}. Cold CGM gas also plays a role in modulating
halo cooling, act as a reservoir for star formation, and thus affects the galactic feedback processes. The CGM therefore has become a central component of studies of galaxy evolution and there structure formation. Over the past decade, new observations have revolutionised the study of the circumgalactic gas surrounding galaxies \cite{Tumlinson11, Borthakur13, Werk14, Heckman17, Lehner18, Muzahid18, Prochaska19, Chen2020, Peroux2020}. These studies show that the CGM is multiphase (with cold, warm, and hot gas) and makes up a large fraction of the baryon and metal budget. More recently, some works focused on the material closer to the galaxy (i.e. $<$0.5 R$_{\rm vir}$), referring to it as the inner CGM.

The CGM is increasingly recognised for its significant role in driving the evolution of galaxies. Hydrodynamical simulations (e.g., IllustrisTNG \cite{naiman18, pillepich18b, nelson18a, marinacci18, springel18}, EAGLE \cite{crain15, schaye2015, mcalpine2016}, SIMBA \cite{dave16}, FIRE \cite{FaucherGiguere16}) also show the complexity of the CGM. A large fraction of the CGM seems to be bound to the galaxy and may have circulated multiple times through the galaxy (Figures~\ref{f:intergalactic_transfer}). Recently, several groups have implemented different solutions to improve the spatial resolution in the lower density CGM of hydrodynamical simulations and to systematically vary the properties of simulated halos. An increasing number of physical processes, such as galactic winds and jets are being incorporated, essential to increase the realism of the simulations. Cosmic rays, turbulence, and magnetic fields likely also impact CGM physics significantly \cite{pakmor17, vandeVoort2021, Ramesh2023}. Full radiation hydrodynamics simulations in large boxes  are also now becoming feasible, providing insight into the role that radiation and thermal pressure play in regulating the hydrogen and helium reionization processes and the structure of the Lyman-$\alpha$ forest. Finally, efficient emulators that utilise GPU acceleration or machine learning techniques are now being developed for sampling the large parameter space relevant for cosmological analyses \cite{VillaescusaNavarro21, Appleby23, Gebek23}.

%%%%%%%%%%%%%%%%%

\clearpage
\begin{trailer}{Hands-on to analyzing cosmological galaxy formation simulations}

\subsection*{[Hands-on \#2] Galaxy population relations and integral properties}

When a physical quantity that you are interested in is already in the group catalogs (or can be computed from quantities in the catalogs), this is a good starting point.

We consider a "galaxy gas fraction" defined as $f_{\rm gas} = M_{\rm gas} / (M_{\rm gas} + M_{\star})$. For a quick look, we load the first 5 subhalos of TNG100-1 at $z=0$, and print this gas fraction.

\begin{tcolorbox}
\begin{verbatim}
basePath = 'sims.TNG/TNG100-1/output/'
snap = 99
\end{verbatim}
\end{tcolorbox}

\begin{tcolorbox}
\begin{verbatim}
for i in range(5):
  sub = il.groupcat.loadSingle(basePath, snap, subhaloID=i)
  gas_mass = sub['SubhaloMassInHalfRadType'][0] # units?
  stars_mass = sub['SubhaloMassInHalfRadType'][4] # units?
  fgas = gas_mass / (gas_mass + stars_mass)
  print(i, fgas)

0 0.097318634
1 0.0105384765
2 0.014834739
3 0.0027430148
4 0.0047753155
\end{verbatim}
\end{tcolorbox}

Q: Within what radius (or "aperture") is this gas fraction computed?

Q: Why is the first $f_{\rm gas} \sim 10\%$, but the other four are $\lesssim 1\%$?

\subsection*{Exercise}

Make a scatterplot of the gas fraction of galaxies as a function of galaxy stellar mass: use `loadSubhalos()`. Overplot a line representing a running mean (or median). What is your interpretation of the typical behavior with mass?

\subsection*{Exercise}

What if we were interested in the "halo gas fraction" $f_{\rm gas} = (M_{\rm gas} / M_{\rm total})$ instead? This would tell us about the gaseous content of the CGM, rather than of the galaxy itself. What aperture/definition would you want to use? What field from the group catalogs would be most appropriate? Make a scatterplot of halo gas fraction as a function of (i) galaxy stellar mass, (ii) total halo mass.

\subsection*{Exercise}

Until now, we have been plotting all subhalos in the group catalog -- that is, both "centrals" and "satellites", where a satellite is another haloe within the Virial radius of the central. It is always a good idea to \textbf{consider centrals and satellites separately}. Look at the list of halo fields, and subhalo fields, in the group catalog. Which fields provide the links between the two? (There are "links" in both directions).

\begin{enumerate}
\item Create a three lists of subhalo indices (or "IDs"): (a) all subhalos, (b) central subhalos only, and (c) satellite subhalos only.
\item Print out the list of central subhalo IDs. What do the gaps represent?
\item Plot the "satellite fraction" of subhalos as a function of mass.
\item Repeat the gas fraction plot from above, (over)plotting centrals and satellites separately. What is your interpretation?
\end{enumerate}

\subsection*{Exercise (Challenge)}

We have not yet seen the ``merger trees''. We can explore how we follow a subhalo through time, and see how it evolves. We will use the `SubLink` merger tree.

\begin{enumerate}
\item Look at the documentation for the \textsc{il.sublink.loadTree()} function.
\item Consider TNG100-1 at $z=0$. Select all central subhalos whose **parent halo** has $11.5 < M_{\rm 200c} / \rm{M}_\odot < 11.6$. How many are there?
\item For the first $N=10$ of these, load the "main progenitor branch" (MPB) of each. (You should do this in a loop.)
\item Make a plot of gas fraction versus time (i.e. snapshot number, or redshift if possible), overplotting all $N$ halos. Label each, in the legend, with the halo ID.
\item Add the median relation, taking the median across the $N$ halos, as a solid black line. What is your interpretation?
\end{enumerate}

\end{trailer}
\clearpage

%%%%%%%%%%%%%%%%%
%%%%%%%%%%%%%%%%%
%SECTION 2
%%%%%%%%%%%%%%%%%
%%%%%%%%%%%%%%%%%

\section{How to Observationally Probe the Baryons?}
\label{sec:obs}

Only a minority of the baryons are scrutinized by observations of starlight from galaxies. The remaining baryons reside in low-density gas which is challenging  to detect in emission because of the low-density of the gas \cite{wijers2019, augustin2019}. The baryons and metals
associated with the majority of the gas that lies beyond the regions excited by hot stars are often studied using resonant absorption lines in the spectra of background sources. Indeed, it is possible to study distant galaxies through the absorption they produce in the light of yet more distant quasars (Figure~\ref{f:spectrum}, top panel). As this absorption is sensitive to the amount and chemical composition of interstellar gas in the foreground object, it is possible to deduce how galaxies have changed between the time at which they are observed and the present day. This is a particularly powerful technique when using bright background sources such as quasars, gamma ray bursts (GRBs) \cite{bolmer2019}, fast radio burst (FRBs) \cite{prochaska2019}, or even other galaxies \cite{cooke2015, peroux2018}. While most background sources provide information on the atomic or molecular gas densities, the FRBs allow us to estimate the electron column density along the line-of-sight. Indeed, the significant dispersive effect of ionised gas on radio waves allow us to quantify the delay known as the dispersion measure (DM). This effect is mainly due to the electrons along the line-of-sight and thus approximately proportional to the column density of free electrons \cite{Cook23}. The important advantage of absorption techniques over emission from galaxies is the ability to reach low gas densities. Moreover, the sensitivity of this technique is solely sensitive to the background source brightness and thus redshift independent thus offering a powerful tool to study the cosmic evolution of the baryon and metal content across the Universe.

\subsection{Physics of Quasar Absorbers}
\label{subsec:abs}

\subsubsection{Classes of Quasar Absorbers}

\begin{figure}[t]
\centering
\includegraphics[width=0.70\linewidth]{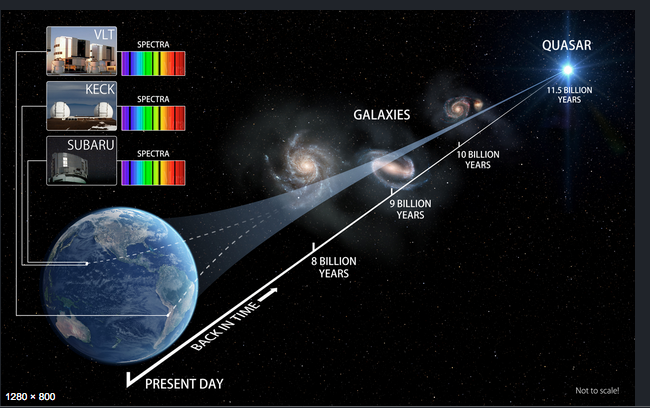}
\includegraphics[width=0.75\linewidth]{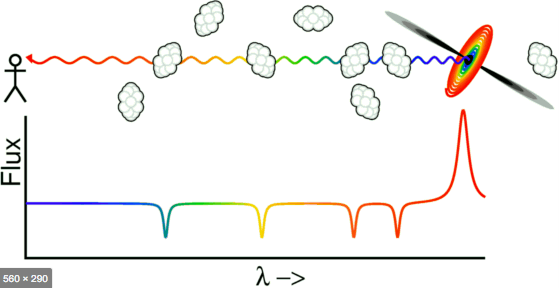}
\caption{Quasar absorption technique. {\bf Top:} Using unrelated high-redshift quasar is a powerful mean to study in absorption the material between the background source and the observer. Since their detection is independent of their luminosity and morphology, these absorbers provide a unbiased method with which to study the circumagalactic medium of early galaxies. {\bf Bottom:} Sketch of a typical quasar spectrum showing flux as a function of wavelength. Bluewards of the \lya\ emission line ($z_{\rm em}$), the absorption systems with wavelength ($z_{\rm abs}$) are caused by \lya\ transitions. }
\label{f:spectrum}
\end{figure}

Figure~\ref{f:spectrum} (bottom panel) displays an illustration of the absorption technique and the sketch of a quasar spectrum. The emission lines are intrinsic to the quasar itself, while the many absorption lines are probing material between the background source and the observer. Bluewards of the \lya\ emission line ($z_{\rm em}$), the absorption systems with wavelength ($z_{\rm abs}$) are caused by \lya\ transitions. Metal lines are routintely detected as well. Since their detection is independent of their luminosity and morphology, these absorbers provide a unbiased method with which to study early galaxies.

\lya\ absorbers are sub-divided into four classes according to their column
density that is the number of hydrogen atoms per unit area along the
line-of-sight between the observer and the quasar (commonly expressed
in atoms cm$^{-2}$). The column density is equivalent to a surface density and is calculated in {\it Hands-on \#3}. Therefore a low column density cloud could either
be a small cloud with high density or a large cloud with low
density. Quasar absorbers thus probe media spanning the range from voids through to halos and disks of both dwarf and normal (proto)galaxies.

\begin{enumerate}

\item {\bf Damped Lyman-$\alpha$ systems} (hereafter DLAs) have N(HI)
$> 2 \times 10^{20}$ atoms cm$^{-2}$. This definition is somewhat
historical since damped wings appear for lower column densities
($N$(H~{\sc i})~$>$~10$^{19}$~cm$^{-2}$; see Figure~\ref{f:curveofgrowth}). This threshold had been introduced
because these lines are characteristic of local galactic disks
\cite{Wolfe86}. Also at the time, damped Ly$\alpha$ surveys were
performed at low resolution and this threshold made them relatively
unambiguous to pick out. The equivalent width is $W_{\rm
obs}$($z$~$\sim$~2.5)~$>$~17.5~\AA~ for $N$(H~{\sc
i})~$>$~10$^{20}$~cm$^{-2}$. The probability that such a strong
absorption feature is the result of blending is small. 

At large column densities, the optical depth at the Lyman limit is large enough for the interior to be {\it self-shielded} from the external radiation field and thus the gas is predominantly neutral (HI). These quasar absorption lines are therefore a powerful diagnostic tool for investigating the chemical composition of high-redshift galaxies (Figure~\ref{f:metals}). Traditionally, these abundances are derived using the following implicit assumption:

\begin{equation}
    [X/H] = \frac{N(\rm X)}{N({\rm H})} = \frac{N(\rm X II)}{N({\rm HI})}
\label{eqn:metallicity}
\end{equation}

where X refers to any element. In addition, lines of different ionisation state allow us to probe the multiphase gas. Similarly, molecules are observed in DLAs \cite{ledoux03, bolmer2019}. Measuring molecules at high-redshift is important because they dominate the cooling function of neutral metal-poor
gas. 

\vspace{0.5cm}

\item {\bf sub-Damped Lyman-$\alpha$ systems} (hereafter sub-DLAs) have
$10^{19} <$ N(HI) $< 2 \times 10^{20}$ atoms cm$^{-2}$. Since the DLA definition is partly historical, it is clear
that it may introduce a systematic bias in the discussion
of the nature of these quasar absorbers. For this reason, other works
generalise the definition in order to compare the properties of
systems with N(HI) $> 10^{19}$ atoms cm$^{-2}$ \cite{peroux2003a}. 
That definition arises from the fact that to capture the whole HI mass in the Universe, one requires to take all systems above this column density threshold into account \cite{peroux2003a, zafar2013b, Berg19}.

Since their relatively low column density implies that some of the gas might be ionised, the ionisation state of the sub-DLAs has been investigated in details in the literature \cite{peroux2003a, Meiring2009}. The results indicate that ionised fraction varies significantly from one system to another. A different question is whether - despite the sometimes high ionisation fraction - the assumption described by equation~\ref{eqn:metallicity} still holds. Indeed, several studies have demonstrated that this is a valid assumption in the sub-DLA regime \cite{dessauges-zavadsky2003, quiret16}.

\begin{figure}[t]
\centering
\includegraphics[width=0.9\linewidth]{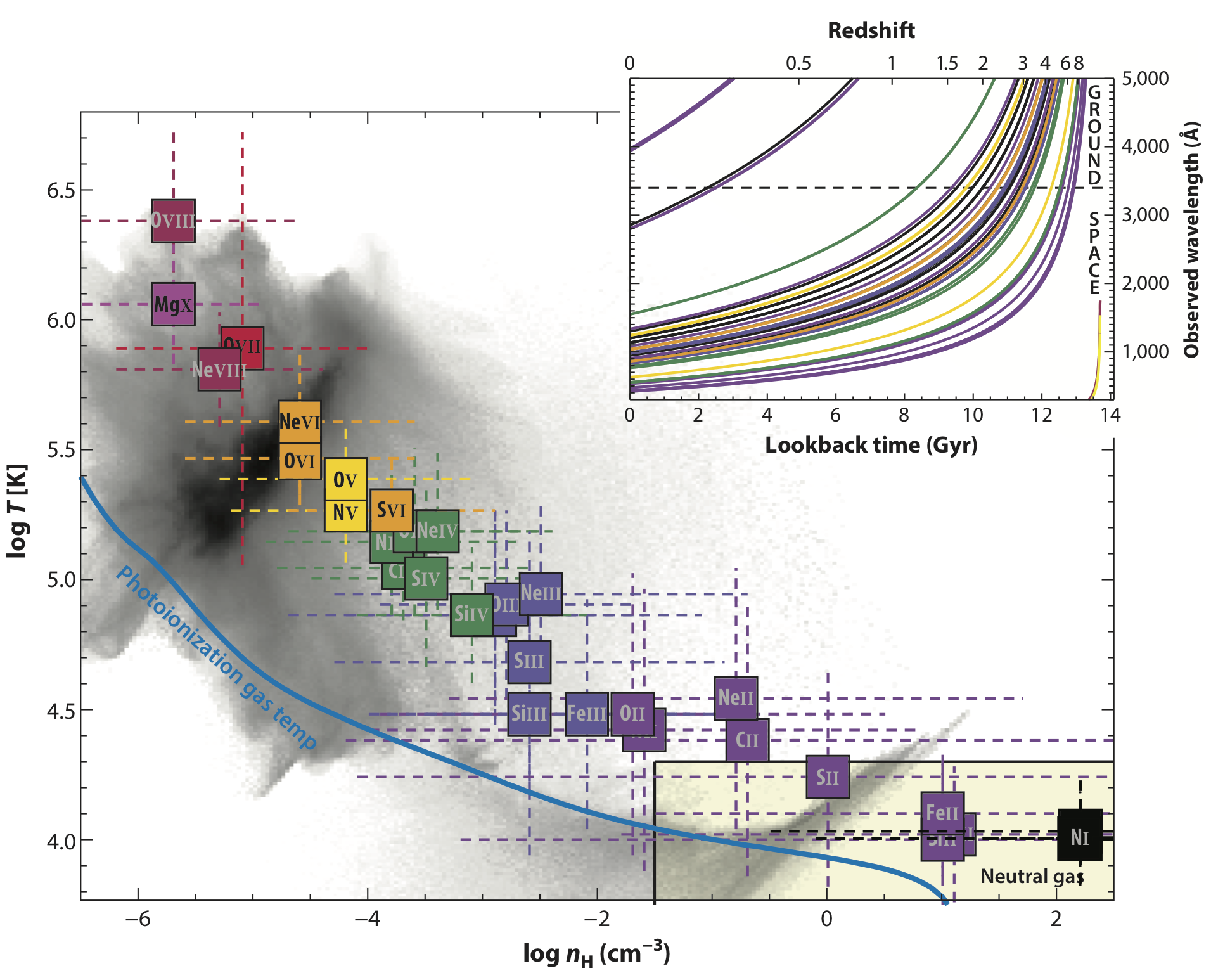}
\caption{A visual rendering of various metal line transitions in a temperature-density plane. This plot highlights the power of quasar absorbers spectroscopy by directly probing multi-phase gas of temperatures and densities ranging several orders of magnitudes \cite{Tumlinson17}. The inset presents the position of the main line diagnostics on a wavelength-redshift plane. This figure emphasizes that the coverage of shorter blue optical wavelengths is key to optimise the number of lines detected.}
\label{f:metals}
\end{figure}

\vspace{0.5cm}

\item {\bf Lyman-limit Systems} (hereafter LLS) have hydrogen column densities N(HI) $> 1.6 \times 10^{17}$ atoms cm$^{-2}$ and are optically thick at the Lyman limit due to the HI photo-ionisation:

\begin{equation}
H^o + \gamma \rightarrow H^+ + e^-
\end{equation}

where the photon energy, 13.6 eV, corresponds to 912 \AA\ rest
wavelength. These absorbers are easily identifiable by their
distinctive break signature in the quasar spectrum. The optical depth, $\tau$, is expressed as
follows:

\begin{equation}
\tau = N(HI) \times \sigma
\end{equation}

where $\sigma$ is the HI photo-ionisation cross-section $6.8 \times
10^{-18}$ cm$^{-2}$. For $\tau_{912} =1$ (optically thick), N(HI)
has to be $= 1.6 \times 10^{17}$ atoms cm$^{-2}$. The optical depth
below $912$ \AA\ is proportional to $\tau_{912}(\lambda / 912)^3$. For
example, if a LLS lies at $z=3$ with a column density N(HI) $=
10^{18}$ atoms cm$^{-2}$, the continuum reaches zero ($\tau_{912} =
6$) at $3648$ \AA\ and the flux only recovers at around $ 2000$ \AA\
($\tau = 1$). The detection of these systems is thus fairly easy, even
in medium resolution spectra. ``Grey'' LLS have optical depth $\tau < 1$ and thus produce a partial break in the quasar spectrum. 

The ionisation state of LLSs can be constrained by measuring the
column density of the same ion in different ionising states
(e.g. Fe$^+$ and Fe $^{++}$) and comparing with photoionisation modelling such as the CLOUDY software package as reviewed in the chapter lead by Michele Fumagalli \cite{Fumagalli24}.
%{\blue as review in Michele's chapter}. 
In addition, LLS provide information on the
ionising photons of the intergalactic medium \cite{Fumagalli2017}. The LLS
are also related to galactic halos as illustrated in Figure~\ref{f:gal_haloes} showing a complex morphological signature with column density \cite{Steidel94, Fumagalli2011, Weng24}.

\begin{figure}[t]
\centering
\includegraphics[width=0.9\linewidth]{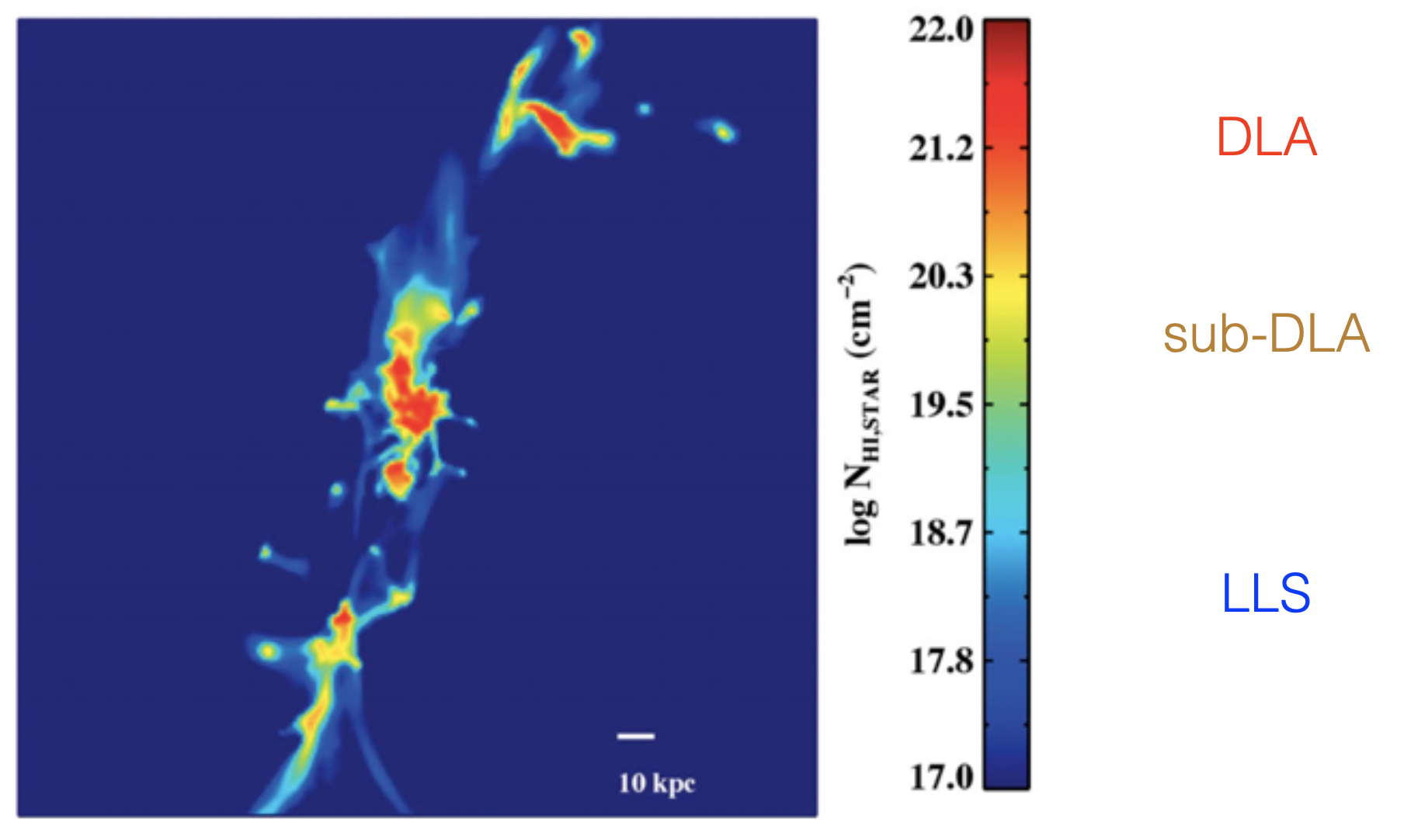}
\caption{Results from cosmological hydrodynamical simulations displaying structures in the multi-scale gas. The gas is colour-coded according to its projected column density as probed along the line-of-sight to a hypothetical background quasar. The figure illustrates the relation between the column density of various quasar absorbers and their physical size  \cite{Fumagalli2011}.}
\label{f:gal_haloes}
\end{figure}

\vspace{0.5cm}

\item Finally, the {\bf Lyman-$\alpha$ Forest} is composed of many
low column density systems ranging from N(HI) $=10^{12}$ to $1.6
\times 10^{17}$ atoms cm$^{-2}$. The absorption in the \lya\ forest is caused not by individual, confined clouds, but by a gradually varying density field characterized by overdense sheets and filaments and extensive, underdense voids which evolve with time. At first, the \lya\ forest was thought to contain pristine material, but traces of metals were detected later on
\cite{Welsh20}. These observations make possible the study of chemical enrichment in the low-density gas including at the epoch at which the first generations of stars formed and then distributed their metals into the surrounding environment \cite{Ferrara00, Skuladottir24}.  

\end{enumerate}

\subsubsection{Characterising Quasar Absorbers in Practice}
\label{subsubsec:col}

\paragraph{\underline {\bf Voigt Profile}}
\noindent

Absorption lines provide a measure of the surface
density or column density of atoms (molecules) between the observer and
the background source (expressed in atoms cm$^{-2}$). The N(HI) column density,
for example, along a sightline passing through gas with a neutral hydrogen
particle density, $n_{\rm H\, I}$ (in atoms cm$^{-3}$) is:
\begin{equation}
    N(X) \equiv \int_0^{+\infty} n_{\rm X} \, ds,
\end{equation}
where $s$ is the path the sightline takes through the gas. While individual absorption measurements are limited to a pencil-beam along the line-of-sight, the large samples now assembled allow us to statistically measure the mean properties of galaxies by combining many lines-of-sight, thereby also minimizing cosmic variance effects \cite{noterdaeme2012, Prochaska05}.

The flux emitted by the background source is absorbed by gas clouds along the line-of-sight. The intensity of the line, $I(\lambda)$ with respect to the intensity in the absence of absorption, $I_0(\lambda)$, is expressed as:
\begin{equation}\label{eqn:i_eqn}
    \frac{I(\lambda)}{I_0(\lambda)} = e^{-\tau(\lambda)}=e^{-N \sigma(\lambda)}.
\end{equation}
where $\tau(\lambda)$, the optical depth due to absorption, is related to the column density, $N$, of the absorbing species and the absorption cross section, $\sigma(\lambda)$.

Two different processes lead to the line broadening that gives
absorption features their characteristic profile:

\begin{enumerate}

\item {\it Lorentzian Profile} - Natural (damping) broadening of an absorption line is due to the intrinsic uncertainty $\Delta E$ in the energy of the upper atomic
level as expressed by the Uncertainty Principle: $\Delta E \Delta t
\sim \hbar$. This leads to a Lorentz profile:

\begin{equation} \label{sigma_eqn}
\sigma_L(\nu) = \left( \frac{\pi e^2}{m_e c} \right) f_{\rm osc}
\frac{\Gamma / 4 \pi^2}{(\nu - \nu_0)^2 + (\Gamma/4\pi)^2}
\end{equation}

where $m_e$ and $e$ are the mass and charge of an electron
respectively, $c$ is the speed of light, $f_{\rm osc}$ is the transition
oscillator strength, $\nu_0$ is the central frequency and $\Gamma$ is
the total damping constant, i.e. the reciprocal of the mean lifetime
of the upper energy state.  

\vspace{0.5cm}

\item {\it Gaussian Profile} - Within the cloud that we are observing via quasar absorbers, the ions
may have a characteristic radial velocity relative to the observer,
resulting in a Doppler-shift. These internal motions can be
characterised as a Gaussian velocity distribution:

\begin{equation}\label{dopp_eqn}
P(v) = \frac{1}{b \sqrt{\pi}} e^{-(v/b)^2}
\end{equation}

where the Doppler width, $b$, is determined by contributions from both
thermal and turbulent motions within the absorbing cloud:

\begin{equation}
b = \sqrt{b_{thermal}^2 + b_{turbulent}^2} = \sqrt{
\frac{2kT}{m_{ion}} + b_{turbulent}^2}
\end{equation}

where $k$ is Boltzmann's constant.

\end{enumerate}

Convolving the natural (Lorentz profile, see equation \ref{sigma_eqn})
and Doppler (Gauss profile, see equation \ref{dopp_eqn}) broadening
produces a Voigt profile with an optical depth, $\tau$:

\begin{equation}
\tau(\nu) = \frac{\sqrt{\pi} e^2}{m_e} \frac{N f_{\rm osc}}{b \nu} H(a,u)
\end{equation}

where:

\begin{equation}
H(a,u) = \frac{a}{\pi} \int_{-\infty}^{+\infty} \frac{e^{-y^2}dy}{(u -
y)^2 + a^2} 
\end{equation}

and:

\begin{equation}
a = \frac{\Gamma}{4 \pi \Delta \nu} ; \hspace{10mm} u = \frac{c(\nu -
\nu_0)}{\nu_0 b} ; \hspace{10mm} y = v/b 
\end{equation}

\begin{figure}[t]
\centering
\includegraphics[width=\linewidth]{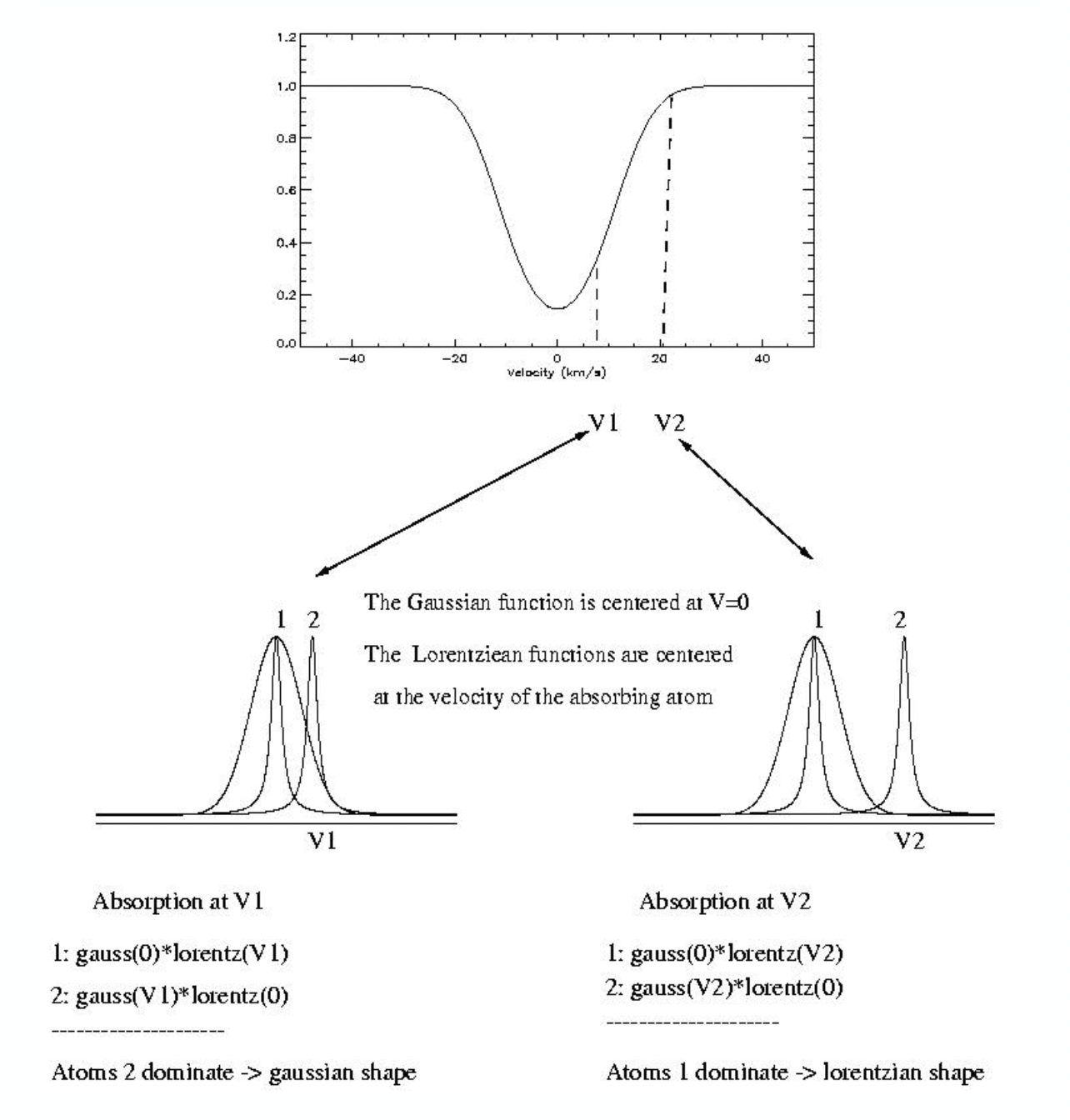}
\caption{Decomposition of a Voigt profile. The Voigt profile is the result of the convolution of Lorentzian and Gaussian functions. At V1, the Lorentzian function falls off more slowly at large $\nu$ than the Gaussian profile that describes Doppler broadening. It is thus the latter which dominates the
absorption profile. However, at V2, the prominent damping wings
completely dominate the outer parts of the line profile leading to a
Lorentzian shape. This corresponds to very high column density
absorbers, the so-called Damped Lyman-$\alpha$ systems \cite{Charlton00}. }
\label{f:voigt_profile}
\end{figure}

The core of the Voigt function is thus Gaussian, while the extended
wings of the profile are Lorentzian. This is illustrated by
Figure~\ref{f:voigt_profile}. Damping wings dominate at very high column densities for DLAs.

The optical depth at the line centre is then given by:

\begin{equation}
\tau(\lambda_0) = \frac{\sqrt{\pi} e^2}{m_e c} \frac{N f_{\rm osc} \lambda_0}{b} =
1.497 \times 10^{-15} \frac{N(\rm{cm}^{-2}) f_{\rm osc} \lambda_0
  (\rm{\AA})}{b(\rm{km s^{-1}})}
\end{equation}

Absorption lines in quasar spectra are commonly fitted with
theoretical Voigt profiles, although this section shows that this is
based on the assumption that the velocity distribution of the atoms is
described by a Gaussian function. Interestingly, departure from Voigt profile is expected at the largest column densities from additional state levels atom assumption \cite{zavarygin2018}.

%%%%%%%%%%%%%%%%%
\clearpage
\begin{trailer}{Hands-on to analyzing cosmological galaxy formation simulations}

\subsection*{[Hands-on \#3] Snapshot Data i.e. ``Observables: predicting gas absorption \& emission''}

In addition to the group catalogs, we can also directly analyze the particle/cell data in the snapshots (gas, stars, dark matter, and black holes). Note: keep in mind that loading all particles/cells for one snapshot will generally require too much memory. (Each JupyterLab instance is limited to 10 GB of memory usage). We stick to a small simulation for now (e.g. TNG50-4), and load the full details on gas positions and masses. We can use these to create the simplest type of visualization: a (weighted) 2D histogram, showing the large-scale structure (i.e. cosmic web) of the box.

\begin{tcolorbox}
\begin{verbatim}
basePath = 'sims.TNG/TNG50-4/output/'
snap = 99
\end{verbatim}
\end{tcolorbox}

\begin{tcolorbox}
\begin{verbatim}
gas = il.snapshot.loadSubset(basePath, snap, 'gas', 
                    fields=['Masses','Coordinates'])

# also load metadata from this snapshot from the 'header'
header = il.groupcat.loadHeader(basePath, snap)
\end{verbatim}
\end{tcolorbox}

\begin{tcolorbox}
\begin{verbatim}
header

{'BoxSize': 35000.0,
 'FlagDoubleprecision': 0,
 'Git_commit': b'unknown',
 'Git_date': b'unknown',
 'HubbleParam': 0.6774,
 'Ngroups_ThisFile': 2,
 'Ngroups_Total': 25257,
 'Nids_ThisFile': 1340081,
 'Nids_Total': 15550740,
 'Nsubgroups_ThisFile': 503,
 'Nsubgroups_Total': 22869,
 'NumFiles': 11,
 'Omega0': 0.3089,
 'OmegaLambda': 0.6911,
 'Redshift': 2.220446049250313e-16,
 'Time': 0.9999999999999998}
\end{verbatim}
\end{tcolorbox}

To create the histogram, we can use \textsc{scipy.binned\_statistic\_2d} which is similar to \textsc{np.histogram2d}, except that it can compute other statistics in addition to counts and sums, such as the mean or median value of a third quantity, per pixel.

\begin{tcolorbox}
\begin{verbatim}
from scipy.stats import binned_statistic_2d

x = gas['Coordinates'][:,0]
y = gas['Coordinates'][:,1]
weights = gas['Masses']
statistic = 'sum'

nPixels = [600,600]

minMax = [0, header['BoxSize']]

grid,_,_,_ = binned_statistic_2d(x, y, weights, 
                                 statistic, bins=nPixels, 
                                 range=[minMax,minMax])
\end{verbatim}
\end{tcolorbox}

Our grid has the sum of gas mass in each pixel (in "code units"). We can plot it.

\begin{tcolorbox}
\begin{verbatim}
grid.shape

(600, 600)
\end{verbatim}
\end{tcolorbox}

\begin{tcolorbox}
\begin{verbatim}
fig, ax = plt.subplots()

extent = [0, header['BoxSize'], 0, header['BoxSize']]

plt.imshow(np.log10(grid), extent=extent, cmap='magma')
ax.set_xlabel('x [code units]')
ax.set_ylabel('y [code units]')
plt.colorbar(label='Gas Mass Per Pixel [log code units]');
\end{verbatim}
\end{tcolorbox}

\begin{figure}
\centering
\includegraphics[width=0.98\textwidth]{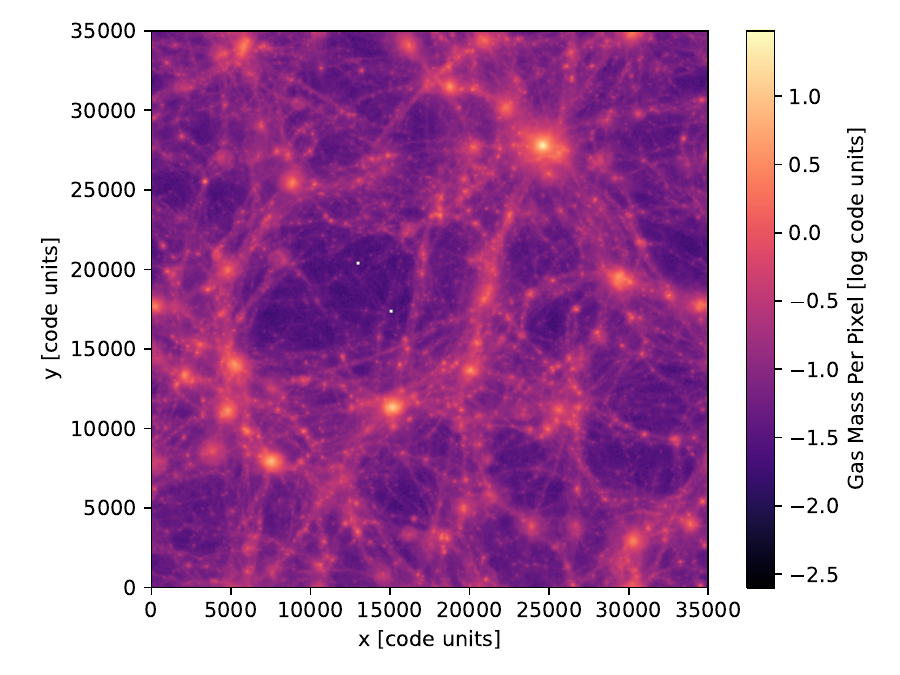}
\caption{\textbf{Hands-on plot:} distribution of gas mass in the TNG50-4 simulation at $z=0$. Tracing out structures including voids, filaments, and nodes, it reveals the `cosmic web' of large-scale structure.}
\label{fig_handson_2}
\end{figure}

Note that these "masses per pixel" are projected along the line of sight, through the whole box depth. We defined the line-of-sight direction as the $\hat{z}$-axis of the simulation box, when we chose to bin in $\hat{x}$ and $\hat{y}$ (and ignore the $z$-coordinate).

\subsection*{Exercise (Absorption \#1)}

Project the gas distribution along a different direction, and compare the two. Try to visualize one of the three other particle types, using the same technique, and compare to the gas.

\subsection*{Exercise (Absorption \#2)}

Above, the values of our projected map (i.e. the colorbar) are "masses per pixel", in code units, i.e. code mass units divided by (code length units) squared.

\begin{enumerate}
\item Convert this to solar masses, then normalize the resulting projected map (of mass) by the pixel area in $\rm{kpc}^2$, to obtain $M_\odot / \rm{kpc}^2$, and visualize it.
\item Convert each gas cell mass to a total number of hydrogen atoms (assuming $X_H = 0.75$), and normalise by the map pixel areas in units of $\rm{cm}^2$. Visualise it. This is the hydrogen column density along each line of sight through the simulation! Plot a "column density distribution function" (i.e. a 1D histogram of $N_{\rm H}$). What is your interpretation? Try changing the number of pixels in your projected map -- this effectively changes the area over which you average each column density (in reality, for e.g. a quasar absorption spectrum, this area would be very small). Does your $N_{\rm H}$ distribution depend on this (numerical) choice?
\item Absorption lines in real spectra are only sensitive to particular transitions of particular species, not total hydrogen. As a first look, We convert our $N_{\rm H}$ from above into $N_{\rm HI}$, the atomic hydrogen column density. We do not have this information directly in the snapshots, but what we do have is the neutral hydrogen fraction (where neutral hydrogen is the sum of atomic plus molecular). For now, assume that all neutral hydrogen is atomic. Load and use this extra snapshot field to compute a $N_{\rm HI}$ distribution -- how does it compare?
\end{enumerate}

\subsection*{Exercise (Absorption \#3)}

We said above that loading the entire snapshot of a high-resolution simulation is difficult (impossible in the Lab). However, the snapshot data is actually organized halo-by-halo, such that it is fast to load the particles/cells of a single halo or subhalo. Instead of the \textsc{il.snapshot.loadSubset()} function, you can use the \textsc{il.snapshot.loadHalo()} or \textsc{il.snapshot.loadSubhalo()} functions.

\begin{enumerate}
\item Switch back to TNG100-1. Use the group catalogs to select a halo with total mass ($M_{\rm 200c}$ value) similar to our own Milky Way. Keep this \textbf{Halo ID} i.e. the index of that halo in the group catalogs.

\item Load just the gas belonging to this halo, and a $N_{\rm HI}$ map and $N_{\rm HI}$ column density distribution function. How do they compare to the full box case?
\end{enumerate}

\subsection*{Exercise (Emission \#1)}

\begin{enumerate}
\item Switch back to TNG100-1. Use the group catalogs to select a halo with total mass ($M_{\rm 200c}$ value) similar to our own Milky Way. Keep this \textbf{Halo ID} i.e. the index of that halo in the group catalogs.

\item Load the positions and masses of all gas in this halo. How many gas cells are there? Visualize the gas density distribution.

\item Calculate the temperature of each gas cell in this halo (snapshots store only the "internal energy" $u$, not the temperature, so you will have to load the required fields, and create a function which derives it from these fields). See the link for full description of this equation:

$$ T = (\gamma - 1) * u / \rm{k_B} * (\rm{UnitEnergy} / \rm{UnitMass}) * \mu $$

\item Plot the distribution of gas temperatures in this halo (i.e. 1D histogram). How many different phases do you see? (What is the virial temperature of this halo)?

\item Visualize the gas temperature distribution (as a 2D image, as above). Hint: use \textsc{binned\_statistic\_2d} but with \textsc{statistic='mean'}.
\end{enumerate}

\subsection*{Exercise (Emission \#2)}

Physical quantities such as gas density or gas temperature are not directly observable. However, we can calculate observables. For example, the emission from the gas due to a specific physical process.

One nice example is free-free (bremsstrahlung) emission from a very hot ($T \gtrsim 10^7 \rm{K}$) gas, as would be found in the intracluster medium of galaxy clusters. \cite{Navarro95} (Eqn. 6) give an approximate expression for the X-ray luminosity of a single parcel of gas at a given temperature $T_{\rm keV}$ (in keV), density $\rho$, mean molecular weight $\mu$, and with a mass of $m_{\rm gas}$,

$$ L_{\rm X} = 1.2 \times 10^{-24} (\mu m_{\rm p})^{-2} m_{\rm gas} \rho T_{\rm keV}^{1/2} \,\,\,\, \rm{erg/s} $$

\begin{enumerate}
\item Use the 10$^{\rm th}$ most massive halo in TNG300-1 at $z=0$. Load the necessary gas fields, and compute the $L_X$ emission from each cell in $\rm{erg/s}$.
\item What is the total X-ray luminosity of the halo? Is it reasonable? See \cite{Anderson15} Fig. 5.

\item Visualize the X-ray surface brightness (use \textsc{binned\_statistic\_2d} with \textsc{statistic='sum'}) and normalize by the pixel area to obtain units of $\rm{erg/s/kpc^2}$.
\end{enumerate}

\end{trailer}
\clearpage

%%%%%%%%%%%%%%%%%

\paragraph{\underline {\bf Equivalent Width}}
\noindent

\begin{figure}[t]
\centering
\includegraphics[width=\linewidth]{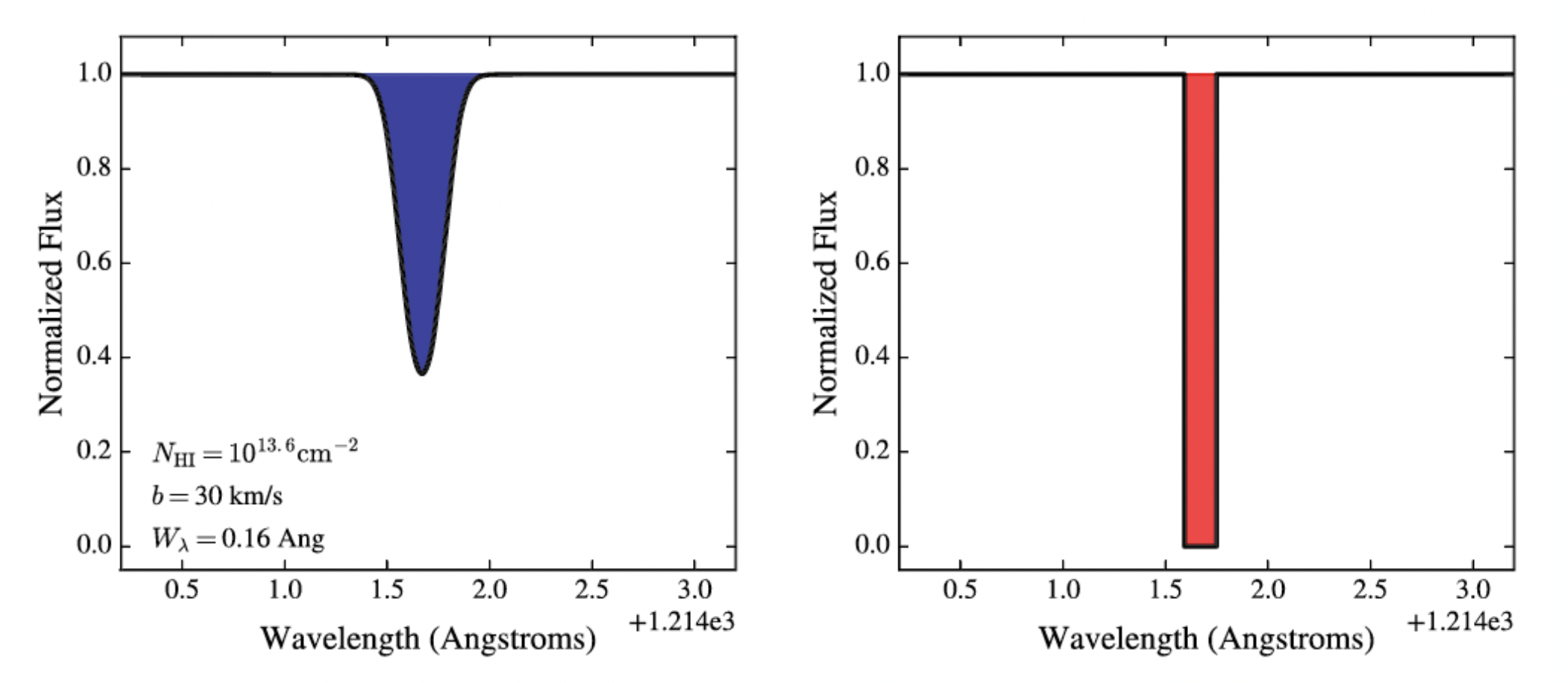}
\caption{The equivalent width of the absorption line. This quantity effectively measures the equivalent area being absorbed in a spectrum. This figure illustrates the equivalent width of a typical absorption line in a quasar spectrum on the left and the equivalent width of the absorption line on the right \cite{Prochaska19}. }
\label{f:EW}
\end{figure}

The {\it equivalent width} of the absorption line effectively measures the area being absorbed in a spectrum. Figures~\ref{f:EW} illustrates the definition of the equivalent width of a typical absorption line. The rest-frame equivalent  width, $W_{\rm rest}$, is defined as:
\begin{equation}
W_{\rm rest} = \int \frac{I_0(\lambda) - I(\lambda)}{I_0(\lambda)} \, d\lambda_{\rm rest}
    = \int (1 - e^{-\tau(\lambda)}) \, d\lambda_{\rm rest},
\label{eqn:EW}
\end{equation}
where the observed equivalent width is:
\begin{equation}
W_{\rm obs}(\lambda) =  W_{\rm rest}(\lambda) \times (1+z_{\rm abs}) 
\end{equation}
The equivalent width of an absorption line is thus independent of the
spectral resolution since it is an integral over $\lambda$. This technique is then useful when several under-resolved lines are available  (e.g., using low-resolution spectroscopy).

\paragraph{\underline {\bf Curve-of-Growth}}
\noindent

\begin{figure}[t]
\centering
\includegraphics[width=\linewidth]{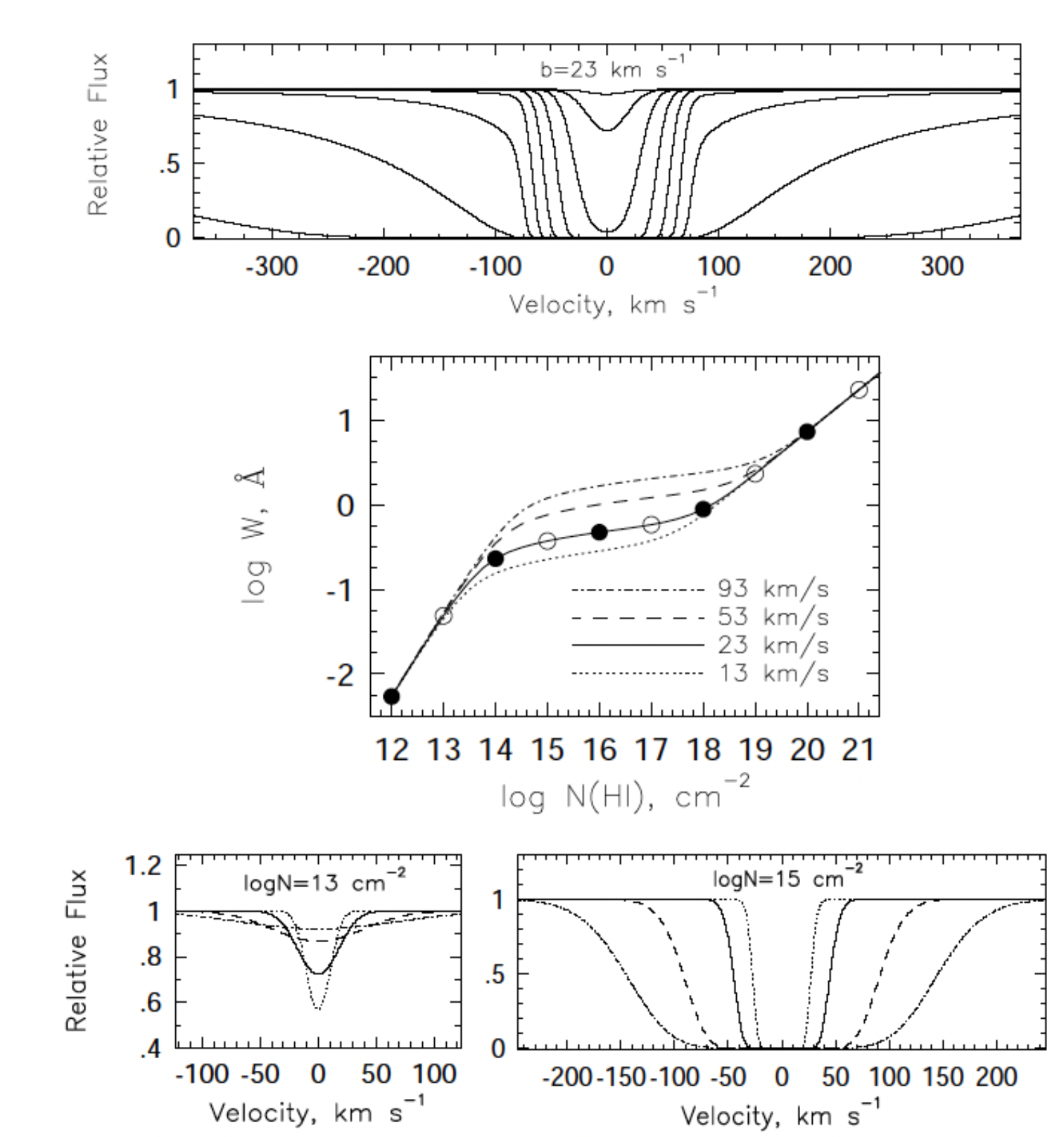}
\caption{Illustration of the
different regimes of the curve-of-growth. The middle panel shows the
curve-of-growth for the HI Lyman-$\alpha$ transition, relating the
equivalent width, $W(\lambda)$, of the absorption profile its column
density, N(HI). The different curves represent four different values
of the Doppler parameter. The upper panel shows absorption profiles with column densities matching the points on the middle panel, while the bottom panel shows absorption profiles with corresponding Doppler parameter values. For N(HI) $\ga 10^{19}$
atoms cm$^{-2}$, the profile develops damping wings, which dominate
the equivalent width (the so-called DLA and sub-DLA
regimes) and thus allows for reliable column density measurements
\cite{Charlton00}.}
\label{f:curveofgrowth}
\end{figure}

In practice, at medium spectral resolution, the column density of the absorbing species is then derived from measured equivalent widths by the classical {\it curve-of-growth} technique \cite{draine2011}. The {\it curve-of-growth} relates the equivalent width of the
absorbers with its column density $N$. The equivalent width is
traditionally used although we note that measuring the Full Width Half Maximum,
FWHM, of the line would be more appropriate as it is less dependent
upon the continuum position. The HI
\lya\ curve-of-growth is shown in Figure~\ref{f:curveofgrowth}. There
are three distinct regimes:

\begin{enumerate}

\item \textbf{The Linear Part.} The lines in this regime are
unsaturated and correspond to absorbers with small column densities
(N(HI) $< 10^{13}$ atoms cm$^{-2}$). Because the feature is optically
thin, the equivalenth width is not dependent on the Doppler parameter
$b$. Instead, the column density can be expressed as:

\begin{equation}
N = 1.13 \times 10^{20} \frac{W(\lambda)}{\lambda_0^2 f_{\rm osc}}
\end{equation}

\vspace{0.5cm}

\item \textbf{The Flat Part.} The lines in this regime are saturated
and dominated by the Doppler contribution. Their column density, $N$, depend on the Doppler parameter $b$
at a given equivalent width $W(\lambda)$:

\begin{equation}
W(\lambda) \sim \frac{2 b \lambda_0}{c}\sqrt{\ln \left
( \frac{\pi^{0.5}e^2 N \lambda_0 f_{\rm osc}}{m_e c b} \right)}
\end{equation}

In order to reliably determine the column density of such absorption
systems, higher-order Lyman series lines which have smaller oscillator
strength ($f_{\rm osc}$) and thus lie on the linear part of the curve-of-growth, need to be observed.

\vspace{0.5cm}

\item \textbf{The Damping Part.} The lines in this regime are
saturated and dominated by the Lorentzian damping wings. They correspond to high column densities (N(HI) $ \ga
10^{19}$ atoms cm$^{-2}$) and their equivalent width is proportional
to the column density independently of $b$-value:

\begin{equation}
N(HI) =  1.88 \times 10^{18} \times W_0^2(\rm{\AA})  \rm{cm}^{-2} 
\end{equation}

\end{enumerate}

\subsection{Quasar Absorbers in a Cosmological Context}
\label{subsec:fct}

\begin{figure}[t]
\centering
\includegraphics[width=0.9\linewidth]{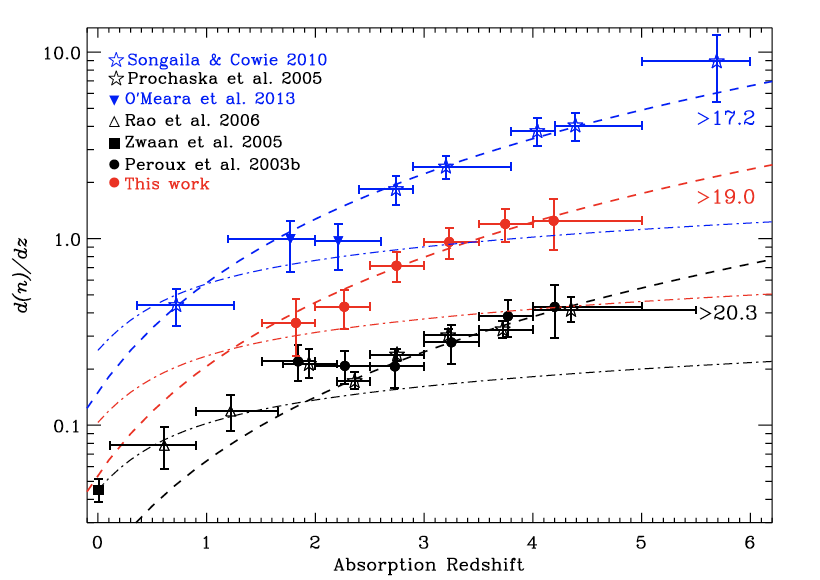}
\caption{Number density, $dn/dz$, of quasar absorbers as a function
of redshift and their fits (dashed lines). The black dot-dashed curve represents a non-evolving population in a nonzero
$\Lambda$-Universe. The red and blue dot-dashed curves are non-evolution
curves scaled to a factor of 2.3 and 5.6, respectively \cite{zafar2013a}.}
\label{f:dn}
\end{figure}

\subsubsection{Number Density}
\noindent

The number density of quasar absorbers is the number of absorbers,
$n$, per unit redshift $dz$, i.e., $dn/dz = n(z)$. If $l(z)$ is the
mean distance in redshift from a lyman limit system to a quasar, then
$n(z)= 1 / l(z)$ is the number density per unit redshift along this
line- of-sight. This is a directly observable quantity, although, its
interpretation is dependent on the geometry of the Universe. Indeed,
the evolution of the number density of absorbers with redshift is the
intrinsic evolution of the true number of absorbers combined with
effects due to the expansion of the Universe.

The number density has been traditionally modelled using a power law with slope $\gamma$:

\begin{equation}
n(z)=n_o(1+z)^{\gamma}
\end{equation}

where $n_0$ is the number of absorbers at $z = 0$. \cite{Lanzetta91b} produced the first significant statistical analysis
of DLA number density and the mass density of HI in DLAs. If the
absorber population is not evolving, in a standard Friedmann cosmology
with no cosmological constant term $\Lambda$, $n(z)$ is given by

\begin{equation}
n(z) = n_0(1 + z)(1 + 2q_0z)^{-0.5}
\end{equation}

where

\begin{equation}
q_0 = \frac{1}{2} \Omega_M - \Omega_{\Lambda}
\end{equation}

For a non-evolving population, the index $\gamma$ is equal
to 1 for $q_0 = 0$ and 0.5 for $q_0 = 0.5$. \cite{zafar2013b} used an enlarged sample of quasar absorbers of various column density and showed that the departure from non-evolution beyond z$\sim$1 is evident for the all classes of quasar absorbers ({Figure~\ref{f:dn}}).

\subsubsection{Survey's Sensitivity}

\begin{figure}[t]
\centering
\includegraphics[width=\linewidth]{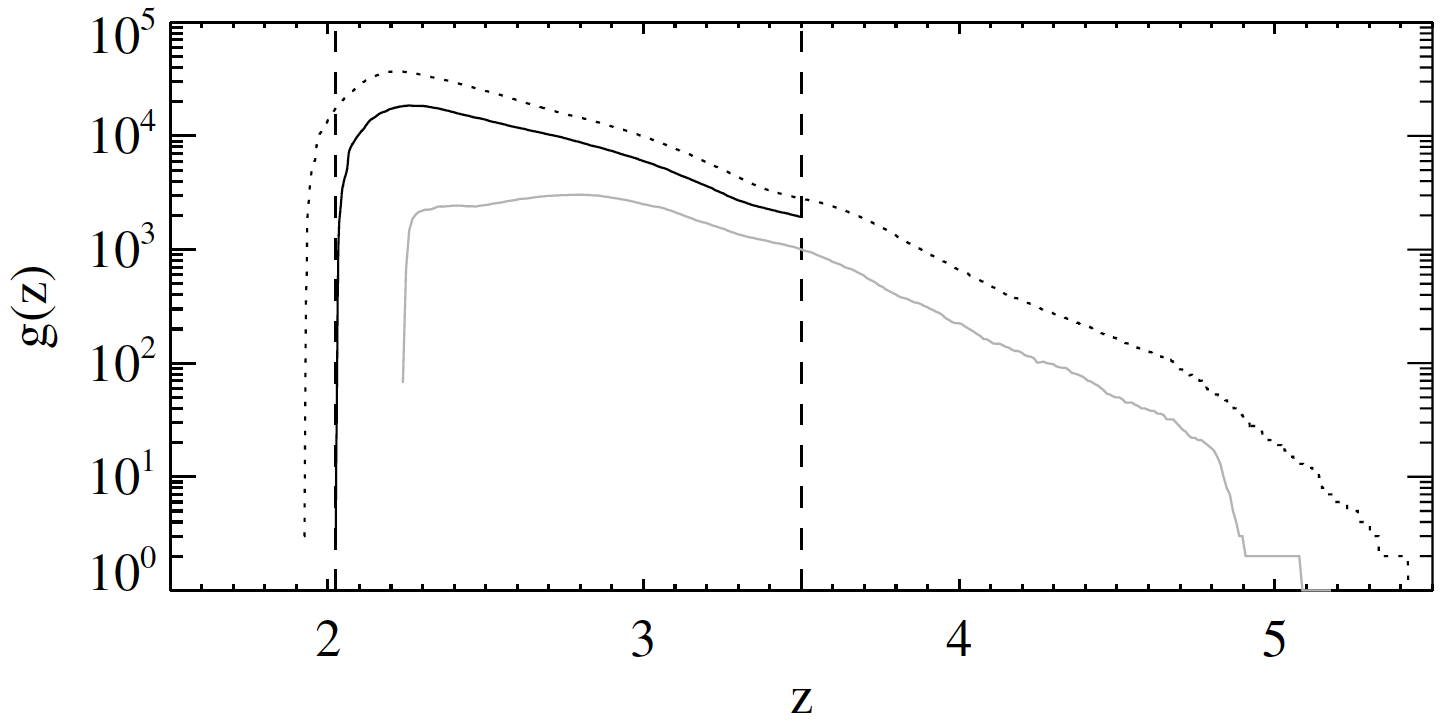}
\caption{Survey sensitivity function. The $g(z)$ function shows the cumulative number of lines-of-sight along which a DLA system could be detected if there were one. The data shown are the Sloan realeases up to DR9 \cite{Noterdaeme12}. }
\label{f:gz}
\end{figure}

Because DLAs are rare it requires probing many quasar lines of
sight to find them. Figure~\ref{f:gz} shows the cumulative number of lines of
sight along which a DLA {\it could} have been detected if there were one in the Sloan survey. This survey sensitivity, $g(z)$, is defined by:  

\begin{equation}
g(z) = \sum H (z^{max}_{i} - z) H (z - z^{min}_{i})
\end{equation}

where H is the Heaviside step function, where a DLA at redshift $z$ can be observed between $z^{min}_{i}$ and $z^{max}_{i}$. Although DLAs have
a low number density per unit redshift compared with lower column
density systems, they contain most of the neutral
hydrogen mass in the Universe.

\subsubsection{Column Density Distribution}

\begin{figure}[t]
\centering
\includegraphics[width=0.8\linewidth]{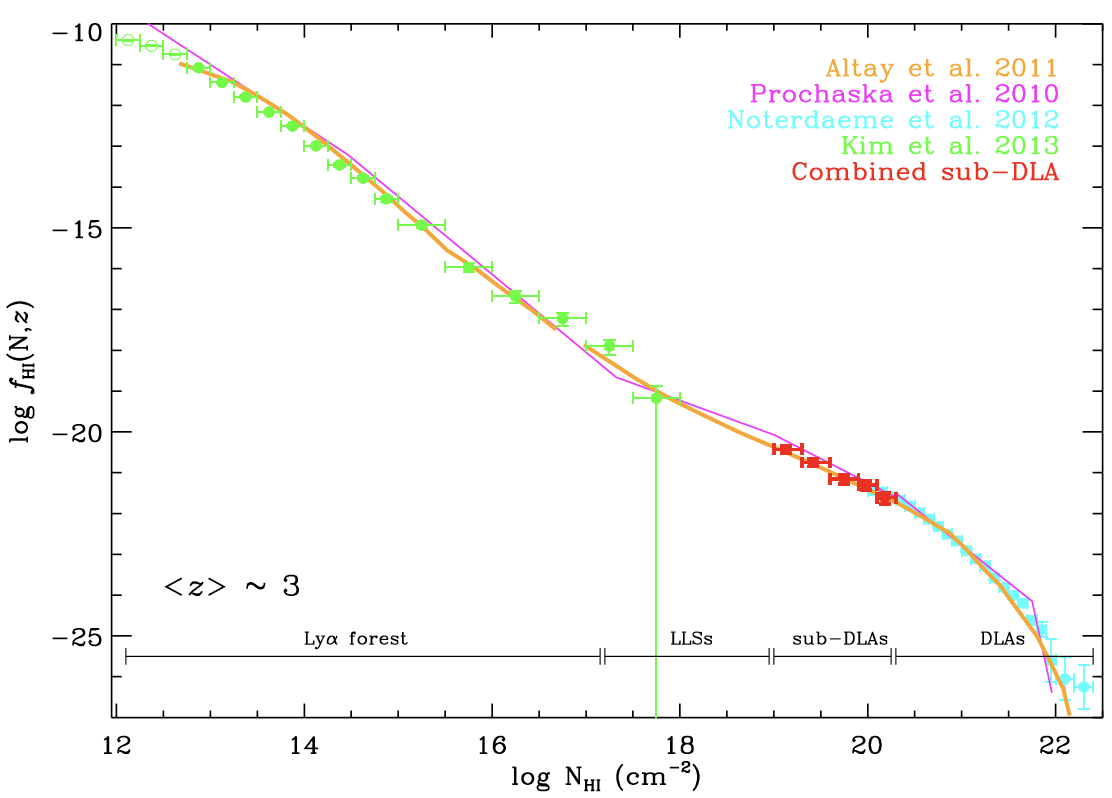}
\caption{Differential column density distribution, 
f$_{\rm NHI}$(N, z), plotted against
log N(HI). The flattening of f$_{\rm NHI}$(N, z) in the sub-DLA regime
is present in the observations. The yellow line indicates results from simulations which show a remarkable agreement with the observations of the distribution of absorbers column densities \cite{Fumagalli2011, vandevoort2012, Rahmati2014}.}
\label{f:fN}
\end{figure}

The column density distribution describes the evolution of quasar
absorbers as a function of column density (Figure~\ref{f:fN}). It is defined as:

\begin{equation}
f(N, z) dN dX = \frac{n}{\Delta N \sum_{i=1}^{m} \Delta X_i} dN dX
\end{equation}

where $n$ is the number of quasar absorbers observed in a column
density bin $[N, N+\Delta N]$ obtained from the observation of $m$ quasar
spectra with total absorption distance coverage $\sum_{i=1}^{m} \Delta
X_i$. Traditionally, the column density distribution was used for HI, although recently such studies have extended to H$_{\rm 2}$ as well as dust columns \cite{Szakacs22, Peroux23}.

The distance interval, $dX$, is used to correct to co-moving
coordinates and thus depends on the geometry of the
Universe. We
derive $X(z)$ for a non-zero $\Lambda$-Universe. Following
\cite{Bahcall69}, we introduce the variable:

\begin{equation}
X(z)= \int_{0}^{z} (1 + z)^2 \left[\frac{H_0}{H(z)}\right] dz
\label{eqn_Xz}
\end{equation}

where

\begin{equation}
H(z) = \left(\frac{8}{3} \Pi G \rho (1 + z)^2 (z + \frac{1}{2q_0}) -
\frac{\Lambda}{3} \left[\frac{(1 + z)^2}{q_0} + z^2 + 2z
\right]\right)^{1/2}
\end{equation}

For convenience, we will define:

\begin{equation}
A= \frac{8}{3} \Pi G \rho 
\end{equation}

and remind the reader that:

\begin{equation}
\Omega_M=\frac{A}{H_0^2}; \hspace{1.5cm} \Omega_{\Lambda} =
\frac{\Lambda}{3 H_0^2}; \hspace{1.5cm} \Omega_k=1 - \Omega_M -
\Omega_{\Lambda}
\end{equation}

In addition,

\begin{equation}
q_0 = \frac{1}{2} \Omega_M - \Omega_{\Lambda}
\end{equation}

or

\begin{equation}
q_0 = \frac{1}{2} \frac{A}{H_0^2} - \frac{\Lambda}{3 H_0^2} =
\frac{1}{2 H_0^2} (A- \frac{2 \Lambda}{3})
\end{equation}

and thus

\begin{equation}
H(z) = \left[A (1+z)^2\left(z+\frac{H_0^2}{A- \frac{2 \Lambda}{3}}
\right) - \frac{\Lambda}{3} \left( \frac{(1+z)^22H_0^2}{A- \frac{2
\Lambda}{3}}+z^2+2z\right) \right]^{1/2}
\end{equation}

\begin{equation}
H(z) = \left[A z (1+z)^2 - \frac{\Lambda}{3} \left(z+(z+2)\right) +
\frac{A (1+z)^2 H_0^2}{A- \frac{2 \Lambda}{3}}- \frac{\Lambda}{3}
\frac{(1+z)^2 H_0^2}{A-\frac{2 \Lambda}{3}}\right]^{1/2}
\end{equation}

\begin{equation}
H(z) = \left[A z (1+z)^2 - \frac{\Lambda}{3} \left(z+(z+2)\right) +
H_0^2 (1+z)^2 \right]^{1/2}
\end{equation}

\begin{equation}
H(z) = H_0 \left[\Omega_M z (1 + z)^2 - \Omega_{\Lambda} (z(z+2)) +
(1+z)^2\right]^{1/2}
\end{equation}

\begin{equation}
H(z) = H_0 \left[(1 + z)^2 (1 + z\Omega_M) - z (2 + z)
\Omega_{\Lambda}\right]^{1/2}
\end{equation}

Including this result in equation~\ref{eqn_Xz} leads to:

\begin{equation}
X(z) = \int_{0}^{z} (1 + z)^2 \left[(1 + z)^2 (1 +
z\Omega_M) - z (2 + z) \Omega_{\Lambda}\right]^{-1/2}dz
\end{equation}
\label{eqn_dist_int}

\vspace{1cm}

\subsection{Mass Density}
\label{subsec:omega}

As described earlier, the ratio of the baryonic density to the critical density of the Universe, $\Omega_{\rm baryons}$, is well constrained. It is important to establish whether the total amount of baryons is matched by the sum of the contribution of the detected baryonic components. To estimate the contribution from atomic gas, one can calculate the contribution of quasar absorbers to the baryonic mass in units of the current critical mass density, $\rho_{crit}$, by integrating the observed column density distribution:

\begin{equation}
\Omega_{\rm HI}(z) = \frac{H_0 \mu m_H}{c \rho_{crit}}
\int_{N_{min}}^{\infty} N f(N,z) dN
\end{equation}
\label{eqn_omega}

where $\mu$ is the mean molecular weight of the gas which is taken to
be 1.3 (75\% hydrogen and 25\% helium by mass), $m_H$ is the hydrogen
mass and $N_{min}$ is the low end of the HI column density range being
investigated. The critical density at z=0 is expressed as:

\begin{equation}
\rho_{crit} = \frac{3 H_0^2}{8 \pi G}
\end{equation}
\label{eqn_omega}

where G is the Gravitational constant.

If the fit to the column density distribution is made with a power law with index $\beta < 2$, most of the mass is in the highest column density systems (DLAs). Indeed, the integral diverges unless an artificial upper limit to the column density distribution, $N_{max}$ is introduced, since

\begin{equation}
{\rm Mass} (HI)=\int_{N_{min}}^{N_{max}}N f(N)
dN=\left[\frac{1}{2-\beta}N^{2-\beta}\right]_{N_{min}}^{N_{max}} =
\frac{N_{max}^{2-\beta} - N_{min}^{2-\beta}}{2-\beta}
\end{equation}

for $\beta < 2$. 

For example, assuming $N_{max} = 21.5$ and $\beta=1.5$ leads to:

\begin{equation}
\Bigl[{\rm Mass} (HI)\Bigr]^{21.5}_{20.3} = 8.41 \times 10^{10}
\end{equation}
 
While the mass below the DLA definition is:

\begin{equation} 
\Bigl[{\rm Mass} (HI)\Bigr]^{20.3}_{17.2} = 2.75 \times 10^{10}
\end{equation}

Thus, DLA absorbers with N(HI) $> 10^{20.3}$ atoms cm$^{-2}$ contain
at least 75\% of the neutral hydrogen (HI) mass, but this result is
strongly dependent on the chosen high-column density cut-off
(i.e. assuming $N_{max} = 22.0$ will lead to a mass fraction of
neutral hydrogen of 85\% in the DLA range). Alternatively, using a $\Gamma$-law brings several advantages including i) representing the data better; ii) requiring fewer parameters than several broken power laws and iii) addressing the divergence
problem of the power law form for the mass integral.

In the DLA region, it is common practice to estimate the total HI by summing
up directly the individual column densities:

\begin{equation}
\int_{N_{min}}^{\infty} N f(N,z) dN = \frac{\sum N_i(HI)}{\Delta X}
\end{equation}

where $\Delta X$ is the distance interval as previously defined. 

The errors in $\Omega_{\rm HI}$ are difficult to estimate accurately
without knowing $f(N,z)$. Other have estimated the fractional
variance in $\Omega_{DLA}$ by comparing the observed distribution of
$f(N,z)$ with the equivalent Poisson sampling process. This gives

\begin{equation}
\Bigl( {\Delta\Omega_{\rm HI} \over \Omega_{\rm HI}} \Bigr )^2 =
\sum_{i=1}^p N_i^2 / \Bigl(\sum_{i=1}^p N_i\Bigr )^2  
\label{eqnerr}
\end{equation}

and $1/\sqrt{p}$ fractional errors if all the column densities
included in a bin are equal \cite{Peroux2003, zafar2011}.

%%%%%%%%%%%%%%%%%
%%%%%%%%%%%%%%%%%
%SECTION 3
%%%%%%%%%%%%%%%%%
%%%%%%%%%%%%%%%%%

\section{Cosmic Evolution}
\label{sec:cosmo}

\subsection{Evolution of Cold Gas}
\label{subsec:baryons}

\begin{figure}[t]
\centering
\includegraphics[width=\linewidth]{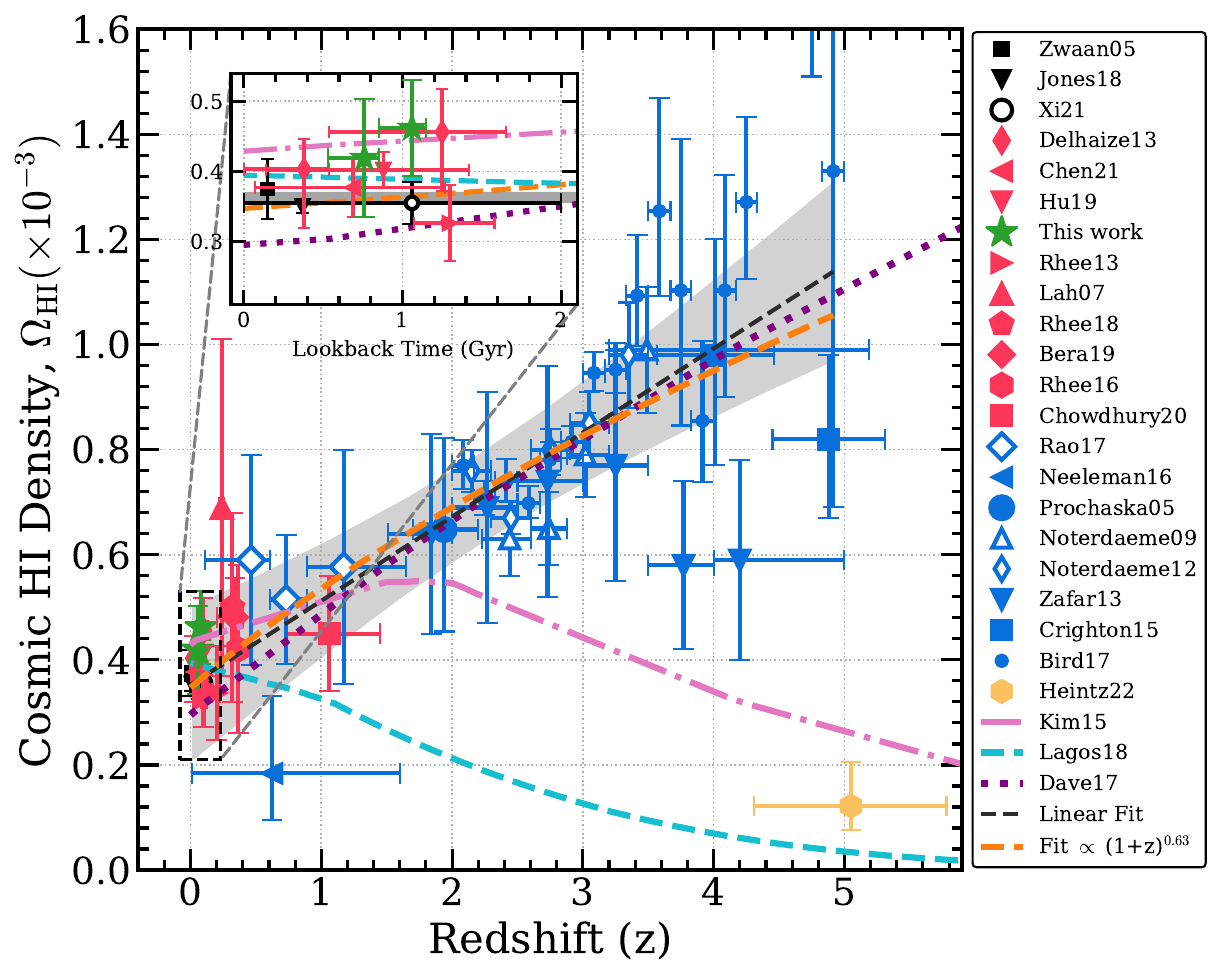}
\caption{Cosmic evolution of the neutral gas density, \OmegaGas$ \equiv
\rhogas / \rhocrit$. The crosses display
the results from individual surveys from the literature. The low-redshift measurements ($z\la0.4$) are
based on 21-cm emission measurements while the higher-redshift values are estimated from \lya\ absorbers.  The power-law fit is shown with the dashed black line. Overall, the gas density is well-constrained and indicates only a mild evolution with redshift \cite{rhee2013}. }
\label{f:omegaHI}
\end{figure}

\begin{figure}[t]
\centering
\includegraphics[width=\linewidth]{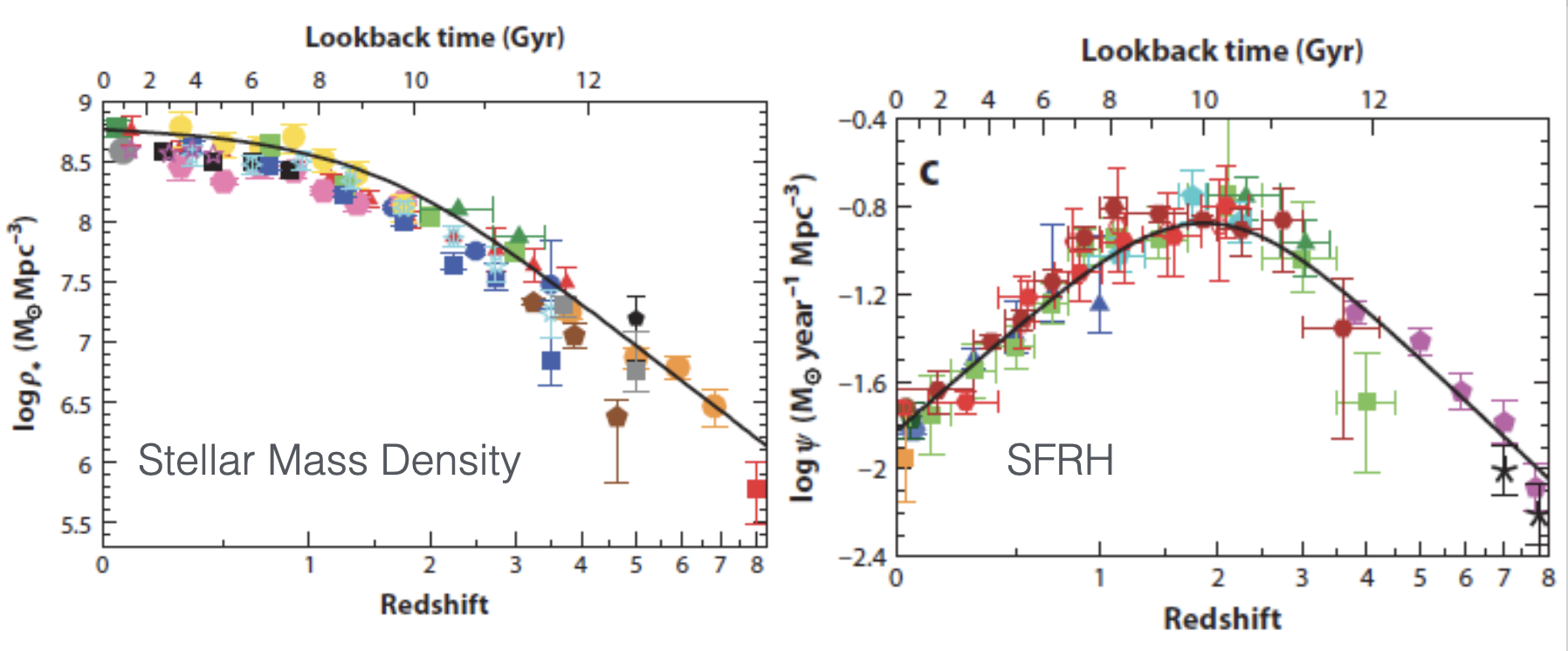}
\caption{The stellar component of the baryonic matter. {\bf Left:} The evolution of the stellar mass density. The data points with symbols are observations. The solid line
shows the global stellar mass density obtained by integrating the best-fit instantaneous star-formation rate
density with a return fraction, the fraction of the stellar mass that is immediately returned to the gas when massive stars explode, R = 0.27. {\bf Right:} The rate of star formation per unit of cosmological volume peaks at a redshift z$\sim$2 (the "epoch of galaxy assembly"), stays high up to z$\sim$1 (Universe age: 2-6 Gyr), then dramatically decreases by more than an order of magnitude from z$\sim$1 until today. 
Naive view would have SFR increases as number of stars increases with cosmic time, which rises the question of the physical process driving this decrease \cite{Madau14}.}
\label{f:stars}
\end{figure}

The cosmic evolution of neutral atomic gas, HI, is now well-constrained. 
The first measurement of the cosmological mass density was made in
1991 by Lanzetta et al. who used a combination of their own spectra
together with data from \cite{Sargent89}. Today the amount of atomic gas mass has been robustly measured from 21-cm emission at radio frequencies at low-redshift and from using \lya\ quasar absorbers at high-redshift (Figure~\ref{f:omegaHI}). An important finding, is that the column density distribution demonstrates that most of this mass comes from the highest column densities, which likely trace the ISM of galaxies. Yet, we are currently short of observational proxy for atomic HI gas at high-redshift and routinely refer to CO, as a tracer of H$_2$, to probe the neutral gas content of galaxies. This shortcoming constitutes a major challenge to our understanding of the gas reservoirs in the early Universe. Another important point is that at $z\la 2.5$, the total increase in stellar density (Figure~\ref{f:stars}, left panel) exceeds the atomic gas consumption. 

\begin{figure}[t]
\centering
\includegraphics[width=\linewidth]{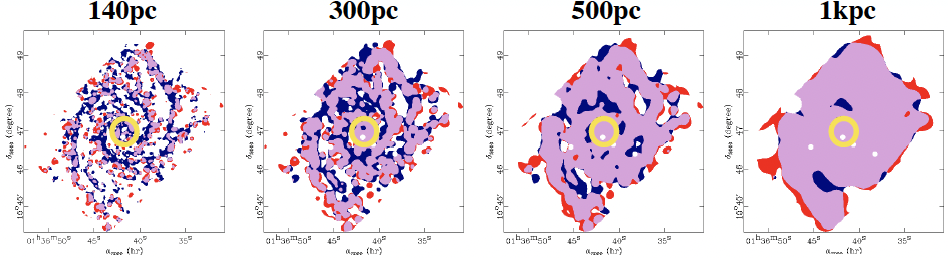}
\caption{Spatial distribution of gas tracers in a nearby objects. Galaxy maps showing regions with overlapping CO and H$\alpha$ emission (lavender), CO only (dark blue) and H$\alpha$ only emission (red) at various spatial resolutions for NGC 628. The Atacama Large Millimeter/submillimeter Array (ALMA) in particular has demonstrated the importance of resolving the spatial distribution of various tracers to capture a complete picture of the gas structure in galaxies \cite{Schinnerer19}.}
\label{f:PHANGS}
\end{figure}

\begin{figure}[t]
\centering
\includegraphics[width=\linewidth]{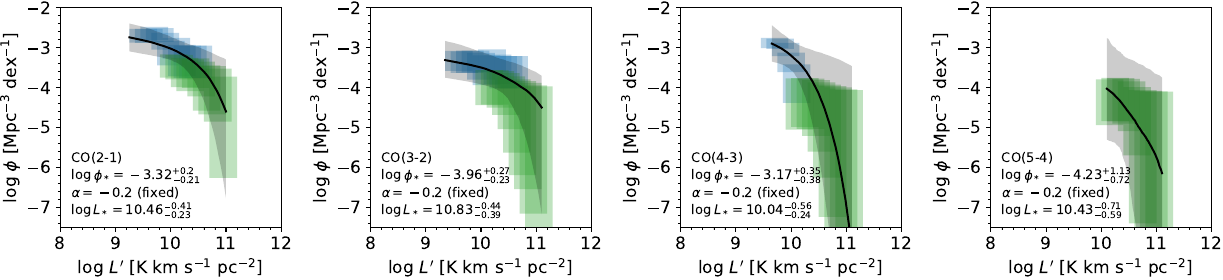}
\caption{CO luminosity function evolution with redshift. Despite the small primary beam of ALMA, there have been a number of major surveys looking blindly for CO emitters \cite{walter2014, decarli2016, Boogaard23}.}
\label{f:CO_LF}
\end{figure}

\begin{figure}[t]
\centering
\includegraphics[width=0.9\linewidth]{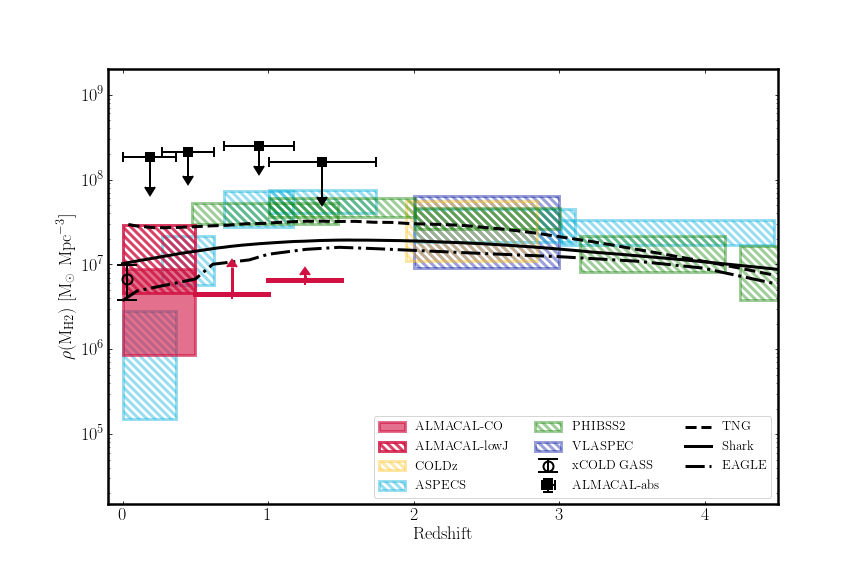}
\caption{The evolution of molecular gas mass density with redshift. The filled box represents the results from samples of CO emitters, including ALMACAL-CO (red boxes). Black squares are the limits from a molecular absorption-line search in ALMACAL by \cite{klitsch2019} (marked as ALMACAL-abs in the legend). The lines mark the predictions of evolution from various simulations \cite{Hamanowicz23}.}
\label{f:omegaH2}
\end{figure}

Probing the molecular phase of the gas is nonetheless key, as this the most direct fuel for star formation. Recent observations of CO emitters have proven valuable. The Atacama Large Millimeter/submillimeter Array (ALMA) in particular has demonstrated the importance of resolving the spatial distribution of these tracers to achieve a complete understanding of the  gas in galaxies (Figures~\ref{f:PHANGS}). Early deep surveys have enabled to compute the CO luminosity function evolution with redshift (Figure~\ref{f:CO_LF}). As a result, observational measurements of the molecular gas mass density are now possible despite being still in their infancy \cite{decarli2016, decarli2019, Hamanowicz23}. Current results indicate a rapid decrease at lower redshift (Figures~\ref{f:omegaH2}). 

Despite many progresses on the observational front, it is still challenging to simulate the colder phase of gas in a full cosmological context because of both the complexity of the physics involve and because of the wide dynamical scales involved \cite{Maio22, Crain23, Casavecchia24}. This challenge is sometimes refer to as "molecular cosmology". A comparison of $\Omega_{\rm HI}$ and $\Omega_{\rm H_2}$ shows that there is more atomic neutral gas than molecular gas at any point in time. Additionally, the molecular gas mass density cosmic evolution mirrors the evolution of star formation rate history (Figure~\ref{f:omegaH2}), as the molecular gas is rapidly consumed by making stars.

\begin{figure}[h]
\centering
\includegraphics[width=0.9\linewidth]{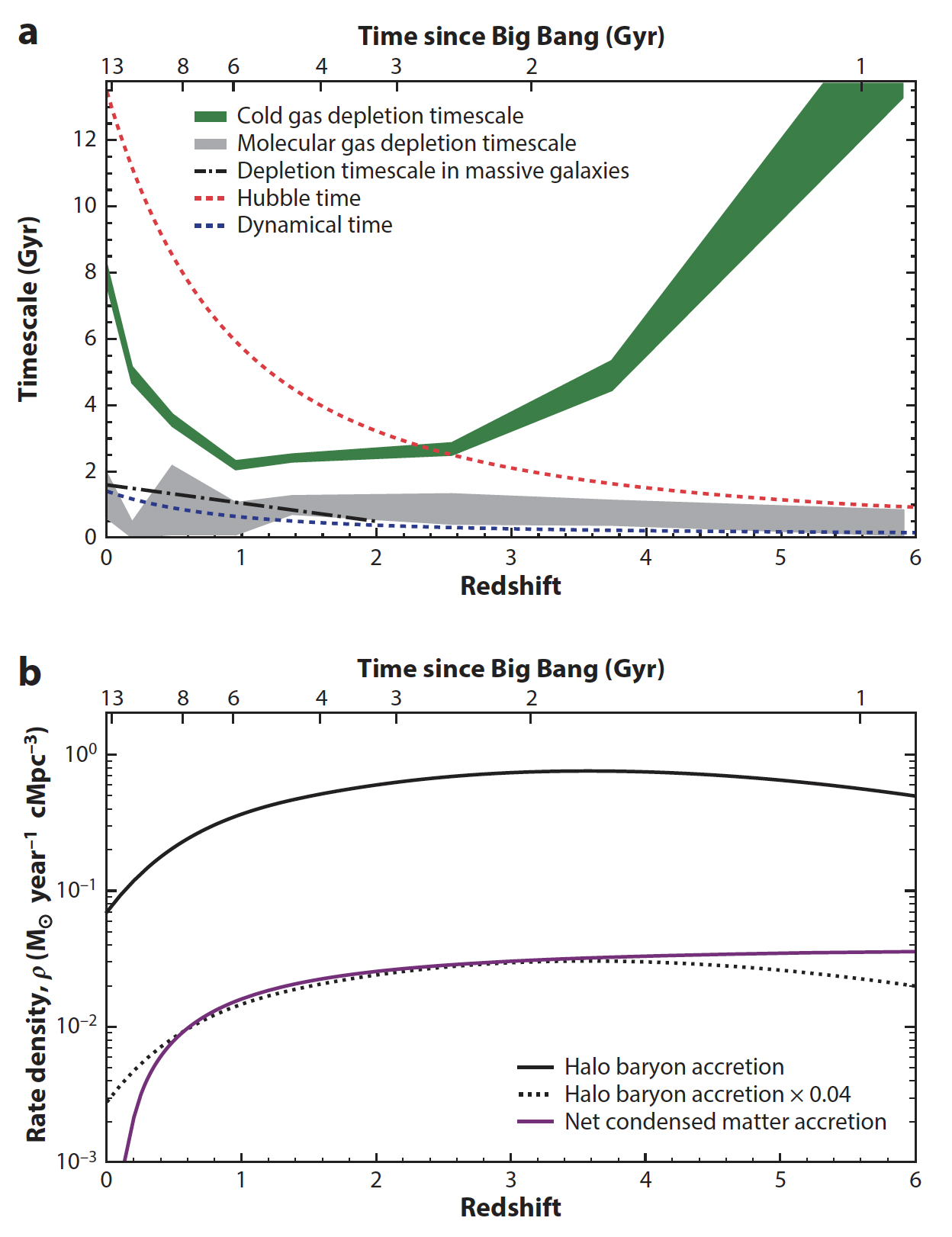}
\caption{{\bf Top:} Depletion timescales. This quantity describes the time it takes for gas to convert into stars. The depletion timescale of molecular gas is evolving  with cosmic time indicating a universal physical process of conversion of molecular gas into stars on global scale. {\bf Bottom:} Net accretion rate. This function represents the gas intake into the system required to describe the observed gas and stellar mass densities. These results indicate that the z $<$ 2 decline
of the star-formation history is driven by the lack of molecular gas supply due to a drop in net gas accretion
rate, which is itself driven by the decreased growth of the dark matter halos. These observational findings are in
remarkable agreement with the gas regulator model. \cite{PerouxHowk20}.}
\label{f:net_acc}
\end{figure}

Importantly, it is valuable to probe the gas cycle through various phases in the so-called baryons cycle. The molecular gas depletion timescale, which describes the time it takes to convert the gas into stars, is almost constant with redshift \cite{PerouxHowk2020, Tacconi2020}. These results indicate that the physical processes  of conversion of molecular gas into stars is universal on large scales (top panel of Figure~\ref{f:net_acc}). At low-redshift, the time it takes to consume molecular gas is comparable with the dynamical time, i.e. the time it takes for the material to fall  from the halo virial radius down to the galaxy. This is the reason why the decrease observed in the star formation history at low-redshifts is related to the decrease in the gas reservoir.

These observed quantities are the basis of a computation of the net accretion rate \cite{PerouxHowk20, Tacconi2020, Walter2020}. This is the rate of accretion (or conversion) from ionised reservoirs given the observed evolution of mass density in condensed matter (Figures~\ref{f:net_acc}). It is important to note that this describes both the {\it motion and transformation} of gas. The net accretion rate decreases continuously with time, and drops significantly at low-redshift. When compared with the baryon mass accretion rate, i.e. the total matter scaled to the cosmic baryon fraction, the efficiency of conversion is found to be only 4\%, indicating that the conversion of accreted baryons into cold gas is an inefficient process. Together, these results indicate that globally the growth of condensed matter in the Universe scales principally with the dark matter accretion rate onto halos. Therefore, the observed decrease of gas accretion is due to the decreased in dark matter halos growth \cite{PerouxHowk20}.

\begin{figure}[h]
\centering
\includegraphics[width=0.65\linewidth]{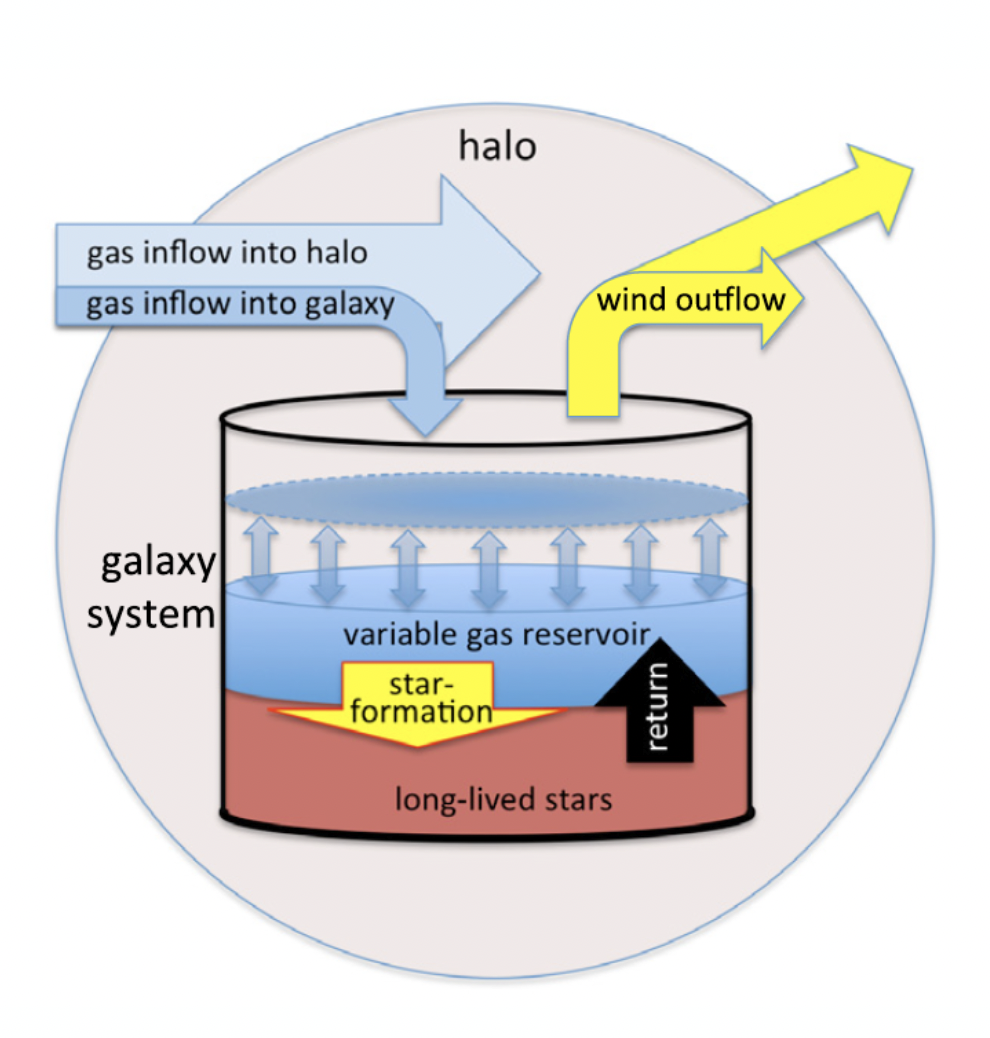}
\caption{Bathtub model of regulation of galaxy formation. This gas-regulated model describes galaxies as systems in a slowly evolving equilibrium between inflow, outflow and star formation. At early times, gas accumulates and the star formation is essentially limited by the gas reservoir. At later times, galaxies reach a steady state in which star formation is regulated by the net accretion rate \cite{Lilly13}. }
\label{f:bathtub}
\end{figure}

Additionally, these observations are consistent with the bathtub/regulator model \cite{bouche2010, lilly2013, peng2014} which describes galaxies as systems in a slowly evolving equilibrium between inflow, outflow and star formation (Figure~\ref{f:bathtub}). At early times, gas accumulates and the star formation is essentially limited by the gas reservoir. At later times, galaxies reach a steady state in which star formation is regulated by the net accretion rate.

\subsection{Evolution of Metals}
\label{subsec:metals}

\begin{figure}[t]
\centering
\includegraphics[width=0.9\linewidth]{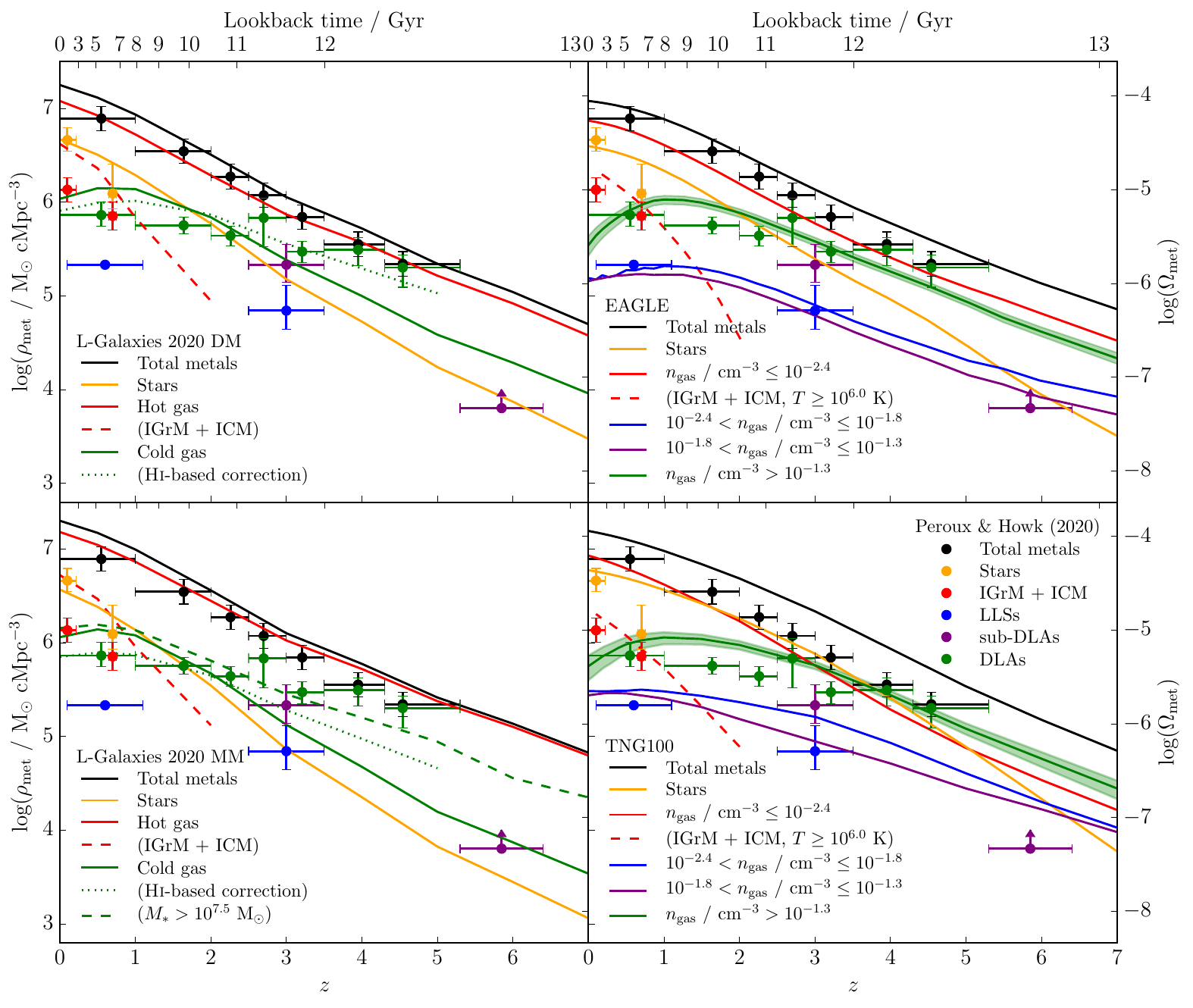}
\caption{Cosmic metal mass density. The metallicity of neutral gas is now well-constrained based on statistically significant samples and includes a self-consistent correction for dust-depletion based on multi-element analysis \cite{jenkins2009, decia2016}. Results indicate a mild evolution of the HI-weighted metallicity of 1 dex over 10 Gyr. Surprinsingly, the neutral gas contains all expected metals at z=3 which cannot be reproduced self-consistently by simulations. These findings therefore remain an open topic of investigation \cite{Yates21}.}
\label{f:omegaMetals}
\end{figure}

Metals refer to elements heavier than helium which have been made in stars within galaxies. Two decades ago, \cite{Pettini1999} and other subsequent works noted the paucity of metals compared with measurements in the available data at z $\sim$ 2. Their expectations were based on some of
the earliest estimates of the star formation rate density evolution of the Universe, measurements of metals in neutral gas, and stellar metals from high-redshift galaxies. This {\it "missing metals problem"} spurred many follow-up works about the global distribution of metals in the Universe \cite{Ferrara00, bouche2007a}.

Iron (Fe) abundance is often used to measure stellar metallicities and it is convenient to use the same indicator in quasar absorbers in order to allow for direct comparison. FeII lines are usually present in large numbers in quasar spectra and have the advantage that they exhibit a
range of rest wavelengths. A challenge with studying metal abundances is the complication due to dust depletion, the process by which particles are removed from the gas phase via condensation onto grains \cite{jenkins2009, decia2016}. Although the abundances of Zn and Fe track each other closely down to metallicities of $\la 0.01 Z_{\odot}$ (where $Z_{\odot}$ refers to the
solar abundance) in Galactic stars, in the local inter-stellar medium
an overabundance of Zn relative to Fe is often observed. This is due
to differential depletion onto grains, such that whilst Fe is usually
heavily depleted, very little Zn is seen to be incorporated into dust
\cite{Pettini97}. For these reasons, Zn (as well as Cr) have traditionally been adopted as the metallicity indicator of choice for quasar absorbers
metal abundances.  It therefore follows that the relative abundances
of [Zn/Cr] and [Zn/Fe] will provide an estimate of the fractions of
these refractory elements which are missing from the gas-phase.  Such
studies show that at $z_{abs} > 1.5$, DLAs are generally metal poor,
typically 1/10 of solar, with small amounts of dust depletion. Today, the metallicity of neutral gas is now well-constrained based on statistically significant samples and includes a self-consistent correction for dust-depletion now based on multi-element analysis \cite{jenkins2009, decia2016}. Observations indicate little sign of metallicity evolution when column density weighted dust-depletion corrected abundances are considered ({Figure~\ref{f:omegaMetals}}). 

These observables are compared with the total metal production by star formation, modulo the fraction of stellar mass immediately returned to gas when massive stars explode. We stress that the expected amount of metals are little dependent on assumptions of the Initial Mass Function (IMF) because the metal production rate is directly related to the mean luminosity density where massive stars dominate the metal production. The metal mass density also indicates that the neutral gas metal component dominates over the ionised gas. At z$<$1, the contribution from metals in groups and clusters is important, but the stars are the dominant contributors to the metal budget. The estimates of the fractional amount of metals in stars rely on assumption of the yields to calculate the total expected amount of metals. These uncertainties are extensively discussed in e.g. \cite{Peeples2014}.

Recent revisits of the missing metals problem calculating the fractional contribution to the total expected amount of metals have been performed \cite{PerouxHowk20}. Surprisingly, the neutral gas contains all expected metals at z=3 and cannot be reproduced self-consistently by simulations. These findings therefore remain an open topic of investigation \cite{Yates21}. Remarkably, at z$>$2.5, all metals are in the low-ionised gas. At low-redshift, there is a diversity of contributors, most of which are catalogued \cite{Lehner18}. At 1$<$z$<$2, the likely contributors (including the ionised gas traced by the Lyman-$\alpha$ forest and the hot gas in group and clusters, as well as stars) are not yet catalogued: future UV/X-ray missions (such as XRISM, HUBS, LEM, LUVOIR, Lynx, Athena) are required to close this census. Overall, these results indicate that the expected metal content of the Universe is likely accounted for,  in contrast with 20 years ago \cite{Pettini1999}.

\subsection{Evolution of Dust}
\label{subsec:dust}

\begin{figure}[t]
\centering
\includegraphics[width=0.8\linewidth]{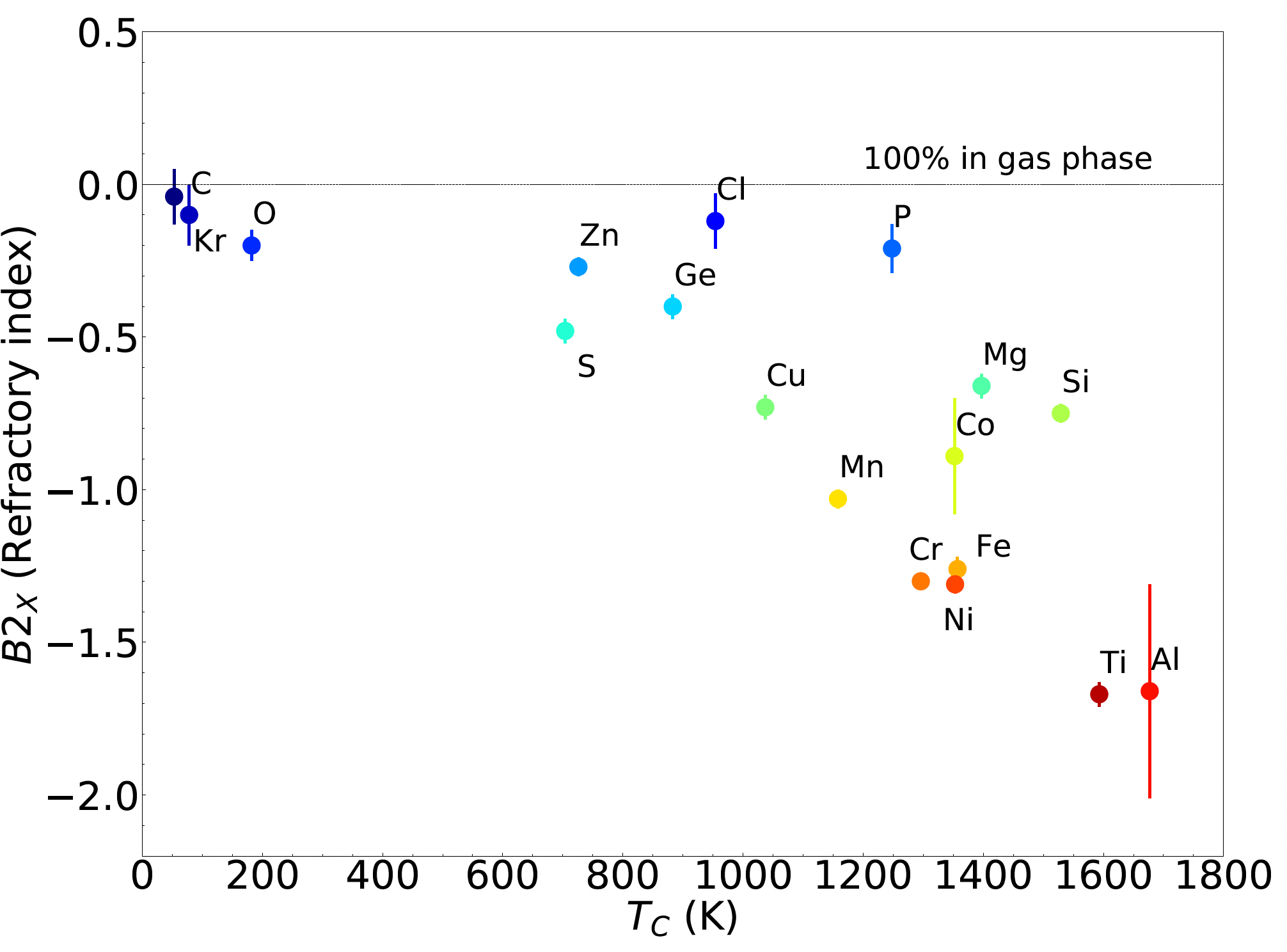}
\caption{Differential elemental depletion. This plot displays the refractory index describing the fractional amount of metals lock onto dust grains as a function of condensation temperature for various elements. This is the basis of the multi-element method which is used for correcting dust depletion to estimate the amount of the metals locked into dust grains in neutral gas \cite{Konstantopoulou2022}. }
\label{f:Condensation_temp}
\end{figure}

The dust is essentially the solid-phase of metals. Indeed, a large fraction of metals is locked into solid-phase dust grains. Dust has a strong influence on observational properties of galaxies but
also on formation of the molecules which in turns are critical to star formation. Hence,
the cosmic evolution of dust mass is a fundamental measure of galaxy evolution.
Multi-element method \cite{jenkins2009, decia2016} have made use of the fact that the condensation temperature varies from element to element to correct depletion in order to estimate the amount of the metals locked into dust grains in neutral gas (Figure~\ref{f:Condensation_temp}). In turns, these works provide us with robust estimates of the dust content of the neutral gas up to high-redshift in a consistent manner.

\begin{figure}[t]
\centering
\includegraphics[width=\linewidth]{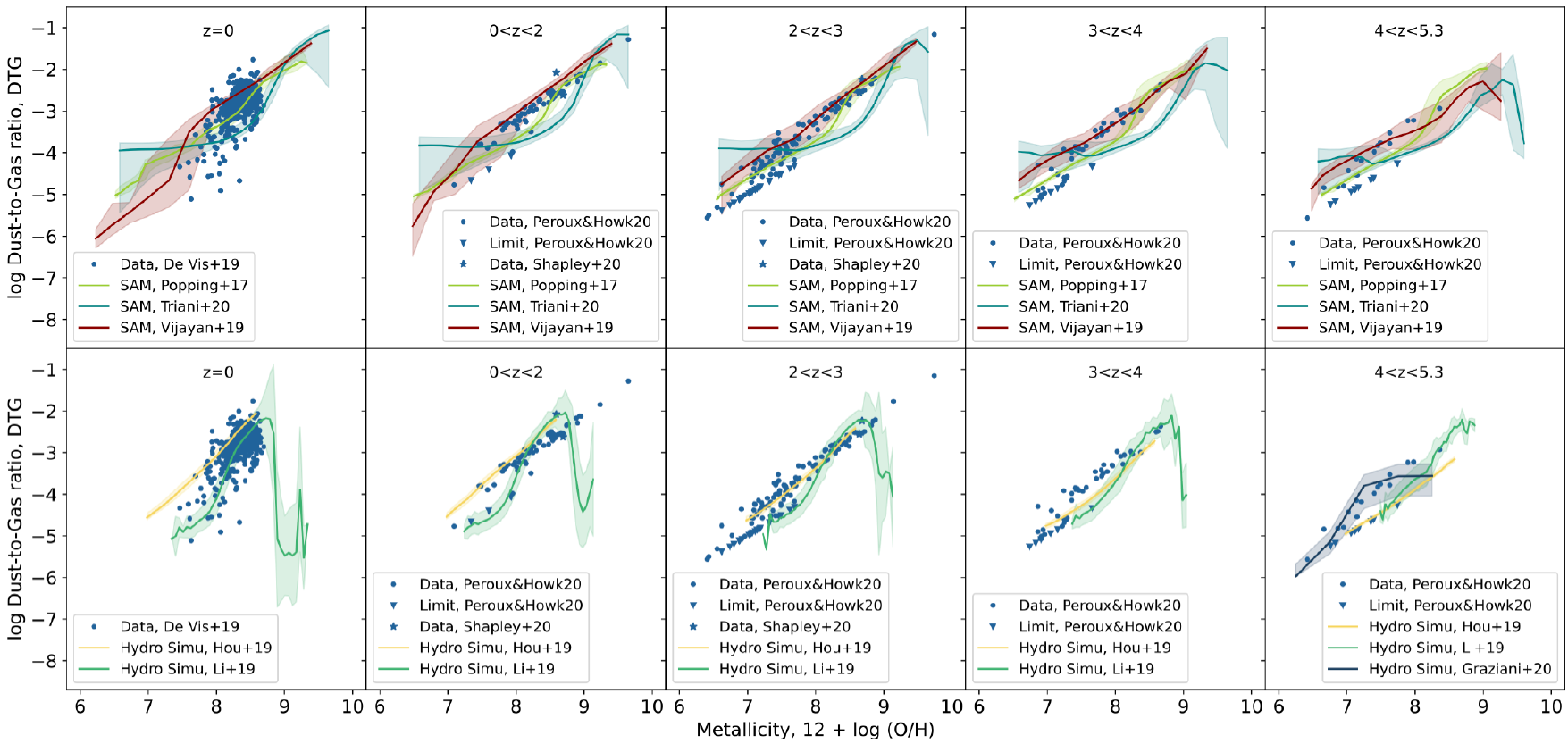}
\caption{Observed and modeled evolution of dust-to-gas ratios with metallicity. The data points are observations from local emission galaxies \cite{de-vis2019} and quasar absorbers at z$>$0. The lines in the top
row of this figure depict three different semi-analytical models \cite{popping2017, Triani20, vijayan2019}, while the lines in the bottom row represent three different
hydro-dynamical models \cite{hou2019, li2019, graziani2019}. The models reproduce the observed DTGs with varying degree of success \cite{Popping22}.}
\label{f:DTG}
\end{figure}

Properties of dust in neutral gas vs. metallicity is traced by the dust-to-gas ratio, or DTG, which is the fraction of the interstellar mass incorporated into grains. Recent results find that the DTG is a strong function of metallicity and it follows the trend of higher metallicity for lower-redshift galaxies (Figure~\ref{f:DTG}). In addition, new measurements extend at lower metallicity, indicating a change in dust assembly in that regime \cite{PerouxHowk2020}. Importantly, the fit to the DTG-metallicity relation provides a refined tool for robust dust-based gas mass estimates inferred from millimeter dust-continuum observations \cite{Popping22}. Over more than 10 Gyr in cosmic time, DTG increases by $\sim$1dex, as a result of the increase in the mean metallicity. 

The dust-to-metal ratio,
DTM, is the  fraction of all the metal mass bound into dust. This quantity is decreasing with decreasing metallicity. Indeed, the DTM values decrease and have an increased scatter, reflecting the complex dust chemistry at work in low-metallicity environment. Further, the column density (or surface density) of the dust can be computed providing new constraints on simulations predicting resolved dust mass functions through 2D projection \cite{Peroux23}.

\begin{figure}[t]
\centering
\includegraphics[width=\linewidth]{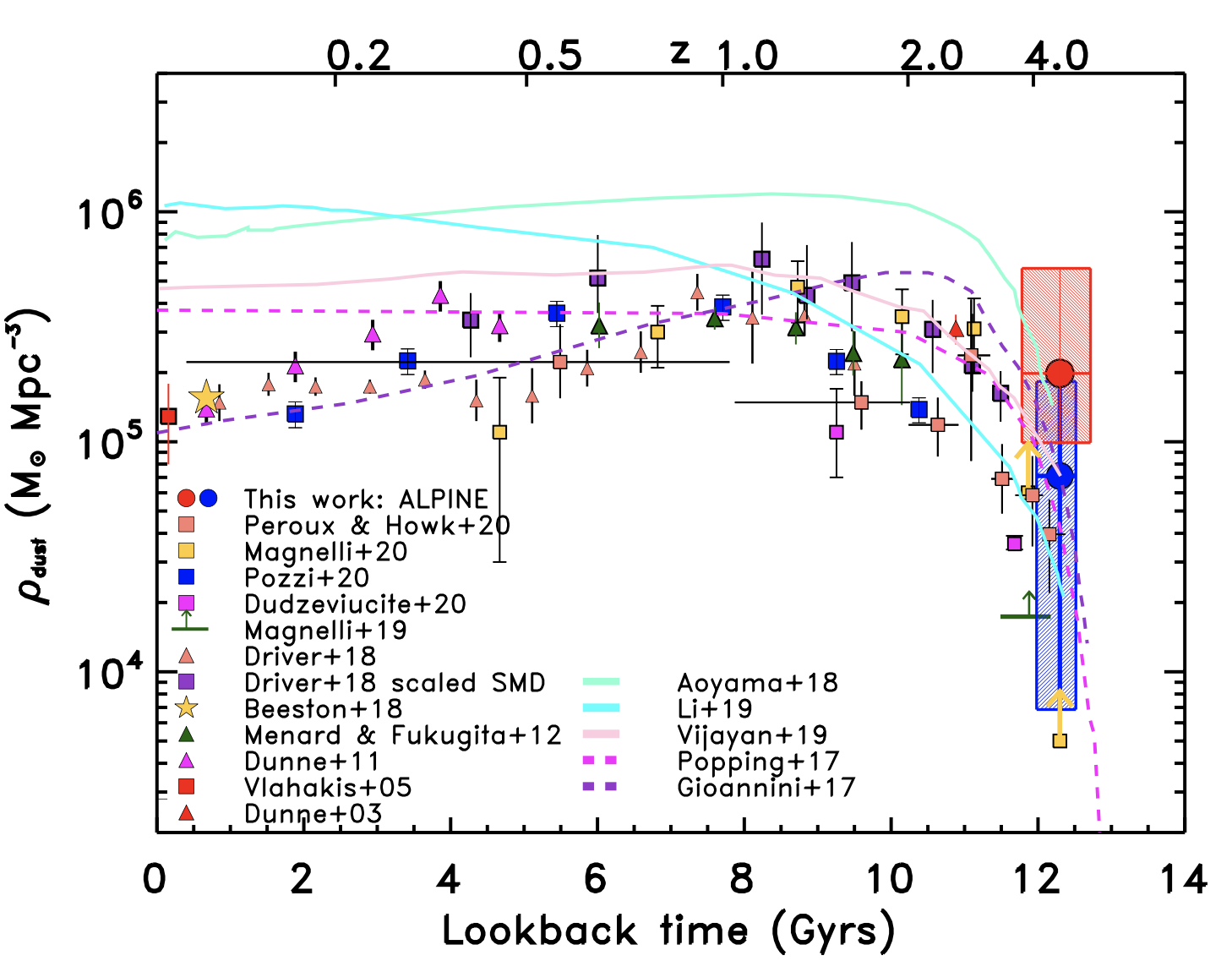}
\caption{Cosmic evolution of dust mass density. The dust in neutral gas (pink squares) provides constraints up to z=5.5 \cite{PerouxHowk20}. Altogether, these measurements put new constraints to the next generations of hydrodynamical simulations (displayed as lines) incorporating dust physics to understand the galaxy contributors to the global build-up of dust \cite{Pozzi21}.
}
\label{f:omegaDust}
\end{figure}

Several methods have been used to derive \OmegaDust\ (Figure~\ref{f:omegaDust}), as described recently in \cite{Eales24}. \cite{PerouxHowk20} provide estimates of the cosmic evolution of dust mass density based on quasar absorbers:
\begin{equation}
\OmegaDust \equiv \rho_{\rm dust} / \rhocrit =  \langle \dtg \rangle \, \Omega_{\rm gas},
\label{eqn:omegadust}
\end{equation}
The dust in neutral gas can be measured up to z=5.5. These estimates provide new constraints to the next generations of hydrodynamical simulations incorporating dust physics to understand the galaxy contributors to the global build-up of dust.

%%%%%%%%%%%%%%%%%
%%%%%%%%%%%%%%%%%
%SECTION 4
%%%%%%%%%%%%%%%%%
%%%%%%%%%%%%%%%%%

\section{The Galactic Baryon Cycle}
\label{sec:CGM}
So far in this chapter, the focus has been on the global quantities of the baryon cycling. This section instead describes observational techniques and theoretical findings to probe the physical properties of this exchanges on galaxy scales. Of particular interest is the study of the physical processes by which gas travels into, through, and out of galaxies. Specifically, making quantitative assessment of the transport of mass, metal, energy and momentum in and out of galaxies would advance our understanding of structure formation and provide a key input component to hydrodynamical cosmological simulations.

\subsection{Observational Techniques}

\begin{figure}[t]
\centering
\includegraphics[width=0.75\linewidth]{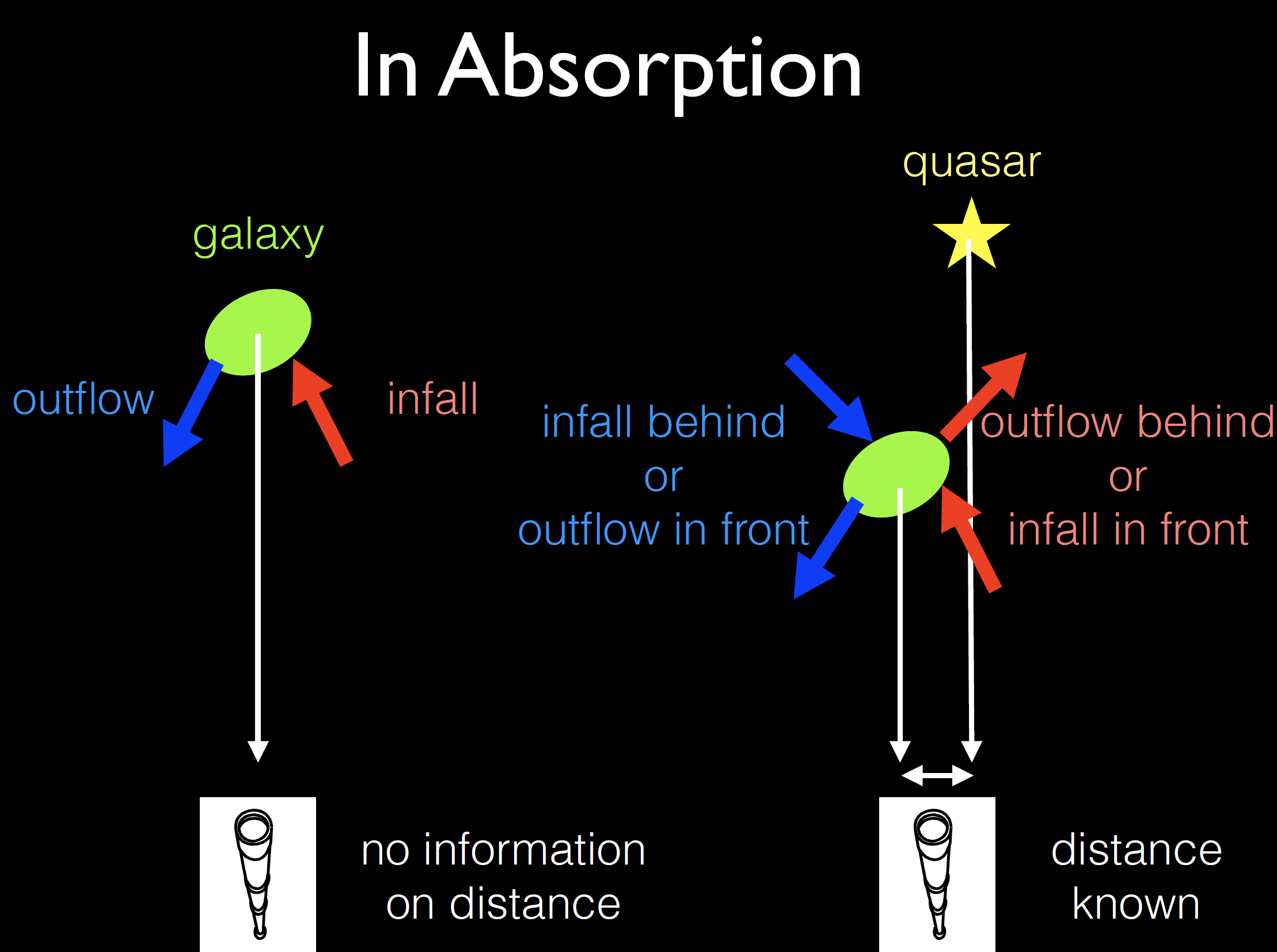}
\caption{Two techniques to observationally probe gas flows around galaxies. On the left, the so-called "down-the-barrel" absorption along the line-of-sight to a background galaxy. Such studies, provide definit information on the direction of the gas flows. When bright background sources are used (as illustrated on the right), complementary information on the distance between the CGM gas and the galaxy becomes available. 
 }
\label{f:DownTheBarrel}
\end{figure}

Observationally probing the physical properties of the gas flows has proven challenging. The CGM was historically first identified and explored using absorption line spectroscopy. The 3D structure around galaxies is now being probed by multiple lines-of-sight or extended arcs of lensed quasars or galaxies \cite{Lopez2018, Lopez2020, Tejos2021}, along with imaging of the cosmic web in emission in a variety of wavebands from radio to X-ray \cite{DeGraaff19, tanimura2019}. Observations indicate that galactic-scale metal-rich outflows with velocities of several hundred kilometers per second are ubiquitous in massive star-forming galaxies at high redshift \cite{pettini2001, shapley2003, veilleux2005, weiner2009}. Early results probing accretion used the so-called "down-the-barrel" absorption along the line-of-sight to a background galaxy. \cite{Rubin12} in particular, stacked several spectra and reported absorption redwards from the systemic redshift of the galaxies which they interpreted as signature of inflows. Such studies however, provide limited information on the distance between the gas flows and the objects, as the measured velocities are a combination of distance and gas flow velocity combined. On the contrary, probing the gas around foreground galaxies using unrelated background sources (quasars or other), {\it does} provide a measure of the distance between the systemic redshift and the background source's sightline as illustrated in Figure~\ref{f:DownTheBarrel}. The challenge then resides in the fact that the direction of the gas flow cannot be identified from the wavelength shift alone. Indeed, blueshifted absorption might be related to infalling gas behind the galaxy or equally to outflowing gas in front of the galaxy. Similarly, redshifted absorption might come from outflowing gas behind the galaxy or infalling gas in front of the galaxy \cite{bouche2010, Peroux2011, shen12}. Finally, depending on the viewing angle, all these processes could be probed simultaneously thus complexifying the interpretation of the signal. Additional information on the orientation of the star-forming disk galaxies is then necessary to help distinguish outflowing from accreting gas. Ideally, one would get a unique diagnostic for differentiating inflows and outflows.

\begin{figure}[t]
\includegraphics[width=0.55\linewidth]{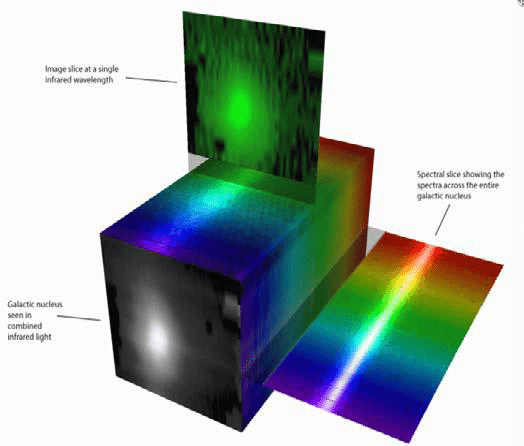}
\includegraphics[width=0.44\linewidth]{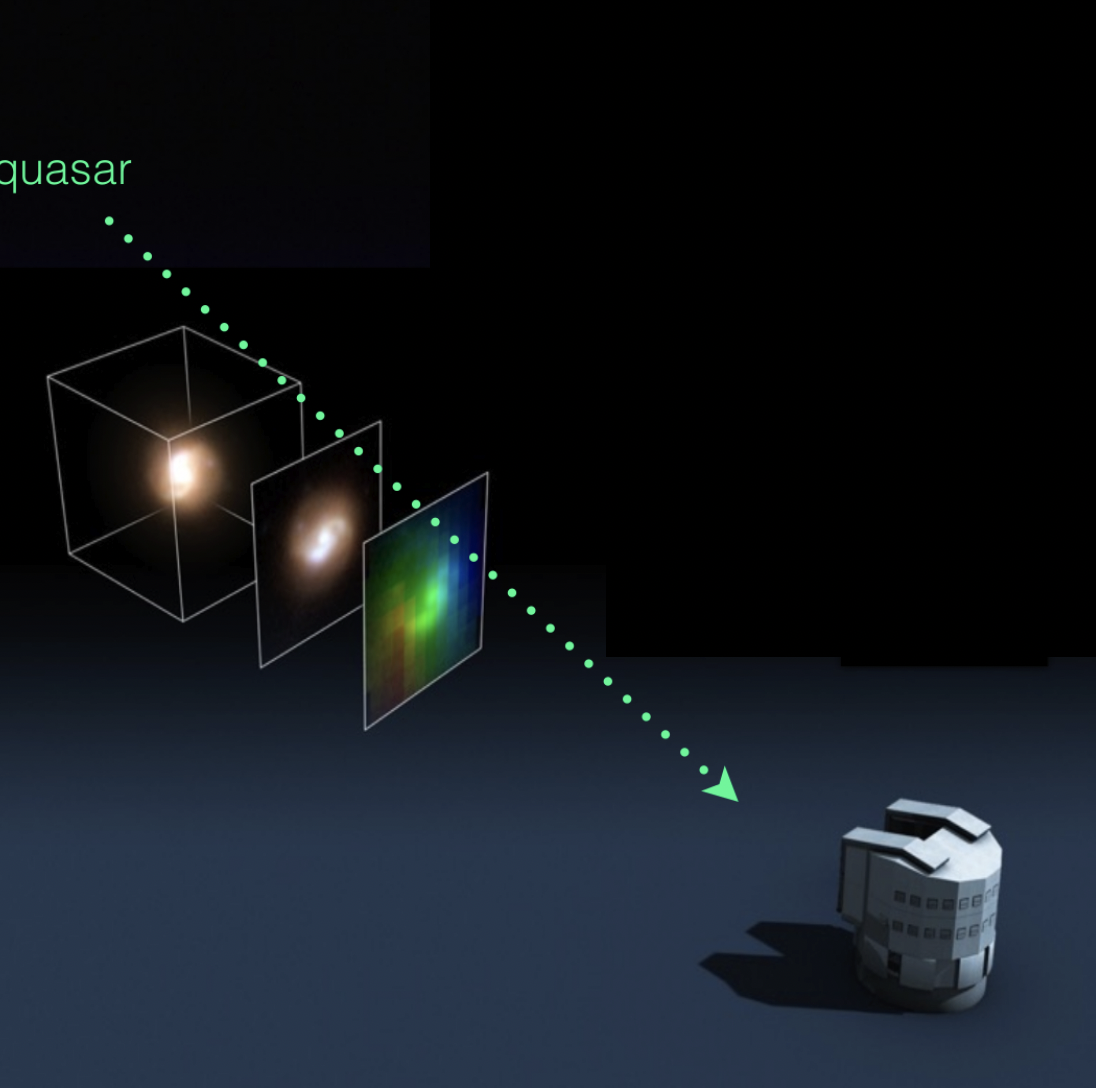}
\caption{A schematic view of Integral Field Unit (IFU) spectrographs. {\bf Left:} The 3D spectroscopy provided by IFU delivers an image where every pixel is a spectrum. {\bf Right:} IFUs are powerful tools for examining the absorption by gas in the CGM. This set-up has revealed extremely powerful in identifying the foreground galaxies in emission thus enabling to relate its physical properties with those of the CGM gas probed in absorption.}
\label{f:IFU}
\end{figure}

Associating quasar absorbers with emitting galaxies have made huge progresses in the last decade. Integral field unit spectrographs (IFUs) have emerged as powerful tools to relate the absorption by gas in the CGM and the foreground galaxy. The 3D spectroscopy provided by IFU delivers an image where every pixel is a spectrum (Figure~\ref{f:IFU}, left panel). This set-up has revealed extremely powerful in identifying the foreground galaxies in emission thus enabling to relate its physical properties with those of the CGM gas probed in absorption as illustrated in Figure~\ref{f:IFU} (right panel). Overall, these
galaxies are a heterogeneous population: they are not just the most
luminous galaxies, but include dwarf and low surface brightness
galaxies. Therefore, strong \lya\ quasar absorbers form a sample of systems unbiased as regards to luminosity, specific morphology, or emission line strength. Thus, they uniquely enable studies of metallicity and HI evolution over a large
redshift range. Interestingly, the Milky Way itself is detected in the spectrum of the low-redshift quasars as DLAs \cite{DeCia21}.

Early efforts started with the Adaptive Optics-equipped near-infrared spectrograph VLT/SINFONI \cite{bouche2007, Peroux2011, peroux2013, schroetter2015, peroux2016}. The potential of this technique for studying the CGM with the wide-field optical spectrograph VLT/MUSE has been further demonstrated \cite{schroetter2016, bouche2016, schroetter2019, zabl2019, muzahid2020, lofthouse2020} as well as with the high spectral resolution optical 3D spectrograph Keck/KCWI \cite{martin19,nielsen2020}. At $z_{\rm abs}<$1, the MUSE-ALMA Halos survey has measured the kinematics of the neutral, molecular and ionised gas of the multi-phase CGM \cite{peroux2017, rahmani2018a, rahmani2018b, klitsch2018, peroux2019, hamanowicz2020, weng23a, weng23b}. Numerous other surveys comfort findings of the detection of multiple galaxies associated with strong quasar absorbers \cite{Lofthouse20, Dutta2020, schroetter2015, zabl2019, Muzahid21, Dutta2023}, see also reports from Michele Fumagalli's chapter of this volume \cite{Fumagalli24}.
%{\blue see also reports in Michele's chapter}. 
Additionally, there has been recently a number of important works revealing the CGM properties on small scales to probe in particular effect of turbulence \cite{Fensch23, Ramesh23a, Ramesh23b, Gronke22, Fielding23, Das24}. By measuring the velocity structure functions (VSFs), one can study supersonic turbulence in local HII regions. \cite{ChenMandy24} have used velocity maps gathered from 3D spectroscopy to calculate a 2$^{\rm nd}$-order VSF and demonstrate it exhibits a power-law slope in agreement with the Kolmogorov expectation.

\subsection{Metal Abundance Determination}
\label{subsec:ab}

In Section~\ref{subsec:metals}, we have reported the global metal mass density of the neutral gas as a function of cosmic time. Here, we now focus on how these metals are distributed spatially around galaxies. A decade ago, larger sample of quasar absorbers indicated a bimodality in its metallicity distribution \cite{Lehner2013}. This puzzling result could be interpreted as signature of gas flows. Indeed, naively one would expect infalling gas to be pristine gas while outflows would be metal-enriched due to star-formation processes taking place inside galaxies. While not reproduced at other redshifts or at different column densities \cite{Berg2023} nor in simulations \cite{Hafen2017}, these interesting results posed the question of whether metallicity of the gas can be used to discriminate accretion from galactic winds in absorption studies.

\begin{figure}[t]
\includegraphics[width=0.5\linewidth]{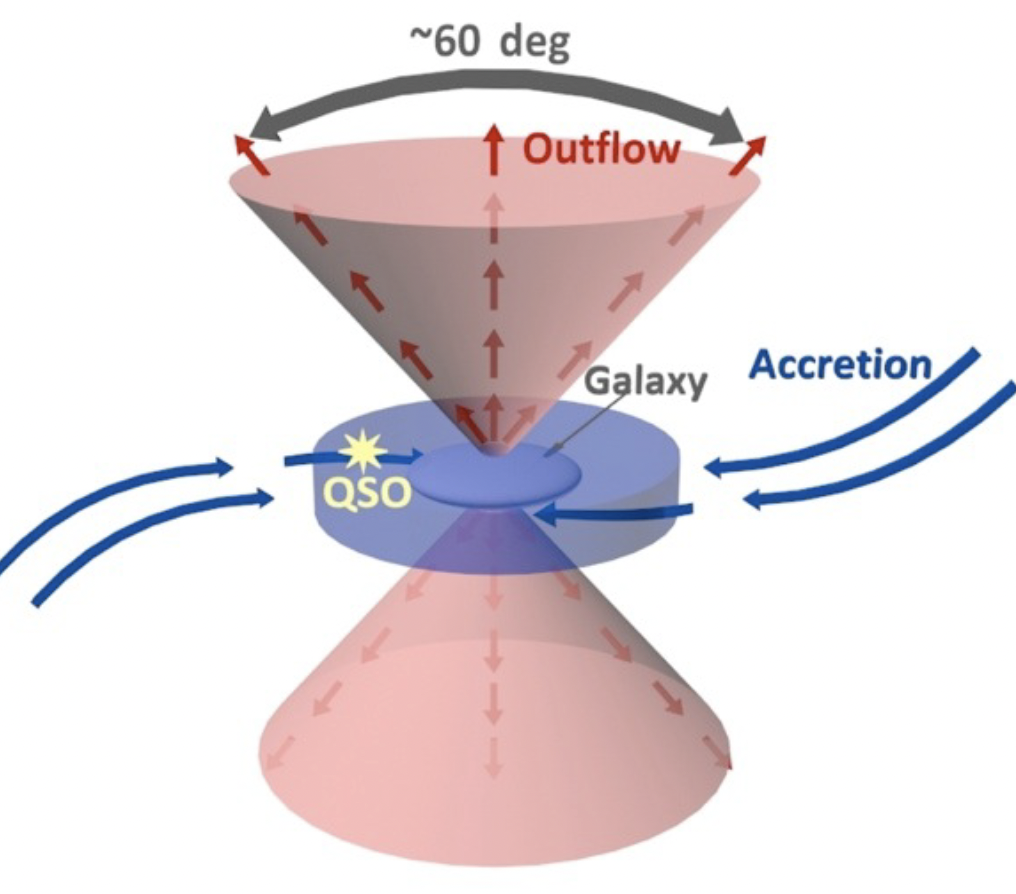}
\includegraphics[width=0.5\linewidth]{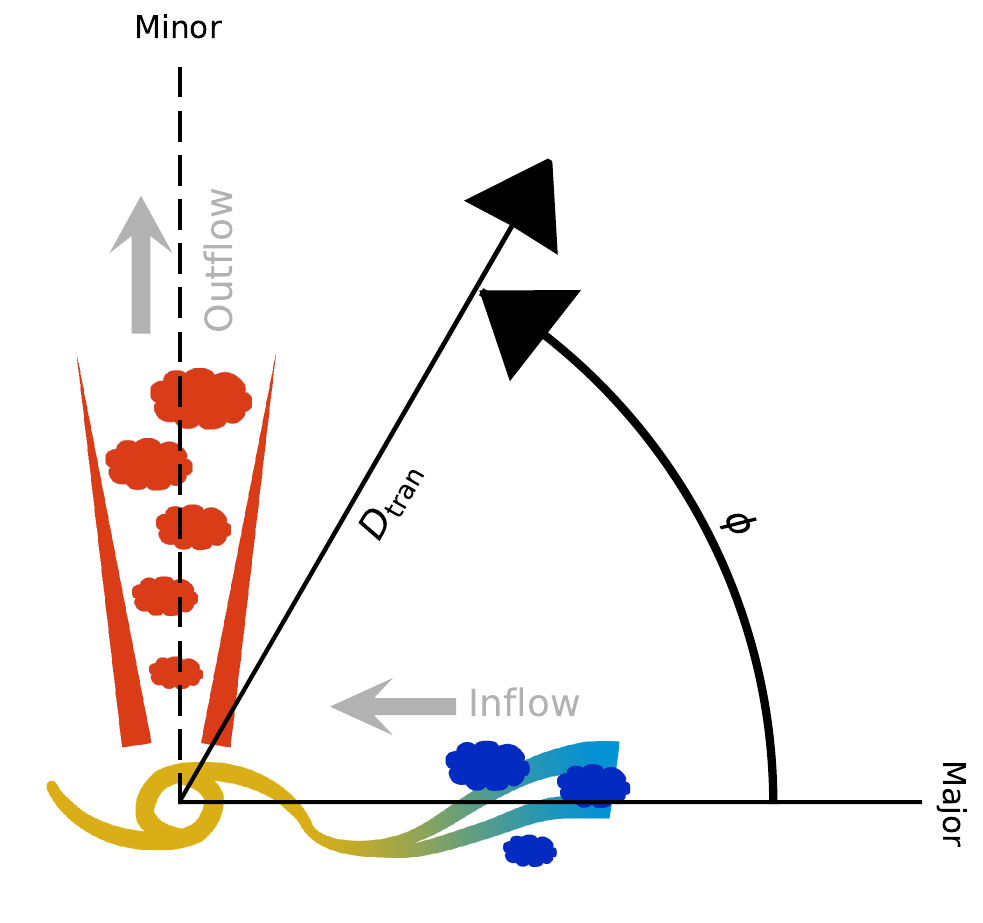}
\caption{A canonical view of the gas flows around galaxies. {\bf Left:} The accreting material is expected to be co-rotating with the central disk in the form of a warped, extended cold gaseous disk, while strong radial/bi-conical outflows are ejected perpendicular to the disk. Gas accretes onto galaxies from the cosmic web filaments, outflowing gas preferentially leaves the galaxy following the path of least resistance, along its minor axis \cite{Bouche17}. {\bf Right:} The azimuthal angle, $\Phi$, is defined as the galiocentric angle with respect to the major axis of the galaxy \cite{Chen21}.
}
\label{f:canonical_view}
\end{figure}

\subsection{Geometrical Argument}
\label{subsec:azi}

Contemporary to these findings, simulations reported evidence of co-rotating gas around galaxies \cite{Stewart11}. In this canonical view, the accreting material is expected to be co-rotating with the central disk in the form of a warped, extended cold gaseous disk, while strong radial/bi-conical outflows are ejected perpendicular to the disk as illustrated in the left panel of Figure~\ref{f:canonical_view}. In particular, the gas kinematics are expected to be offset by about 100 km/s from the galaxy’s systemic velocity and these kinematic signatures of gas accretion should be observable in suitable quasar absorption line systems. These results led to a picture where gas accretes onto galaxies from the cosmic web filaments, while outflowing gas preferentially leaves the galaxy following the path of least resistance, along its minor axis. Where they compete, galactic winds prevent infall of material from regions above and below the disc plane \cite{brook2011,mitchell20a, defelippis20}. As a result, inflowing gas is almost co-planar with the major axis of the galaxy \cite{shen12, vandevoort2012a}.

Observationally, a common approach to characterise these geometrical effects is to measure the `azimuthal angle' between the absorber and the projected major axis of the galaxy \cite{bouche2012a}. A small (large) azimuthal angle aligns with the major (minor) axis of the galaxy. In this way, observations have broadly found that absorption is not isotropic -- it depends on the orientation of the galaxy as illustrated in the right panel of Figure~\ref{f:canonical_view}. \cite{bordoloi2011} found a strong azimuthal dependence of \mgii\ absorption, implying the presence of a bipolar outflow aligned along the disk rotation axis. Concomitantly, \cite{bouche2012a} and \cite{Kacprzak2012} found a bimodal distribution of azimuthal angles hosting strong \mgii\ absorption, suggestive of bipolar outflows contributing to the cold gas cross-section \cite{schroetter2019, martin2019}. Such signatures are also found in hotter gas phases including ionised gas traced by OVI absorption \cite{Kacprzak2015}. Interestingly, these trends are not as clearly seen towards AGN-selected galaxy samples \cite{Kacprzak2015}, implying a stronger connection to stellar feedback.

\begin{figure}[t]
\centering
\includegraphics[width=0.9\linewidth]{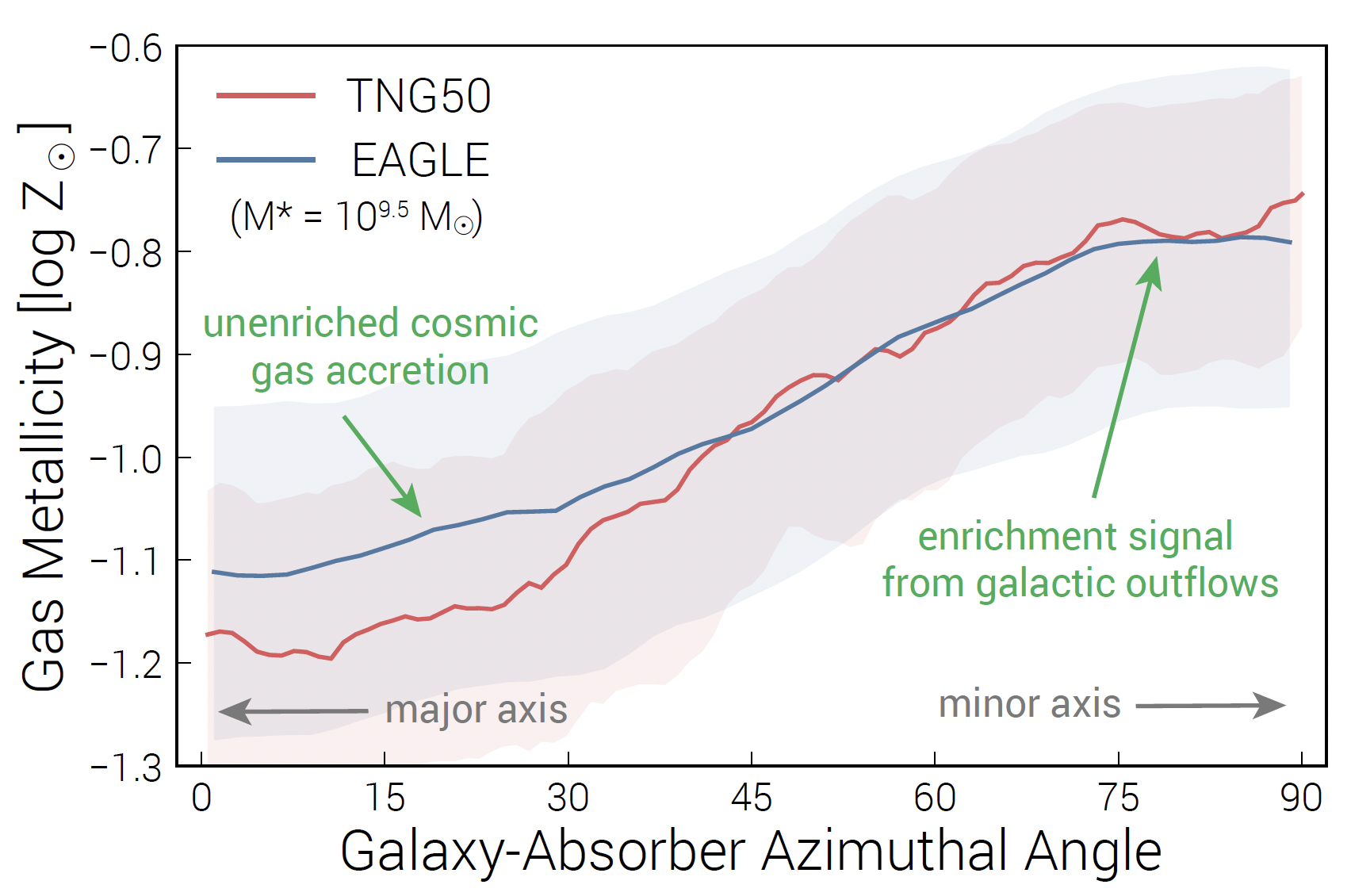}
\caption{Predictions from two cosmological hydrodynamical simulations for the relation between CGM metallicity and azimuthal angle. The two models (TNG50 and EAGLE) indicate that the average metallicity of the CGM is higher along the minor with respect to the galaxies' major axes. Therefore, despite the numerous mixing processes at play, there is overall a significant angular dependence of the CGM metallicity \cite{Peroux2020, vandeVoort2021, Wendt2021}.}
\label{f:metal_azi}
\end{figure}

Pushing these findings one step further, cosmological hydrodynamical simulations indicate that the CGM gas physical properties of normal galaxies vary with angular orientation. Indeed, \cite{Peroux2020} found that some properties of the CGM evolve with azimuthal angle in both  TNG50 and EAGLE simulations (Figure~\ref{f:metal_azi}). These results support early findings of accretion more likely to take place in the disk of the plane, while winds are strongest along the minor axis. It is interesting to note that these results are common to both types of simulations despite them having different physical models for AGN feedback as well as different numerical solvers. Together, there appears to be a global signature of the metal content of the CGM gas with angular orientation even though the gaseous atmosphere of galaxies is the location of numerous mixing processes. These results have also been supported by other studies \cite{vandeVoort2021, Wendt2021}. Observationally, 3D forward modelling is used to recover morphological and kinematical parameters of the gas from the detection of emission lines \cite{peroux2019, Szakacs2021}. Combined with metallicity measurements, such calculations provide a direct measurement of the azimuthal angle and give fresh clues on the gas flow direction. 

Similarly, simulations predicted for a long-time that magnetic fields play a key role in the physical properties of the CGM. Magnetic fields have now been detected in the CGM at few tenth of microGauss level, as traced by Faraday rotation measure \cite{Bockmann23, Mannings23, vandeVoort2021, Ramesh23a, Pakmor23}. Interestingly, it has been reported that magnetic fields are also stronger along the minor versus major axes of galaxies \cite{Heesen23, Ramesh23}.

Collectively, these results suggest a picture whereby the cooler component of the CGM at $\sim 10^4$\,K originates from major axis-fed inflows, and/or recycled gas, together with minor axis-driven outflows. Metallicity and azimuthal angle, two readily accessible observables, can be used to constrain the direction of the gas flow in absorption. The powerful synergy of emission and absorption provides unique combination of the metallicity of the neutral and ionised gas, the geometry of the system and the kinematics of the various gaseous phases. Because the gas in CGM is multiphase, multi-wavelength studies have proven powerful to characterise the neutral, ionised and molecular gas probed at mm wavelengths. In addition, optical spectroscopy breaks the ALMA single emission-line degeneracy for redshift identification. Analyses based on such datasets report a coupling between the neutral, ionised and molecular gas \cite{Szakacs2021}.

\subsection{Mass Loading Factor}
\label{subsec:loading}

Large galaxy samples observed with IFUs increasingly constrain how the physical properties of the CGM vary with angular orientation. Thanks to detailed kinematics studies, it is possible to reconstruct the orientation of the foreground galaxy with respect to the CGM gas probed in absorption along the line-of-sight to the background quasar. A key component is a measure of the mass loading factor, $\eta$, which compares the gas mass loss through outflows to the consumption in star formation as follows:

\begin{equation}
    \eta = \dot M_{\rm out}/SFR
\end{equation}

For a hollow bi-conical flow, the mass outflow rate is  \cite{bouche2012a, schroetter2015}:

\begin{equation}
 \dot M_{\rm out} \approx \mu \cdot N_{\rm H}(b) \cdot b \cdot V_{\rm out} \cdot \frac{\pi}{2} \cdot \theta_{\rm max}
\label{eqn:Mout}    
\end{equation}

where $\mu$ is the mean mass per hydrogen particle, $b$ the impact parameter, $\theta$ is the cone opening angle (where $\theta_{\rm in}$ is defined from the central axis, and the cone subtends an area of $\pi \cdot \theta_{\rm max}^2$ with $\theta_{\rm in}$ the opening angle of the inner empty cone), $V_{\rm out}$ the outflow velocity and $N_{\rm H}(b)$ the hydrogen column density at the $b$ distance. Typically, the parameters $V_{\rm out}$, $b$ and the cone opening angle are constrained from observations. Recent results report values of the mass loading factor of order unity. These estimates have proven challenging to reproduce in simulations. 

For a 'down-the-barrel' observations of such a wind, the
mass outflow rate, reduces to:

\begin{equation}
\eta \propto \mu N_{\rm H}\, R_0\, V_{\rm out}\, \Omega
\label{eqn:Mout_barrel}    
\end{equation}

where $R_0$ is the radius at which the wind is launched and $\Omega$ the solid angle subtended by an outflow with spherical geometry \cite{Heckman00,Martin05,Martin12}. 

%%%%%%%%%%%%%%%%
%%%%%%%%%%%%%%%%%

\clearpage
\begin{trailer}{Hands-on to analyzing cosmological galaxy formation simulations}

\subsection*{[Hands-on \#4] The baryon cycle: measuring mass flow rates}

So far we have only looked at the distribution of gas in galaxies and halos, and the physical properties of that gas. What about how this gas is moving? We can consider the amount (i.e. rate) of gas inflow, and gas outflow, through the CGM.

The radial mass flux can be computed as

$$ \dot{M} = \frac{\partial M}{\partial t} = \frac{1}{\Delta r} \sum_{i} \left( \frac{\vec{v}_i \cdot \vec{r}_i}{|r_i|} m_i \right) $$

where the subscript $i$ enumerates gas cells with masses $m_i$ in a particular volume of space, which we can take as a spherical shell with some thickness $\Delta r$ from the center of a (central) galaxy. Each gas cell position $\vec{r}_i$ is **relative** to the subhalo center, and the velocity $\vec{v}_i$ is **relative** to the subhalo bulk motion.

The term $\frac{\vec{v}_i \cdot \vec{r}_i}{|r_i|}$ is the radial velocity $v_{\rm rad}$, and if $v_{\rm rad} > 0$ we have outflow, while $v_{\rm rad} < 0$ denotes inflow.

\subsection*{Exercise}

We will again focus first on a single halo.

\begin{enumerate}
\item Pick a simulation and redshift of interest. Pick a halo of interest. (e.g. try the 100$^{\rm th}$ most massive halo in TNG100-1 at $z=0$).
\item Load the needed fields to compute the distance, and radial velocity, of each gas cell.
\item Use the distances to compute a radial mass density profile, i.e. in a number of bins of distance, sum up the total gas mass, and divide by the volume of that spherical shell. Plot the result in $\rm{M}_\odot / \rm{kpc}^3$ as a function of distance in $\rm{kpc}$.
\end{enumerate}

\subsection*{Exercise}

\begin{enumerate}
\item Define a "CGM region", i.e. a minimum and maximum radius.
\item Use the equation above to compute the total inflow rate, and total outflow rate, of gas in the CGM (in $\rm{M}_\odot / \rm{yr}^{-1}$). Which is larger? How do these values compare to the star formation rate of the galaxy itself, the so-called mass loading factor $\eta$?
\item Calculate and (over)plot the radial profiles of inflow rate and outflow rate, from the center of the halo to its virial radius. How does each change with distance from the galaxy?
\end{enumerate}

\subsection*{Exercise (Challenge)}

We have not yet seen the "merger trees". We can explore how we follow a subhalo through time, and see how it evolves. We will use the `SubLink` merger tree.

\begin{enumerate}
\item Look at the documentation for the \textsc{il.sublink.loadTree()} function.
\item Load the "main progenitor branch" (MPB) of your chosen halo. (Careful! Note the difference between halo and subhalo).
\item In a loop, go back through the earlier snapshots (maybe every 10th, to save time).
\item For each earlier snapshot, find the subhalo ID at that snapshot, load its gas data, and compute total inflow and outflow rates as above.
\item Make a plot of inflow and outflow rate versus time (e.g. snapshot number, redshift, or age of the Universe).
\end{enumerate}

\subsection*{Exercise (Challenge)}

\textbf{Note:} we have so far ignored the periodic boundary conditions of the box!

If you loaded any halo (or snapshot data in general) near the edges of the box, the particle/cell \textsc{Coordinates} can "wrap" around an edge. Some would then appear near a value of 0, and some near a value of BoxSize. This should \textbf{always} be corrected e.g. when computing relative distances, such as the radial distance above.

\begin{enumerate}
\item Use the catalog to find a (large) halo near the edge of the simulation box.
\item Load its particle/cells, and make a visualisation. What goes wrong?
\item Write a function to correct for the periodic boundary conditions (see "(A) Restrict particle coordinates to the simulation box" on the Wikipedia page for ``periodic boundary conditions''), and repeat your plot.
\end{enumerate}

\subsection*{Hands-on: what's next?}

Consider how galaxy formation simulations can be useful for your particular science topic or goals? The topics and exercises explored in these four hands-on sections give a starting point, and can be expanded further in order to analyse any other quantities of interest from the simulations.

\end{trailer}
\clearpage

%%%%%%%%%%%%%%%%%

\subsection{Elusive Accretion}
\label{subsec:acc}

\begin{figure}[t]
\centering
\includegraphics[width=0.85\linewidth]{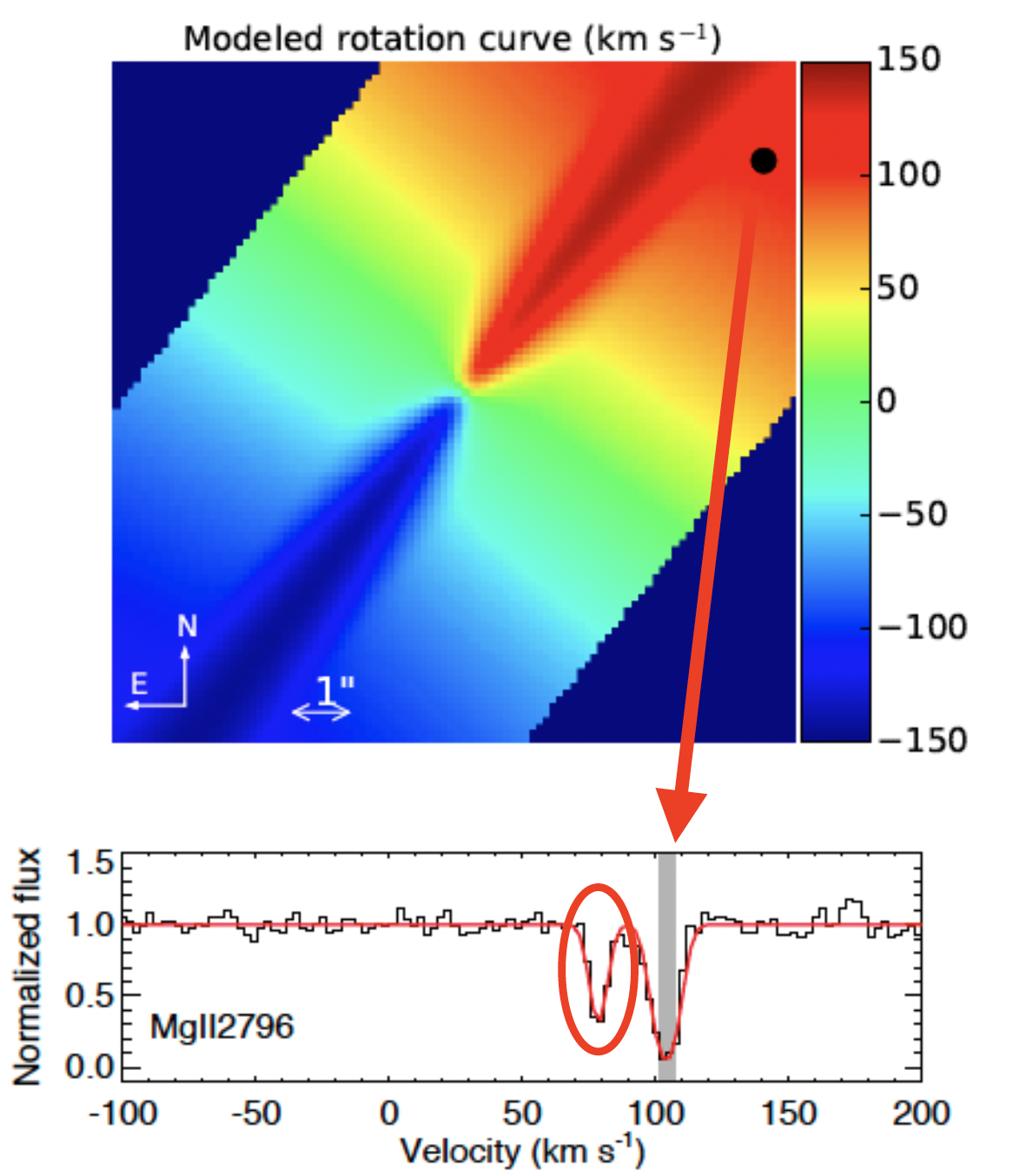}
\caption{Example of possible observed signature of accretion. The top panel displays the extended modelled velocity field of the foreground galaxy, with the quasar line-of-sight skyp position indicated with a black dot. The bottom panel displays the normalised quasar spectra along this sightline with velocity zero set at the systemic redshift of the galaxy. Interestingly, the main absorption MgII component precisely relates to the velocity expected from the rotation of the disk. Conversely, the weaker component at lower velocity is more metal-poor and has three times the specific angular momentum, a possible indication of gas accretion \cite{Rahmani18}.}
\label{f:acc}
\end{figure}

The current standard theoretical model indicates that early galaxies accrete their gas from the intergalactic medium through cosmic web filaments \cite{Dekel2009a}. The accretion results in a  multiphase (hot and cold) gas that enters into the halos and fuel star formation. This gas is further mixed and heated by several physical processes including supernova explosions and photoionisation. 

Observational probes of infall are difficult to obtain, possibly due to the small cross-section or because accretion is swamped by outflows in studies of absorption back-illuminated by the galaxy \cite{Rubin12, rubin2014}. Notably, \cite{Bouche13} report accretion along the filament feeding the galaxy inflowing mass rate comparable to the star formation rate at z$\sim$2.2. Similarly, \cite{Rahmani18} find a possible indication of gas accretion using similar approach (Figure~\ref{f:acc}).

\begin{figure}[t]
\centering
\includegraphics[width=0.85\linewidth]{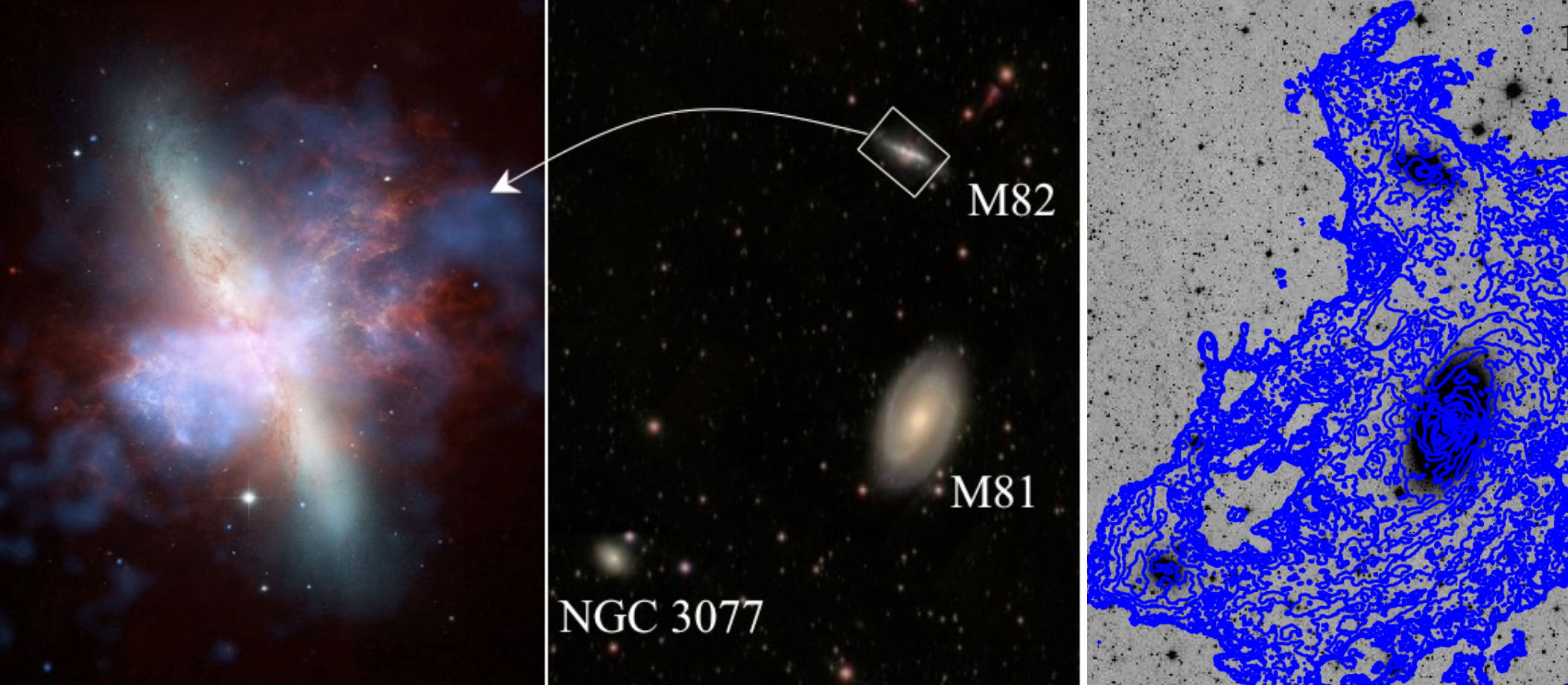}
\caption{Multiwavelength view of M81-M82 local group. The optical image in the center displays seemingly unrelated object. The dazzling new images however show extended HI gas emission spanning the whole space between the galaxies and more \cite{Sorgho19}. Zooming-on M82 as shown on the left image exhibits the school-case example of galactic winds. The neutral gas probed by quasar absorbers is likely a mix-bag of high-redshift equivalent of all these phenomenons.  }
\label{f:mental_pic}
\end{figure}

\vspace{0.5cm}
In conclusion, the mental picture one can draw from these overall findings is similar to the M81-M82 local group (Figure~\ref{f:mental_pic}). At optical wavelengths, this low-redshift analog displays seemingly unrelated group of galaxies (also including NGC3077), while spectacular recent radio observations from MeerKAT indicate extending HI gas in between the systems which could be the remnant of previous interactions and/or tidal streams \cite{deBlok18, Sorgho19}. Additionally, a close-up look at M82 displays the textbook case of bipolar galactic winds. Therefore, the gas observed in absorption is sensitive to these complex phenomena and therefore proves a powerful tool to probe the low density CGM gas.

%%%%%%%%%%%%%%%%%
%%%%%%%%%%%%%%%%%
%SECTION 5
%%%%%%%%%%%%%%%%%
%%%%%%%%%%%%%%%%%
\begin{figure}[t]
\centering
\includegraphics[width=0.85\linewidth]{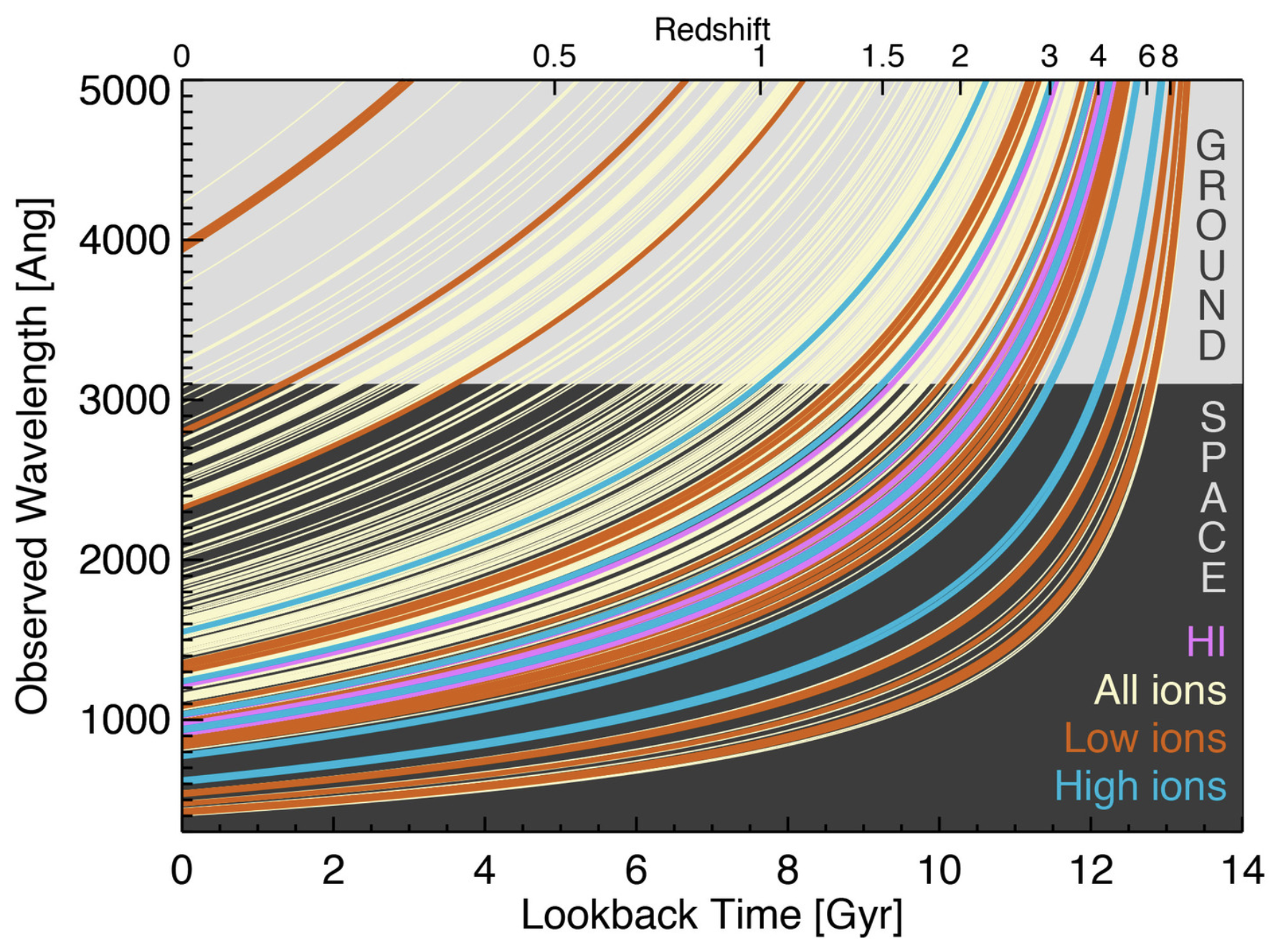}
\caption{Diagnostic ion and hydrogen lines as a function of wavelengths. Most lines fall at ultra-violet wavelengths which will be out of reach in the near infra-red, even once redshifted. This fundamental limitation triggers thoughts for the long-term vision of future international facilities. }
\label{f:UV}
\end{figure}

\section{Future Avenues}
\label{subsec:future}

In this final chapter, some of the emerging and future techniques to further our understanding of the baryon cycle are described. On aspect worth stressing is the pivotal role played by ultra-violet (UV) and optical wavelengths. While the major facility sensitive to UV wavelengths represented by the Hubble Space Telescope has been operated for decades, many of up-coming missions are optimised at near to infrared wavelengths both in space (James Webb Space Telescope, JWST) and on the ground (Extremely Large Telescopes, including the European ELT\footnote{https://elt.eso.org/}, GMT and TMT). Figure~\ref{f:UV} however illustrates the wealth of diagnostic lines available at bluer wavelengths which will be out of reach in the near infra-red, even once redshifted. This fundamental limitation triggers thoughts for the long-term vision of future international facilities.

\subsection{The Hot CGM Gas}
\label{subsec:xray}

\begin{figure}[t]
\centering
\includegraphics[width=\linewidth]{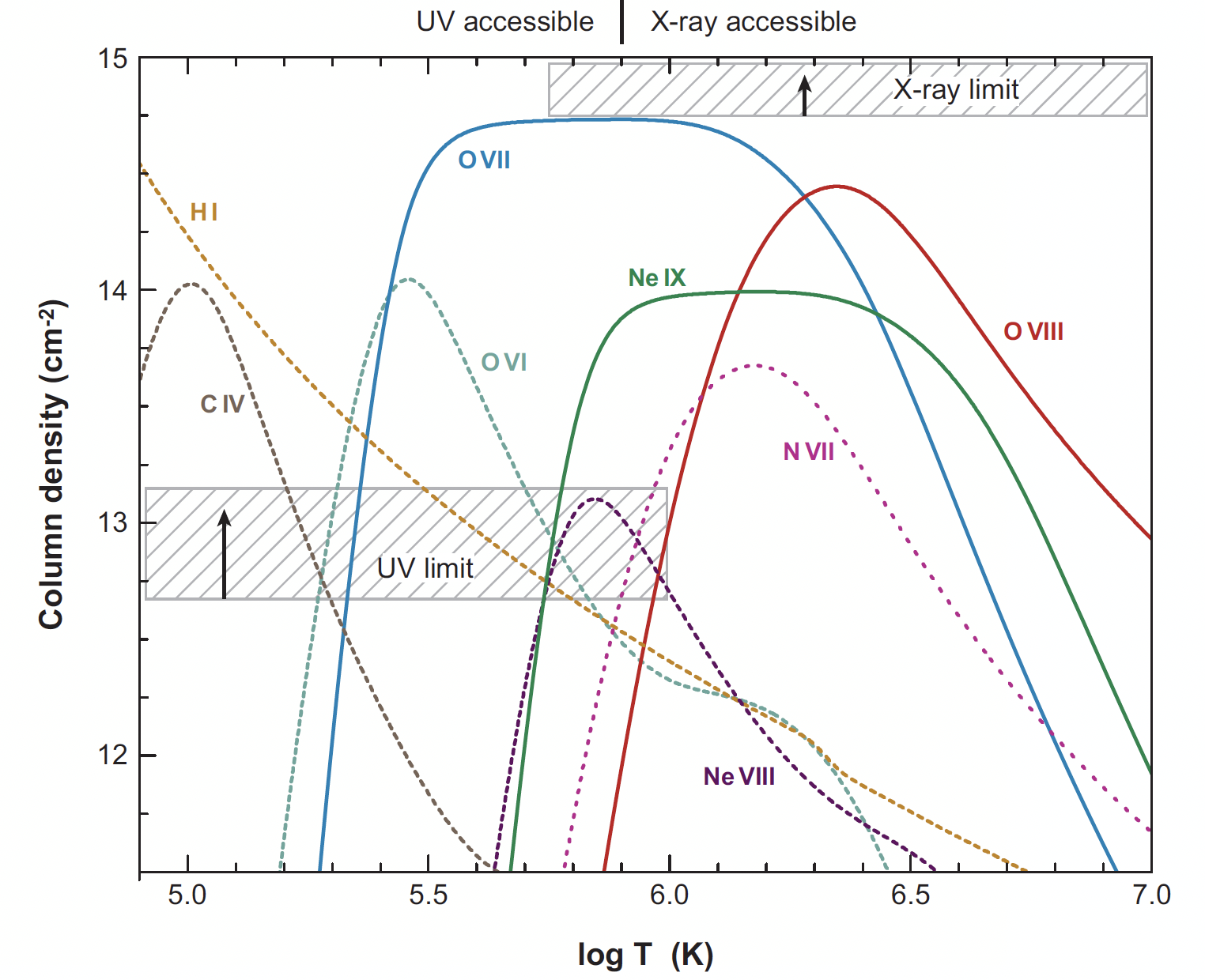}
\caption{Metal and hydrogen column densities as a function of temperature of the gas. These predicted column densities result from a photoionisation modelling  assuming a gas column density of 10$^{19}$ cm$^{-2}$ and metallicity 1/10$^{\rm th}$ solar. The UV and X-ray domains probe different temperatures, UV providing more sensitive constraints today (as illustrated by the ashed box) but the X-ray energies covering the hotter component of the CGM gas.  }
\label{f:Bregman}
\end{figure}

The hot phase of the CGM surrounding galaxies is the dominant baryonic reservoir of dark matter haloes. A major fraction of the baryons at lower redshift is found in hot gas (10$^5$-10$^6$K), including the gas in groups and clusters as well as the warm-hot intergalactic medium \cite{cen1999, dave2001}. The thermodynamical and cooling properties of that gas are largely responsible for the balance between cosmic gas accretion on to, and feedback-driven outflows from, the central galaxy \cite{Donahue22}. One way to observationally detect and characterize the hot CGM is through its X-ray emission. While less sensitive than measurements at UV wavelengths, the X-ray frequencies probe the hotter phase of the gas as illustrated by Figure~\ref{f:Bregman}.

X-rays are a key tool to probe massive clusters \cite{Pratt09, Vikhlinin09, McDonald13}, as well as groups \cite{Lovisari15, Eckert21}. While individual detections of the hot CGM are challenging for Milky Way-mass galaxies \cite{Anderson11, Bogdan13, Li13}, stacking with ROSAT \cite{Anderson15} and recently eROSITA \cite{Chadayammuri22, Comparat22, Zhang24a, Zhang24b} access for the first time this regime. We note that the latter remains challenging measurements related to accurately estimating the PSF and removing signature from X-ray binaries and AGN contamination. Predicting such observational signatures is one of the topics of {\it Hands-on \#3} (emission). More recently the hot gas has been probed through the distortions the gas imprints on the cosmic microwave background spectrum due to thermal and bulk motions of free electrons, the so-called thermal and kinetic Sunyaev-Zel'dovich effects \cite{lim2018, de-graaff2019, tanimura2019}. However, measuring the physical properties of the CGM, such as temperature, abundance, density, and velocity at X-ray energies is difficult at CCD spectral resolution \cite{Kraft22}. This motivates the need for future X-ray imaging spectrographs, including XRISM \cite{XRISM22}, HUBS \cite{Cui20}, LEM \cite{Kraft22} and ultimately Athena \cite{Nandra13}.

\begin{figure}[t]
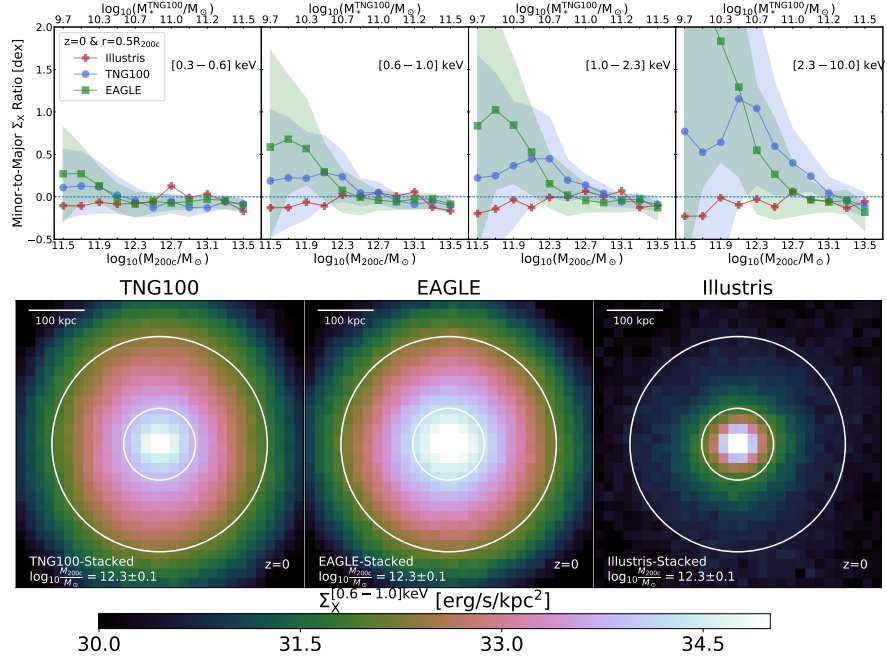

\centering
\includegraphics[width=\linewidth]{Truong21_Panela.png}
\includegraphics[width=\linewidth]{Truong21_Panelb.png}
\caption{Simulations predict an azimuthal signature in the distribution of the hot gas in the CGM of galaxies. The amount of hot gas is larger along the minor axis than along the major axis in EAGLE and TGN100 but not in Illustris (bottom panel). The predictions are also a strong function of the observed energies (top panel) \cite{Truong2021, Truong24}. }
\label{f:hot_gas}
\end{figure}

Interestingly, some simulations predict again an azimuthal signature in the distribution of the hot gas in the CGM of galaxies. Figure~\ref{f:hot_gas} shows that the amount of hot gas is higher along the minor axis than along the major axis in EAGLE and TGN100 but not in Illustris. The predictions are also a strong function of the observed energies \cite{Truong24}.

\subsection{The Power of Statistics}
\label{subsec:stat}

\begin{figure}[t]
\centering
\includegraphics[width=0.65\linewidth]{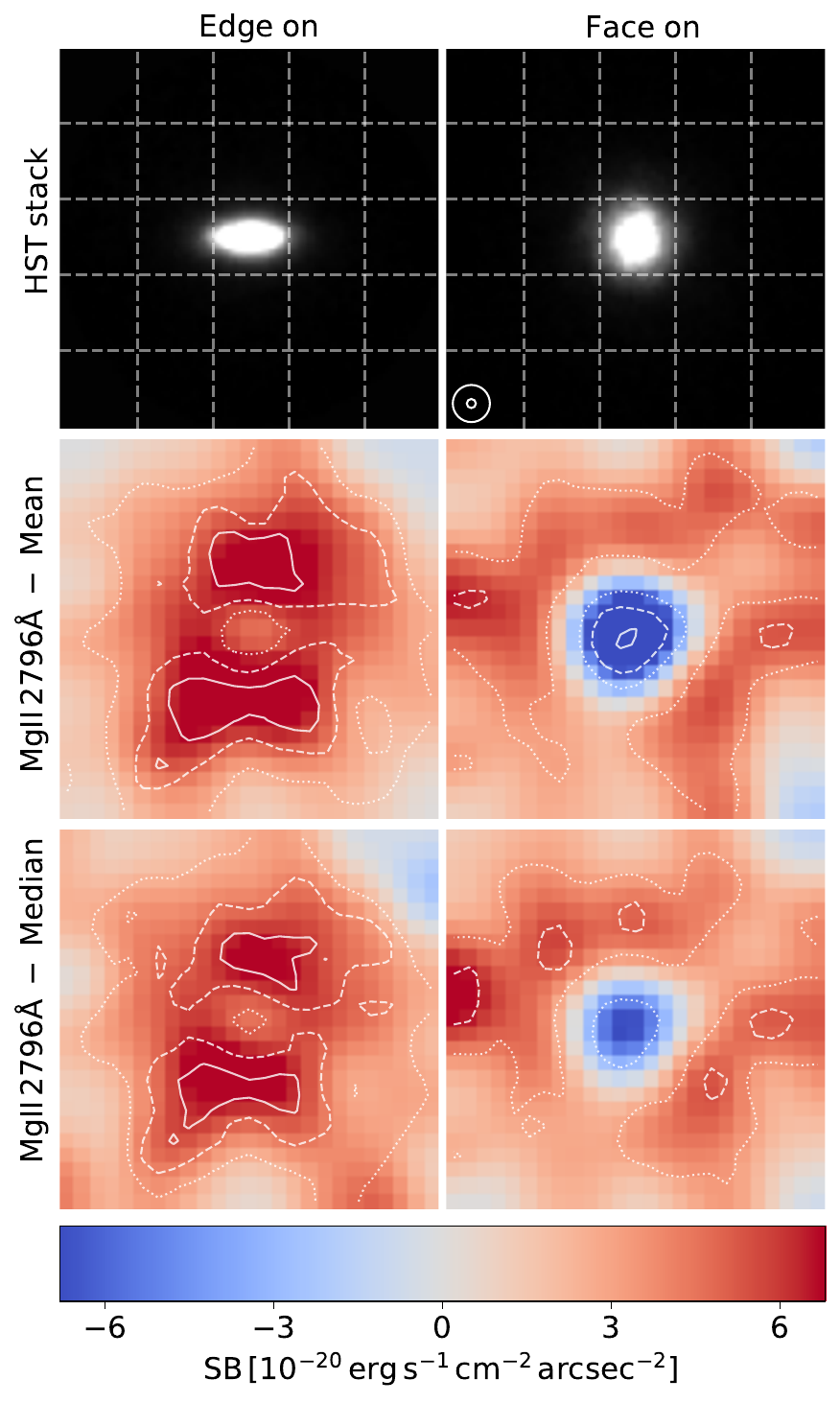}
\caption{The stacked stellar continuum and MgII emission. {\bf Top panels:} The stacked average of HST images  representing the stellar component. The mean ({\bf Middle panels}) and median ({\bf Bottom panels})
stacked MgII 2,796 \AA\ pseudo-narrow band images, smoothed by a Gaussian kernel of
width 0.4". These results display a strong dependence of the detected
signal on the inclination of the central galaxy, with edge-on galaxies clearly showing enhanced MgII emission 
along the minor axis, whereas face-on galaxies show much weaker and more isotropic emission \cite{Guo24}.}
\label{f:Guo24}
\end{figure}

The \lya\ line is readily observable from the ground at z$\ge 1.7$. Historically, extremely large surveys of thousands of quasar absorbers have brought such studies in a new era \cite{noterdaeme2012, bird2017,
parks2018}. By producing a uniform set of data products in numbers of million of quasar spectra, these projects dramatically advanced the field. Today, instruments on 4-m class telescopes provide a new wealth of low and medium-resolution quasar spectra in extremely large numbers, notably the DESI experiment, the WEAVE-QSO survey, and soon the 4MOST instrument \cite{merloni2019, Peroux23}. Clearly, such surveys will require specific approaches, including machine learning techniques, to analyse these large data outputs \cite{Szakacs23}. In the near future, dedicated facilities such as the proposed Maunakea Spectroscopic Explorer (MSE) \cite{Babusiaux19} and Wide-field Spectroscopic Telescope (WST) \cite{Mainieri24} will bring such efforts to 10-m class telescopes.

These large number of quasar absorber-galaxy pairs have enabled to assess the strength of the absorbers by computing the optical depth of the metal doublets for different phases of the gas from cold at T=10$^4$K (MgII) to warm at T=10$^5$K (CIV). The gas is positioned spatially and in velocity space with respect to the associated galaxies \cite{Rudie12, Turner14}. Recent studies reported results from \lya\ stacking \cite{Kimm22, Blaizot23}. In these works, spaxels with similar transversal distance from the quasar sightline are averaged to create the emission map in distance-velocity space. \cite{Chen21} specifically built 2-D maps of the hydrogen optical depth around galaxies have been built based 5000 pairs. The map has been smoothed with a Gaussian kernel with standard deviation of half of the bin size. The authors report an excess feature marginally significant statistically. They find that stacks into distinct azimuthal directions display a level of asymmetry relative to  the total \lya\ flux which is small. Last year, new reports of MgII emission stacking indicated different evidence \cite{Nelson21, Guo24}. The stacking on HST images of narrow MgII pseudo-narrow band images taken from deep VLT/MUSE observations display a clear emission signal which is a strong function of azimuthal angle as illustrated in Figure~\ref{f:Guo24} \cite{Guo24}. Future facilities such as 4MOST will provide a significant increase in the numbers of metal-absorber/galaxy pairs available for such studies with respect to previous works \cite{Peroux23}.

\subsection{IGM Tomography}
\label{subsec:tomo}

\begin{figure}[t]
\centering
\includegraphics[width=0.8\linewidth]{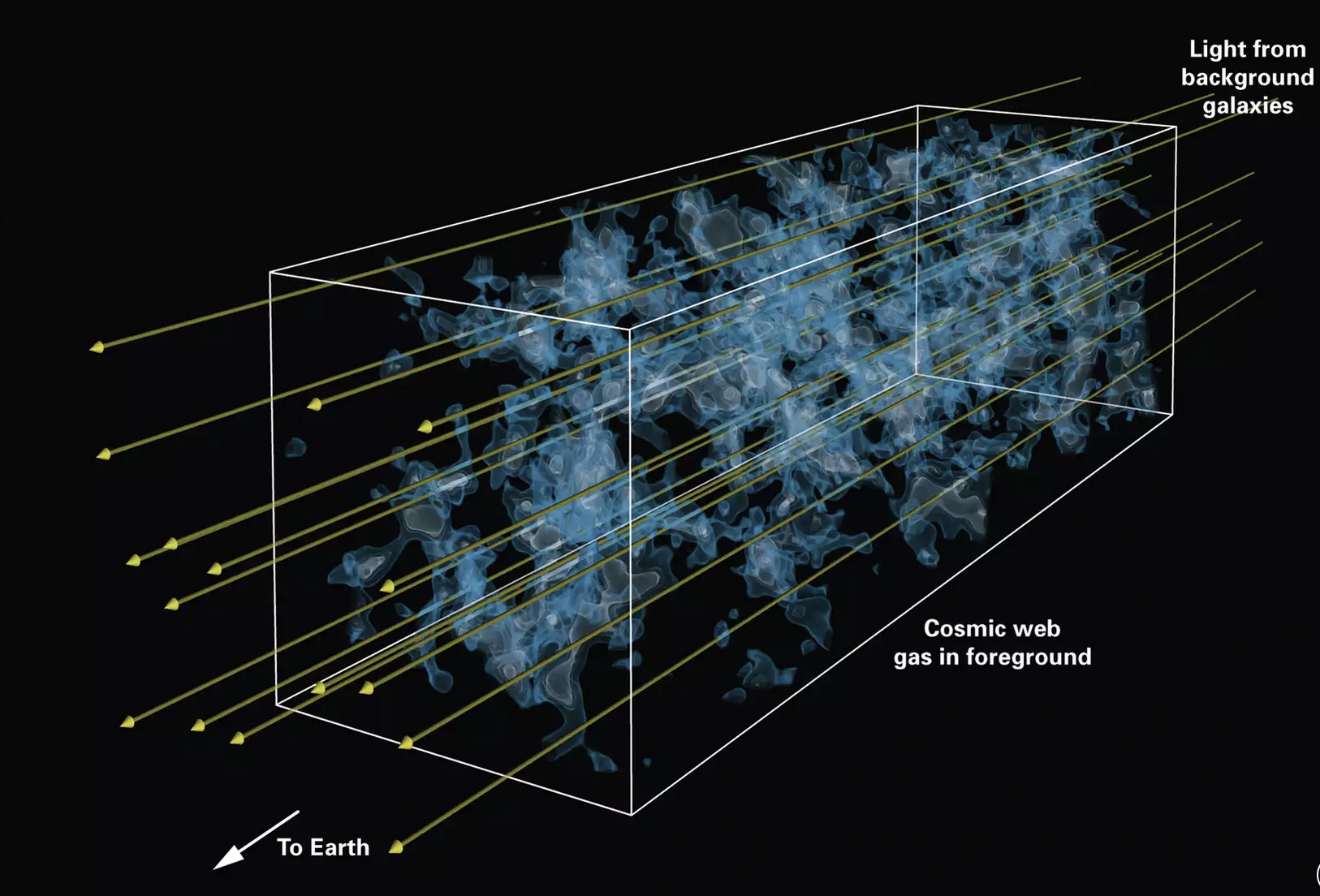}
\caption{Sketch illustrating the tomography technique. In a nutshell, gaseous structure can be traced by the absorption profiles they imprint onto the spectra of densely sampled bright background sources, such as quasars and even the more numerous star-forming galaxies. The new generation of multi-object spectroscopic surveys will offer the prospect to reconstruct robustly the 3D structure of the cosmic web (PFS, ELT/MOSAIC). }
\label{f:tomo}
\end{figure}

Ultimately, mapping the 3D spatial distribution of various phases of the gas would provide key to a full understanding the baryon's budget and small scales gas exchanges. To be able to characterize the connection of this large scale structure to halos with 3D tomographic maps, experiments making use of quasar absorbers have been proposed \cite{pichon2001, lee2016}. In a nutshell, gaseous structure can be traced by the absorption profiles they imprint onto the spectra of densely sampled bright background sources, such as quasars and even the more numerous star-forming galaxies (Figure~\ref{f:tomo}). Upcoming new facilities will provide enough background sources to enable the 3D reconstruction of the filaments in the cosmic web. Amongs these, the Subaru {\it PFS} will provide spectroscopy of $2 < z < 3$ galaxies over a large volume. In the future, {\it ELTs} will offer the collecting area for instruments like {\it MOSAIC} \cite{Pello24} to enlarge these types of studies.

\subsection{Intensity Mapping}
\label{subsec:IM}

\begin{figure}[t]
\centering
\includegraphics[width=\linewidth]{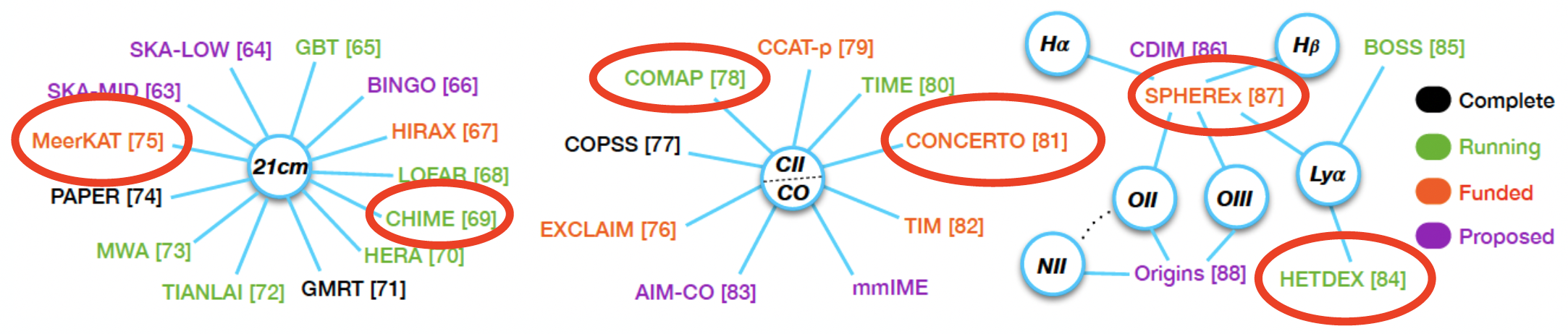}
\caption{Current and future intensity mapping experiments. The figure displays the facilities according to the targeted emission line: 21cm at radio frequencies, [CII] and/or CO in the sub-mm and traditional optical emission lines including Ly$\alpha$, H$\alpha$ and [OIII] \cite{kovetz2017}. On-going experiments are circled in red. In the near future, measurements of the global atomic, molecular and ionised gas mass based on these techniques should become available. }
\label{f:IM}
\end{figure}

Intensity mapping measures the spatial fluctuations in the integrated emission
from spectral lines originating from many individually unresolved objects. The line wavelength dependence can be used to measure the redshift distribution of the line emission along the line-of-sight, but high angular resolution is not required which means the technique can be used to survey large areas. Thus intensity mapping will enable to probe large volumes so as to map the distribution of densities on cosmological scales. The resulting power spectrum for the intensity map is composed of two components: one from the clustering of galaxies on large scales and a second that arises from the scale-independent shot noise, which dominates on small scales.
\cite{kovetz2017, Bernal22} summarize current and future intensity mapping experiments. Figure~\ref{f:IM} describes a number of existing and planned facilities according to the targeted emission line: 21cm at radio frequencies (MeerKAT, CHIME) \cite{chang2010, masui2013}, [CII] and/or CO in the sub-mm (CONCERTO, COMAP) \cite{keating2015, Pullen18, vanCuyck23} and traditional optical emission lines including Ly$\alpha$, H$\alpha$ and [OIII] (HETDEX, SPHEREx). In the near future, measurements of the global atomic, molecular and ionised gas mass using such techniques should become available.

%%%%%%%%%%%%%%%%%
%%%%%%%%%%%%%%%%%
%SECTION 6
%%%%%%%%%%%%%%%%%
%%%%%%%%%%%%%%%%%

\clearpage
\section{Open Questions \& Conclusions}
\label{sec:qu}

\subsection{The Global Baryon Cycle}
\label{subsec:global_qu}

The box below summarises some of the important questions still to be addressed to reach a full understanding the global baryon cycle on cosmological scales. 

\begin{question}{Questions}
\begin{itemize}

\item How to measure neutral gas content at z$>$5.5? How much atomic HI gas do high-redshift galaxies contain? What are the direct tracers of molecular gas? How to characterise the ionised part of baryons?

\item How to close the metal and dust budgets at all redshifts? Can we reproduce them in simulations?

\item How to localise and characterise the physical properties of the hidden baryons?

\item How to relate global and local scales: which galaxies contribute most to the mass densities budget? How does this fraction evolve with cosmic time?

\item How to find pristine gas? Can we identify signature of Population III stars?

\end{itemize}
\end{question}
%\eject%

\subsection{The Galactic Baryon Cycle}
\label{subsec:local_qu}

The box below provides a set of opened questions yet to be answered to draw a complete picture of the local baryon cycle taking place on galaxy scales.

\begin{question}{Questions}
\begin{itemize}

\item What are the physical properties of gas outflows in haloes of all masses?

\item What are the observational evidence for gas accretion?

\item Is there a metallicity and magnetic field vs. azimuthal angle signature as expected from some simulations?

\item How to map the CGM, its gas, metal and dust content?

\item How to observationally quantify the mass, metal, energy and momentum of gas flows?

\item How to estimate the contribution of the hot gas to the baryon budget?

\item How to simulate the coldest phase of the baryons (atomic and molecular) in cosmological context, the so-called "molecular cosmology"?

\end{itemize}
\end{question}

\subsection{Conclusions}
\label{subsec:ccl}

This chapter summarises some of the efforts to advance our understanding of both galaxy formation and cosmology by focusing on the physics of the gas component, bridging the largest to the smallest scales. The life cycle of galaxies is indeed regulated by the gaseous reservoirs of their interstellar and circumgalactic media, through gas collapse processes from large (cosmological) to small (microscopic) scales. During these processes, the gas transitions through a number of distinct phases, from the hot and diffuse to ultimatel a molecular phase. Each of these phases have a unique observational signatures, from the radio and sub-mm through the UV, optical, infrared to the X-ray.

The main observational probes of the intergalactic gas are the absorption lines imprinted in the spectra of unrelated high-redshift background sources due to the scattering of \lya\ photons by intervening intergalactic and circumgalactic neutral hydrogen. Conveniently, the properties of the observed absorption lines are directly related to the physical state and chemistry of the gas. Additionally, the gas fluctuations trace the underlying perturbations of the dark matter and therefore can be used to constrain cosmological model parameters and the nature of the dark matter itself. In addition to the hydrogen component, these techniques probe a wide range of metal ions and molecules. Recently, the observational campaign to map out the distribution of \lya\ and metal absorbers in statistical samples, their evolution and their affiliation with galaxies and larger scale environment has made startling progress. Working closely with  the observations, numerical simulations of galaxy formation and evolution in a cosmological context have equally grown increasingly refined. Of specific interest is the modelling of the diffuse gas near galaxies, the circumgalactic medium, as it is the interface where the intergalactic medium flows into galaxies and is impacted by their winds and jets in a dynamic baryon cycle. 

Today, the wealth of observational information has provided us with an increasingly accurate understanding of this so-called baryon cycle. It is of uttermost important to study the physical processes by which gas travels into, through, and out of galaxies. In the last few decades astrophysicts have been able to put together a solid framework -- based on excellent observations and simulations of structure build-up across cosmic time -- which describes the broad picture of the evolution of galaxies and their surrounding gas. These studies are essential to understand the growth of structures in the Universe and hence impact the fields of astrophysics, cosmology and fundamental physics.

\begin{svgraybox}
{\bf Reading list}:\\
Tumlison, Peeples \& Werk, 2017, ARAA \cite{Tumlinson17}\\
Prochaska, 2019, Saas-Fee Lectures \cite{Prochaska19}\\
Peroux \& Howk, 2020, ARAA \cite{PerouxHowk20}\\
Tacconi, Genzel \& Sternberg, 2020, ARAA \cite{Tacconi2020}\\
Faucher-Giguere \& Peng, 2023, ARAA \cite{FaucherGiguere23}
\end{svgraybox}

\vspace{-0.3cm}
\begin{acknowledgement}
The authors are thankful to the 52$^{\rm nd}$ Saas-Fee Advanced Course organisers and fellow lectures for putting together an exciting program. All participants to the school are acknowledged for their responsive welcome to these lectures and hands-on sessions. We are grateful to Julien, Pierre and the hotel staff for support. Thanks to all for white powder fun. 
\end{acknowledgement}

\newpage
\bibliographystyle{styles/spphys} % - numbers, I'm not big fan
\bibliography{biblio.bib} 
%\printbibliography

\end{document}